\documentclass[a4paper,11pt]{article}
\usepackage[utf8]{inputenc}
\usepackage[margin=2.5cm]{geometry}
\usepackage[style=nature]{biblatex}
\usepackage{xcolor}
\usepackage{graphicx}
\usepackage{amsmath,amssymb}
\usepackage{amsfonts}
\usepackage[nice]{nicefrac}
\usepackage[title]{appendix}
\usepackage{hyperref}
\usepackage{authblk}
\usepackage[misc]{ifsym}

\usepackage{multirow}
\usepackage{booktabs}

\title{Fluid-driven slow slip and earthquake nucleation on a slip-weakening circular fault}

\author[$a$]{Alexis S\'aez$^\star$}
\author[$a$]{Brice Lecampion}
\affil[$a$]{Ecole Polytechnique F\'ed\'erale de Lausanne (EPFL), Institute of Civil Engineering, Gaznat chair on Geo-Energy, CH-1015 Lausanne, Switzerland\\
\begin{center}
    $^\star$E-mail: \href{mailto:alexis.saez@epfl.ch}{alexis.saez@epfl.ch}
\end{center}}

\date{}

\begin{document}

\setlength{\parskip}{5pt}

\maketitle



\begin{abstract}
    Following the work of Sáez \textit{et al}. \cite{Saez_Lecampion_2022} who examined the three-dimensional propagation of injection-induced stable frictional sliding on a constant-friction fault interface separating two identical elastic solids, we extend their model to account for a friction coefficient that weakens with slip. This enables the model to develop a proper cohesive zone besides incorporating a finite amount of fracture energy, both ingredients absent in the former model. To do so, we consider two friction laws characterized by a linear and an exponential weakening of friction respectively. We focus on the particular case of axisymmetric circular shear ruptures as they capture the most essential aspects of the dynamics of unbounded ruptures in three dimensions. It is shown that fluid-driven slow slip can occur in two distinct modes in this model: as an interfacial rupture that is unconditionally stable, or as the quasi-static nucleation phase of an otherwise dynamic rupture. Whether the interface slides in one way or the other depends primarily on the sign of the difference between the initial shear stress ($\tau_0$) and the in-situ residual strength ($\tau_r^0$) of the fault. For ruptures that are unconditionally stable ($\tau_0<\tau_r^0$), fault slip undergoes four distinct stages in time. Initially, ruptures are self-similar in a diffusive manner and the fault interface behaves as if it were governed by a constant friction coefficient equal to the peak (static) friction value. Slip then accelerates due to frictional weakening while the cohesive zone develops. Once the latter gets properly localized, a finite amount of fracture energy emerges along the interface and the rupture dynamics is governed by an energy balance of the Griffith's type. We show that in this stage, fault slip always transition from a large-toughness to a small-toughness regime due to the diminishing effect of the fracture energy in the near-front energy budget as the rupture grows. Moreover, while slip grows likely confined within the pressurized region in prior stages, here the rupture front can largely outpace the pressurization front if the fault is close to the stability limit ($\tau_0\approx\tau_r^0$). Ultimately, self-similarity is recovered and the fault behaves again as possessing a constant friction coefficient, but this time equal to the residual (dynamic) friction value. It is shown that in this ultimate regime, the fault interface operates to leading order with zero fracture energy. On the other hand, when slow slip propagates as the nucleation phase of a dynamic rupture ($\tau_0>\tau_r^0$), fault slip also initiates in a self-similar manner and the interface operates at a constant peak friction coefficient. The maximum size that aseismic ruptures can reach before becoming unstable (inertially dominated) can be as small as a critical nucleation radius equal to the shear modulus divided by the slip-weakening rate, and as large as infinity when faults are close to the stability limit ($\tau_0\approx\tau_r^0$). The former case corresponds to faults that are critically stressed before the injection starts, in which case ruptures always expand much further away than the pressurized region. The larger the critical nucleation radius is with regard to the cohesive zone size, the longer ruptures can accelerate aseismically before becoming unstable. When the nucleation radius is smaller than the cohesive zone size, aseismic ruptures accelerate upon departing from the self-similar response due to continuous frictional weakening over the entire slipping region, undergoing nucleation unaffected by the residual fault strength. Conversely, when the nucleation radius is (much) larger than the cohesive zone size, aseismic ruptures transition towards a stage controlled by a front-localized energy balance and undergo nucleation in a `crack-like' manner. Our results include analytical and numerical solutions for the problem solved over its full dimensionless parameter space, as well as expressions for relevant length and time scales characterizing the transition between different stages and regimes. Due to its three-dimensional nature, the model enables quantitative comparisons with field observations as well as preliminary engineering design of hydraulic stimulation operations. Existing laboratory and in-situ experiments of fluid injection are briefly discussed in the light of our results.
\end{abstract}


\section{Introduction}\label{sec:introduction}

Sudden pressurization of pore fluids in the Earth's crust has been widely acknowledged as a trigger for inducing slow slip on pre-existing fractures and faults \cite{Hamilton_Meehan_1971,Scotti_Cornet_1994,Guglielmi_Cappa_2015,Wei_Avouac_2015}. Sometimes referred to as injection-induced aseismic slip, this phenomenon is thought to play a significant role in various subsurface engineering technologies and natural earthquake-related phenomena. Notable examples of the natural source include seismic swarms and aftershock sequences, often attributed to be driven by the diffusion of pore pressure \cite{Parotidis_Shapiro_2005,Miller_Collettini_2004} or the propagation of slow slip \cite{Lohman_McGuire_2007,Perfettini_Avouac_2007}, with recent studies suggesting that the interplay between both mechanisms may be indeed responsible for the occurrence of some seismic sequences \cite{Ross_Rollins_2017,Yukutake_Yoshida_2022,Sirorattanakul_Ross_2022}. Similarly, low-frequency earthquakes and tectonic tremors are commonly considered to be driven by slow slip events occurring downdip the seismogenic zone in subduction zones \cite{Rogers_Dragert_2003,Shelly_Beroza_2006}, where systematic evidence of overpressurized fluids has been found \cite{Shelly_Beroza_2006,Kato_Iidaka_2010,Behr_Burgmann_2021}, with recent works suggesting that the episodicity and some characteristics of slow slip events may be explained by fluid-driven processes \cite{Warren-Smith_Fry_2019,Zhu_Allison_2020,Perez-Silva_Kaneko_2023}. 
 
Anthropogenic fluid injections are, on the other hand, known to induce both seismic and aseismic slip \cite{Scotti_Cornet_1994,Wei_Avouac_2015,Guglielmi_Cappa_2015}. For instance, hydraulic stimulation techniques employed to engineer deep geothermal reservoirs aim to reactivate fractures through shear slip, thereby enhancing reservoir permeability by either dilating pre-existing fractures or creating new ones. The occurrence of predominantly aseismic rather than seismic slip, is considered a highly favorable outcome, as earthquakes of relatively large magnitudes can pose a significant risk to the success of these projects \cite{Deichmann_Giardini_2009,Ellsworth_Giardini_2019}. Injection-induced aseismic slip can, however, play a rather detrimental role in some cases, as slow slip is accompanied by quasi-static changes of stress in the surrounding rock mass which, in turn, may induce failure of unstable fault patches that could sometimes lead to earthquakes of undesirably large magnitude \cite{Eyre_Eaton_2019}. Moreover, since injection-induced aseismic slip may propagate faster than pore pressure diffusion, this mechanism can potentially trigger seismic events in regions that are far from the zone affected by the pressurization of pore fluids \cite{Guglielmi_Cappa_2015,Bhattacharya_Viesca_2019,Eyre_Eaton_2019}. Fluid-driven aseismic slip may play a similar role in other subsurface engineering technologies than deep geothermal energy, such as hydraulic fracturing of unconventional oil and gas reservoirs \cite{Eyre_Eaton_2019}, oil wastewater disposal \cite{Chen_Nakata_2017}, and carbon dioxide sequestration \cite{Zoback_Gorelick_2012}.

The apparent relevance of injection-induced aseismic slip in the aforementioned phenomena have motivated the development of physical models that are contributing to a better comprehension of this hydro-mechanical problem. The first rigorous investigations on the mechanics of injection-induced aseismic slip focused, for the sake of simplicity, on idealized two-dimensional configurations. Specifically, on the propagation of fault slip under plane-strain conditions considering either in-plane shear (mode II) or anti-plane shear (mode III) ruptures, with a fluid source of infinite extent along the out-of-the-plane direction. Yet these studies have significantly contributed to establish a fundamental qualitative understanding of how the initial state of stress, the fluid injection parameters, the fault hydraulic properties, and the fault frictional rheology affect the dynamics of fluid-driven aseismic slip transients \cite{Dublanchet_2019,Garagash_2021,Viesca_2021,Yang_Dunham_2021}, the applicability of such models remains limited as three-dimensional configurations are expected to prevail in nature. Recently, Sáez \textit{et al.} \cite{Saez_Lecampion_2022} examined the propagation of injection-induced aseismic slip under mixed-mode (II+III) conditions, on a fault embedded in a fully three-dimensional domain. An important finding of this study is that for the same type of fluid source (either constant injection rate or constant pressure in \cite{Saez_Lecampion_2022}), the spatiotemporal patterns of fault slip differ even qualitatively between the three-dimensional model and its two-dimensional counterpart, highlighting the importance of resolving more realistic rupture configurations. 

In the three-dimensional model of Sáez \textit{et al.} \cite{Saez_Lecampion_2022}, the perhaps strongest assumption is the consideration of a constant friction coefficient at the fault interface. This friction model, known as Coulomb's friction, corresponds to the minimal physical ingredient that can produce unconditionally stable shear ruptures. As discussed by Sáez \textit{et al.} \cite{Saez_Lecampion_2022}, a model with Coulomb's friction represents a case in which the frictional fracture energy spent during rupture propagation is effectively zero, without the possibility of developing a process zone in the proximities of the rupture front. In this paper, we eliminate this assumption and therefore extend the model of Sáez \textit{et al.} \cite{Saez_Lecampion_2022} to account for a friction coefficient that weakens upon the onset of fault slip. This incorporates into the model the proper growth and localization of a process zone, with the resulting finite amount of fracture energy. We do so by considering the simplest model of friction that can provide the sought physical ingredients, namely, a slip-weakening friction coefficient \cite{Ida_1972}. We consider the two most common types of slip-weakening friction: a linear and an exponential decay of friction with slip, from some peak (static) value towards a constant residual (dynamic) one.

On the other hand, as shown by \textcite{Saez_Lecampion_2022} for the Coulomb's friction case, a Poisson's ratio different than zero has mainly an effect on the aspect ratio of the resulting quasi-elliptical ruptures which become more elongated for increasing values of $\nu$. The characteristic size of mixed-mode, quasi-elliptical ruptures is nevertheless determined primarily by the rupture radius of circular ruptures, which occur in the limit of $\nu=0$ for such an axisymmetric problem. The case of a null Poisson's ratio is therefore particularly insightful and notably simpler since in that limit, we can leverage the axisymmetry property of the problem to compute more efficient numerical solutions \cite{Bhattacharya_Viesca_2019,Saez_Lecampion_2022,Saez_Lecampion_2023} besides allowing the problem to be tractable analytically to some extent. We therefore focus in this paper on the case of mixed-mode circular ruptures alone. We also note that our model can be considered as an extension of the two-dimensional model of Garagash and Germanovich \cite{Garagash_Germanovich_2012}. While Garagash and Germanovich focused their investigation on the problem of nucleation and arrest of dynamic slip, our work here is concerned primarily with a different phenomenon, namely, the propagation of aseismic slip. Nonetheless, since aseismic slip could correspond indeed to the nucleation phase of an ensuing dynamic rupture, we also examine the problem of nucleation of a dynamic instability under mixed-mode conditions. In fact, by pursuing this route, we provide an extension of the nucleation length of Uenishi and Rice \cite{Uenishi_Rice_2002} to the three-dimensional, axisymmetric case, for both tensile and shear ruptures. Similarly, we extend some other relevant nucleation lengths identified by Garagash and Germanovich \cite{Garagash_Germanovich_2012}. 

We organize this paper as follows. In section \ref{sec:problem-formulation}, we introduce the mathematical formulation of our physical model. In section \ref{sec:two-simplified-models}, we present two simplified models that will be later shown to be asymptotic and/or approximate solutions of the slip-weakening model under certain regimes. In section \ref{sec:scaling-map-regimes}, we introduce the scaling of the problem and the map of possible rupture regimes. In section \ref{sec:stable-ruptures}, we examine in detail the case of ruptures that are ultimately stable. In section \ref{sec:nucleation-phase}, we focus on the case in which aseismic slip corresponds to the nucleation phase of an otherwise dynamic rupture. Finally, in section \ref{sec:discussion}, we provide a brief discussion of recent laboratory and in-situ experiments of fault reactivation by fluid injection in light of our results.

\section{Problem formulation}\label{sec:problem-formulation}

\subsection{Governing equations}\label{sec:governing-equations}

\begin{figure}
    \centering
    \includegraphics[width=14cm]{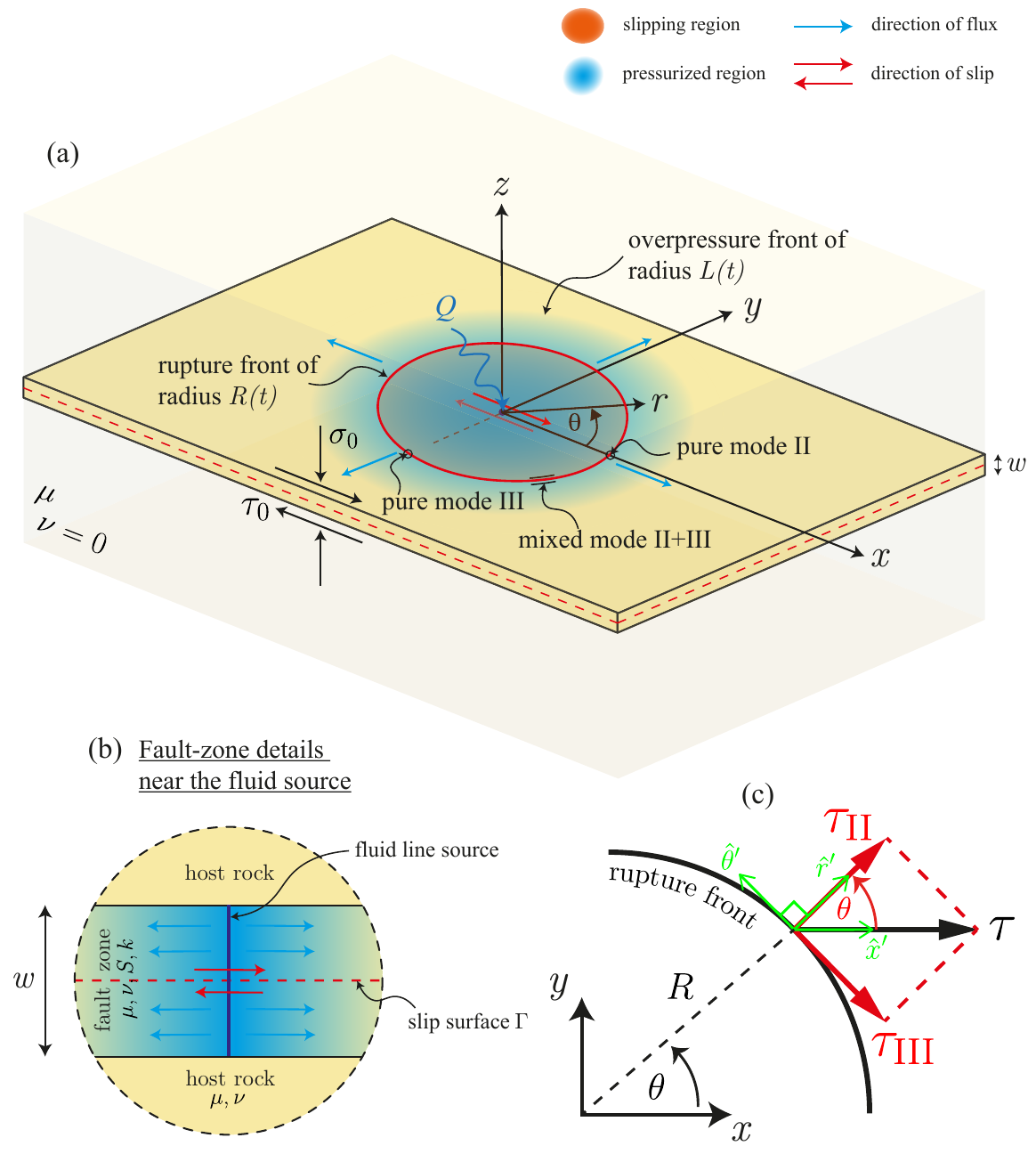}
    \caption{Model schematics. (a, b) Fluid is injected into a permeable fault zone of width $w$ via a line-source that crosses the entire fault zone width. The fault is planar and embedded in an unbounded linearly elastic impermeable host rock of same elastic constants. The initial stress tensor is uniform. The resulting mixed-mode (II+III) shear rupture is circular when Poisson's ratio $\nu=0$. (c) Direction of shear stress $\tau$ along the rupture front and the corresponding mode-II and mode-III components with regard to both the Cartesian and cylindrical coordinate systems.}
    \label{fig:model-schematics}
\end{figure}

Fluid is injected into a poroelastic fault zone of width $w$ that is characterized by an intrinsic permeability $k$ and a storage coefficient $S$, assumed to be constant and uniform (see figure \ref{fig:model-schematics}b). The fault zone is confined within two linearly elastic half spaces of same elastic constants, namely, a shear modulus $\mu$ and Poisson's ratio $\nu$. The initial stress tensor is assumed to be uniform and is characterized by a resolved shear stress $\tau_{0}$ and total normal stress $\sigma_{0}$ acting along the \textit{x} and \textit{z} directions of the Cartesian reference system of figure \ref{fig:model-schematics}a, respectively. We consider the injection of fluids via a line source that is located along the \textit{z} axis and crosses the entire fault zone width. Under such conditions, fluid flow is axisymmetric with regard to the \textit{z} axis and occurs only within the porous fault zone. Moreover, the displacement field induced by the fluid injection is irrotational and the pore pressure diffusion equation of poroelasticity reduces to its uncoupled version \cite{Marck_Savitski_2015}, $\partial p/\partial t=\alpha\nabla^{2}p$, where $\alpha=k/S\eta$ is the fault hydraulic diffusivity, with $\eta$ the fluid dynamic viscosity. Solutions of the previous linear diffusion equation are known extensively for a broad range of boundary and initial conditions \cite{Carslaw_Jaeger_1959}. Here, we focus on the perhaps most practical case in which the fluid injection is conducted at a constant volumetric rate $Q$. For the following boundary conditions: $2\pi r w (k/\eta) \partial p / \partial r=-Q$ when $r\to0$ and $p=p_0$ when $r\to\infty$, with $p_0$ the initial pore pressure field assumed to be uniform, the solution of the diffusion equation in terms of the overpressure $\Delta p(r,t)=p(r,t)-p_0$ reads as (section 10.4, eq. 5, \cite{Carslaw_Jaeger_1959})
\begin{equation}\label{eq:p-solution}
    \Delta p(r,t)=\Delta p_* E_1\left(\frac{r^2}{4\alpha t}\right), \; \text{with} \; \Delta p_{*}=\frac{Q\eta}{4\pi kw },
\end{equation}
where $\Delta p_*$ is the intensity of the injection with units of pressure, and $E_{1}\left(x\right)=\int_{x}^{\infty}\left(e^{-x\xi}/\xi\right)\textrm{d}\xi$ is the exponential integral function. 

Let us define the following characteristic overpressure,
\begin{equation}\label{eq:pc-definition}
    \Delta p_{c}=\frac{Q\eta}{kw},
\end{equation}
which relates to the injection intensity as $\Delta p_c=4\pi\Delta p_*$. A close examination of equation \eqref{eq:p-solution} for the large times in which the line-source approximation is valid, $t\gg r_s^2/\alpha$ with $r_s$ the characteristic size of the actual fluid source, shows that $\Delta p_c$ is in the order of magnitude of the overpressure at the fluid source. Moreover, as discussed in Appendix \ref{appendix:line-source}, the fluid-source overpressure, say $\Delta p_s$, increases slowly (logarithmically) with time, such that for practical applications, one could think in considering $\Delta p_s$ to be rather constant and approximately equal to the characteristic overpressure $\Delta p_c$. Because of its simplicity, we adopt such approximation $\Delta p_s\approx \Delta p_c$ throughout this work, with the implications further discussed in Appendix \ref{appendix:line-source}. Having established that, we note that in this work we consider exclusively injection scenarios in which the characteristic fluid-source overpressure satisfies $\Delta p_c \lessapprox \sigma_0^\prime$, where $\sigma_0^\prime=\sigma_0-p_0$ is the initial effective normal stress. In this way, we make sure that the walls of the fault remain always in contact, thus avoiding hydraulic fracturing. This latter scenario has been notably addressed in the context of slip instabilities by others \cite{Azad_Garagash_2017}.

\begin{figure}
    \centering
    \includegraphics[width=10 cm]{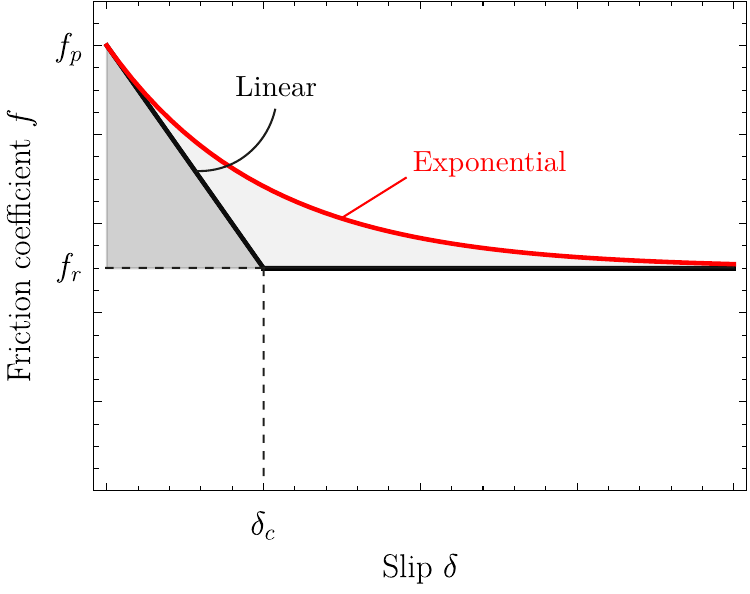}
    \caption{Slip weakening friction law for (black) linear weakening and (red) exponential decay with slip. If the friction coefficient $f$ is multiplied by some constant effective normal stress that is approximately uniform and representative of the one acting along the process zone, then the gray areas times such effective normal stress represent fracture energy $G_c$. Note that $G_{c}^{\text{exp}}=2\cdot G_{c}^{\text{lin}}$ when $\delta_c$ is the same in both models.}
    \label{fig:friction-laws}
\end{figure}

Suppose now that the fault zone possesses a slip surface located at $z=0$ where the totality of fault slip is accommodated (see figure \ref{fig:model-schematics}b). The slip surface is assumed to obey a Mohr-Coulomb shear failure criterion without any cohesion such that the maximum shear stress $\tau$ and fault strength $\tau_s$ satisfy at any position along the slip surface and any time, the following local relation
\begin{equation}\label{eq:shear-strength}
|\tau|\leq \tau_s=f\times\left(\sigma_0^\prime-\Delta p\right),
\end{equation}
where $\Delta p$ is the overpressure given by equation \eqref{eq:p-solution} and $f$ is a local friction coefficient that depends on fault slip $\delta$ \cite{Ida_1972}. There are two common choices for the slip-weakening friction model, namely, a friction coefficient that decays linearly with slip, 
\begin{equation}\label{eq:linear-weakening-friction}
    f(\delta)=
    \begin{cases}
        f_p-\left(f_p-f_r\right)|\delta|/\delta_c & \text{if } |\delta| \leq \delta_c\\
        f_r & \text{if } |\delta| > \delta_c,
    \end{cases}
\end{equation}
and a friction coefficient that decays exponentially with it, 
\begin{equation}\label{eq:exponential-weakening-friction}
    f(\delta)=f_r+\left(f_p-f_r\right)e^{-|\delta|/\delta_c}.
\end{equation}
In the previous equations, $f_p$ is the peak (or static) friction coefficient, $f_r$ is the residual (or kinetic) friction coefficient, and $\delta_c$ is the characteristic `distance' over which the friction coefficient decays from $f_p$ to $f_r$, as displayed in figure \ref{fig:friction-laws}.
Moreover, we assume that the slip surface is fully locked before the injection starts and, as such, the initial shear stress $\tau_0$ must be lower than the in-situ static strength of the fault, $\tau_p^0=f_p \sigma_0^\prime$. 

According to equation \eqref{eq:shear-strength}, the injection of fluid has the effect of reducing the fault strength owing to the increase of pore-fluid pressure which decreases the effective normal stress locally. Such pore pressure increase will be eventually sufficient to activate fault slip when the fault strength equates the pre-injection shear stress $\tau_0$, which marks the onset of the interfacial frictional rupture. Indeed, owing to the line-source approximation of the fluid source, the activation of slip in our model occurs immediately upon the start of the injection as a consequence of the weak logarithmic singularity that the exponential integral function features near the origin (see Appendix C). In the three-dimensional axisymmetric configuration under consideration, the resulting shear rupture will propagate under mixed-mode II+III conditions, where II and III represent the in-plane shear and anti-plane shear deformation modes, respectively. The modes of deformation are schematized in figure \ref{fig:model-schematics}c in terms of the near-front shear stress components. Moreover, as stated in the introduction, we restrict ourselves to the case of circular ruptures alone. Such an idealized case is exact for radial fluid flow when the Poisson's ratio $\nu=0$ \cite{Bhattacharya_Viesca_2019,Saez_Lecampion_2022}. Moreover, in some limiting regimes of the fault response, numerically-derived asymptotic expressions for the aspect ratio of elongated ruptures ($\nu\neq 0$) \cite{Saez_Lecampion_2022} may result useful to construct approximate solutions for the evolution of non-circular rupture fronts using the solution for circular ruptures, at least in the case of Coulomb's friction \cite{Saez_Lecampion_2022}.

By neglecting any fault-zone poroelastic coupling upon the onset of the rupture, the quasi-static elastic equilibrium that relates fault slip $\delta$ to the shear stress $\tau$ acting along the fault, can be written as the following boundary integral equation along the $x$ axis \cite{Salamon_Dundurs_1977,Bhattacharya_Viesca_2019},
\begin{equation}\label{eq:elastic-equilibrium}
    \tau(r,t)=
    \tau_{0}+
    \frac{\mu}{2\pi} \int_{0}^{R(t)}F\left(r,\xi\right)\frac{\partial\delta(\xi,t)}{\partial \xi}\textrm{d}\xi,
\end{equation} 
where the kernel $F\left(r,\xi\right)$ is given by
\begin{equation}\label{eq:kernel-F}
    F\left(r,\xi\right)=\frac{K\left(k(r/\xi)\right)}{\xi+r}+\frac{E\left(k(r/\xi)\right)}{\xi-r},\;\text{with}\quad k(x)=\frac{2\sqrt{x}}{1+x},
\end{equation}
and $K\left( \cdot \right)$ and $E\left( \cdot \right)$ are the complete elliptic integrals of the first and second kind, respectively. Note that in equation \eqref{eq:elastic-equilibrium}, the shear stress $\tau$ can be written as a function in space of the radial coordinate only due to the aforementioned axisymmetry property when $\nu=0$. Equations \eqref{eq:p-solution}, \eqref{eq:shear-strength}, and \eqref{eq:elastic-equilibrium}, plus the corresponding constitutive friction law, either \eqref{eq:linear-weakening-friction} or \eqref{eq:exponential-weakening-friction}, provide a complete system of equations to solve for the spatio-temporal evolution of fault slip $\delta(r,t)$ and the position of the rupture front $R(t)$: our primary unknowns in the problem.

\subsection{Front-localized energy balance}

Under certain conditions and over a certain spatial range, the previous problem may be formulated equivalently through an energy balance of the Griffith type, that is, as a classical shear crack in the theory of Linear Elastic Fracture Mechanics (LEFM) \cite{Broberg_1999}. The idea that frictional ruptures could be approximated by classical shear cracks is relatively old \cite{Ida_1972,Palmer_Rice_1973}. Yet it has been just recently validated by modern experiments concern with the problem of frictional motion on both dry and lubricated interfaces \cite{Svetlizky_Bayart_2019}.

Let us assume that there exists a localization length $\ell_*$ near the rupture front such that the shear stress evolves from some peak value $\tau_p$ at the rupture front $r=R$ to some residual and approximately constant amount $\tau_r$ at distances $r<R-\ell_*$. The localization length $\ell_*$ is sometimes called the process zone size or cohesive zone size for the similarity of the shear rupture problem to the case of tensile fractures \cite{Barenblatt_1962}. Let us further assume that $\ell_*$ is small in comparison to the rupture radius $R$. Under such conditions, we can invoke the `small-scale yielding' approximation of LEFM to shear ruptures \cite{Palmer_Rice_1973}. In particular, during rupture propagation, the influx of elastic energy into the edge region $G$, also known as energy release rate, must equal the frictional fracture energy $G_c$. The energy release rate for an axisymmetric, circular shear rupture is (Appendix B of \cite{Saez_Lecampion_2022})
\begin{equation}\label{eq:G-circular}
    G=\frac{2}{\pi \mu R(t)} \left[\int_{0}^{R(t)}\frac{\tau_0-\tau_{r}(r,t)}{\sqrt{R(t)^2-r^2}}r \textrm{d}r\right]^2.
\end{equation}
In the previous equation, it is assumed that the current shear stress acting on the `crack' faces is the residual strength $\tau_r$. This is consistent with the small-scale yielding approach where the details of the process zone are neglected for the calculation of $G$ \cite{Rice_1968,Palmer_Rice_1973}. On the other hand, the fracture energy $G_c$ corresponds to the energy dissipated within the process zone per unit area of rupture growth, which in the case of a frictional shear crack is equal to the work done by the fault strength $\tau_s$ against its residual part $\tau_r$ \cite{Palmer_Rice_1973},
\begin{equation}\label{eq:Gc-definition}
    G_c=\int_0^{\delta_*} \left[\tau_s(\delta)-\tau_r(\delta_*)\right]\text{d}\delta,
\end{equation}
where $\delta_*$ is the accrued slip throughout the process zone assuming, again, that there exists a proper localization length $\ell_*$. 

Let us now recast the Griffith's energy balance, $G=G_c$, in a way that will be more convenient for analytic derivations. For a mixed-mode shear rupture, the energy release rate can be expressed as $G=K_{\text{II}}^2(1-\nu)/2\mu+K_{\text{III}}^2/2\mu$ \cite{Irwin_1958}, where $K_{\text{II}}$ and $K_{\text{III}}$ are the mode-II and mode-III stress intensity factors. Here, $K_{\text{II}}$ and $K_{\text{III}}$ are understood as the intensities of the singular fields of LEFM that emerge as intermediate asymptotics at distances $\ell_*\ll r\ll R$. Moreover, since $\nu=0$, we can conveniently define 
\begin{equation}\label{eq:K-equivalent}
    K^2=K_{II}^2+K_{III}^2=2\mu G,\quad \text{and} \quad K_c=\sqrt{2\mu G_c},
\end{equation}
where $K$ is an `axisymmetric stress-intensity factor' and $K_c$ an `axisymmetric fracture toughness'. Note that if one considers the singular terms of the mode-II and mode-III shear stress components acting nearby and ahead of the rupture front, say $\tau_{\text{II}}$ and $\tau_{\text{III}}$ respectively (see figure \ref{fig:model-schematics}c), then $K$ represents the intensity of the square-root singularity associated with the absolute (maximum) shear stress $\tau$ (acting along the $x$-direction of our Cartesian reference system) which relates to the in-plane shear and anti-plane shear stress components as $\tau^2=\tau_{\text{II}}^2+\tau_{\text{III}}^2$.

By combining equations \eqref{eq:G-circular} and \eqref{eq:K-equivalent}, we can rewrite the Griffith's energy balance in the sought Irwin's form, $K=K_c$, which yields
\begin{equation}\label{eq:K-Kc}
    \frac{2}{\sqrt{\pi R(t)}} \int_{0}^{R(t)}\frac{\tau_0-\tau_r(r,t)}{\sqrt{R(t)^2-r^2}}r \textrm{d}r=K_c.
\end{equation}

Let us now consider some details about the calculation of $G_c$ in equation \eqref{eq:Gc-definition}. Upon the arrival of the rupture front at a certain location over the fault plane, the fault strength $\tau_s$ weakens according to equation \eqref{eq:shear-strength} due to both the decrease of the friction coefficient $f$ and the increase of overpressure $\Delta p$. However, the near-front processes associated with energy dissipation for rupture growth in equation \eqref{eq:Gc-definition} are for the most part related to the frictional process only. This can be readily seen after closely examining equation \eqref{eq:p-solution} for the spatio-temporal evolution of $\Delta p$. Equation \eqref{eq:p-solution} introduces the well-known diffusion length scale $L(t)=\sqrt{4\alpha t}$ in the problem, which is itself a proxy for the radius of the nominal area affected by the pressurization of pore fluid (see figure \ref{fig:model-schematics}a). 
We note that if the overpressure front $L$ is either in the order of or much greater than the rupture radius $R$, then $\Delta p$ varies smoothly over the process zone ---whose length $\ell_*$ is at this point by definition much smaller than $R$. On the other hand, if $L$ is much smaller than $R$, then $\Delta p$ varies abruptly over the slipping region but highly localized near the rupture center over a small zone that is far away from the process zone, such that the overpressure within the process zone is negligibly small. Hence, the overpressure has the simple role of approximately setting the current amount of effective normal stress within the process zone, which can be reasonably taken as uniform at a given time and evaluated at the rupture front, $\Delta p(R,t)$. The fracture energy $G_c$ can be thus approximated as
\begin{equation}\label{eq:Gc-approximation}
    G_c\approx\left[\sigma_0^\prime-\Delta p(R,t)\right]\int_0^{\delta_*} \left[f(\delta)-f_r\right]\text{d}\delta.
\end{equation} 
Note that in addition to the separation of scales between the diffusion of pore pressure and the frictional weakening process just discussed, the previous equation assumes that the friction coefficient itself must effectively evolve throughout $\ell_*$ (or equivalently over $\delta_*$) towards an approximately constant residual value $f_r(\delta_*)=f_r$. This latter is guaranteed in the linear-weakening model \eqref{eq:linear-weakening-friction} as $\delta_*=\delta_c$, and it seems a good approximation for the exponential-weakening case \eqref{eq:exponential-weakening-friction} at some $\delta_*>\delta_c$. Assuming this latter separation of scales too, the residual strength of the fault can be then written as
\begin{equation}
    \tau_r(r,t)=f_r\times\left[\sigma_0^\prime-\Delta p(r,t)\right].
\end{equation}
Substituting the previous equation into \eqref{eq:K-Kc} leads after some manipulations to an energy-based equation for the rupture front $R(t)$,
\begin{equation}\label{eq:final-energy-balance}
    \underbrace{\frac{2}{\sqrt{\pi}}\frac{f_{r}\Delta p_{*}}{\sqrt{R}}\int_{0}^{R}\frac{E_1\left(r^2/L^2\right)}{\sqrt{R^2-r^2}}r\textrm{d}r}_{K_{p}} +
    \overbrace{\frac{2}{\sqrt{\pi}}\left[\tau_{0}-f_{r}\sigma_{0}^{\prime}\right]\sqrt{R}}^{K_{\tau}}
    =
    K_c(R,t),
\end{equation}
where the explicit dependence of the rupture front $R$ and overpressure front $L$ on time $t$ has been omitted for simplicity.

Equation \eqref{eq:final-energy-balance} is an insightful form of the near-front energy balance, equivalent to the one obtained by Garagash and Germanovich \cite{Garagash_Germanovich_2012} in two dimensions. It shows that the instantaneous position of the rupture front is determined by the competition between three distinct `crack' processes that are active during the propagation of the rupture. The first term on the left-hand side $K_p$ is an axisymmetric stress intensity factor (SIF) associated with the equivalent shear load induced by the fluid injection alone, which continuously unclamps the fault. The second term on the left-hand side $K_\tau$ is an axisymmetric SIF due to a uniform shear load that is equal to the difference between the initial shear stress $\tau_0$ and the residual strength of the fault without any overpressure, $f_r\sigma_0^\prime$, commonly known as stress drop. Note that this term may be either positive or negative, whereas the first term is always positive
. Finally, the right-hand side of the near-front energy balance is associated with the fracture energy $G_c$ that is dissipated within the process zone or, in the Irwin's form, the fracture toughness $K_c$. Note that in our model, the fracture energy may generally depend on the rupture radius $R$ and time $t$ due to variations in overpressure (see equation \eqref{eq:Gc-approximation}). 

\section{Two simplified models}\label{sec:two-simplified-models}

\subsection{The constant friction model}\label{sec:Coulomb-friction}

The constant friction model is the simplest idealization of friction that produces fluid-driven stable frictional ruptures. It has been extensively studied in two-dimensional configurations for different injection scenarios \cite{Viesca_2021,Saez_Lecampion_2022} and more recently in the fully three-dimensional case \cite{Saez_Lecampion_2022}. Here, we briefly summarize the main characteristics of the circular rupture model \cite{Saez_Lecampion_2022}, whose results will be later put in a broader perspective in relation to the slip-weakening model.

S\'aez \textit{et al}. \cite{Saez_Lecampion_2022} showed that fault slip induced by injection at a constant volumetric rate is self-similar in a diffusive manner. The rupture radius $R(t)$ thus evolves simply as
\begin{equation}\label{eq:lambda-definition-1}
    R(t)=\lambda L(t),
\end{equation}
where $L(t)=\sqrt{4\alpha t}$ is the diffusion length scale and nominal position of the overpressure front, and $\lambda$ is the so-called amplification factor for which an analytical solution was derived \cite{Saez_Lecampion_2022} from the condition that the rupture grows with no stress singularity at its front,
\begin{equation}\label{eq:propagation-condition-Coulomb-friction}
    \int_0^1 \frac{E_1\left(\lambda\eta\right)}{\sqrt{1-\eta^2}}\eta\text{d}\eta = \mathcal{T},
\end{equation}
which leads after evaluating analytically the corresponding integral to
\begin{equation}\label{eq:analytical-solution-Coulomb-friction}
2-\gamma+\frac{2}{3}\lambda^{2}{}_{2}F_{2}\left[\begin{array}{cc}
1 & 1\\
2 & \nicefrac{5}{2}
\end{array};-\lambda^{2}\right]-\ln(4\lambda^{2})=\mathcal{T},
\end{equation}
where $\gamma=0.577216...$ is the Euler-Mascheroni's constant and $_{2}F_{2}\left[\;\right]$ is the generalized hypergeometric function. Note that $\lambda$ is function of a sole dimensionless number, the so-called stress-injection parameter
\begin{equation}\label{eq:T-definition}
    \mathcal{T}=\frac{f_{\text{cons}}\sigma_0^\prime-\tau_0}{f_{\text{cons}}\Delta p_*},
\end{equation}
where $f_{\text{cons}}$ is the constant friction coefficient. 

The parameter $\mathcal{T}$ is defined as the ratio between the amount of shear stress that is necessary to activate fault slip $f_{\text{cons}}\sigma_0^\prime-\tau_0$, and $f_{\text{cons}}\Delta p_*$ which quantifies the intensity of the fluid injection. $\mathcal{T}$ can vary in principle  between $0$ and $+\infty$ \cite{Saez_Lecampion_2022}. However, for practical purposes $\mathcal{T}$ is upper bounded. Indeed, as described previously, $\Delta p_*=\Delta p_c/4\pi$ with $\Delta p_c$ taken as approximately equal to the overpressure at the fluid source. Hence, one can estimate the minimum amount of overpressure that is required to activate fault slip as $f_{\text{cons}}\Delta p_c\approx f_{\text{cons}}\sigma_0^\prime-\tau_0$. Substituting the previous relation into equation \eqref{eq:T-definition} leads to an approximate upper bound for $\mathcal{T}\lessapprox10$, where we have approximated the factor $4\pi$ by $10$. The lower and upper bounds of $\mathcal{T}$ are associated with two end-member regimes that were first introduced by Garagash and Germanovich \cite{Garagash_Germanovich_2012}. When $\mathcal{T}$ is close to zero, $\tau_0\to f_{\text{cons}}\sigma_0^\prime$ and thus the fault is critically stressed or about to fail before the injection starts. On the other hand, when $\mathcal{T}\approx10$, the fault is `marginally pressurized' as the injection has provided just the minimum amount of overpressure to activate fault slip. 

The asymptotic behavior of $\lambda$ for the limiting values of $\mathcal{T}$ is particularly insightful. For critically stressed faults ($\mathcal{T}\ll 1$), the amplification factor turns out to be large ($\lambda\gg 1$) and thus the rupture front outpaces largely the fluid pressure front ($R(t)\gg L(t)$), whereas for marginally pressurized faults ($\mathcal{T}\sim 10$), the amplification factor is small ($\lambda\ll 1$) and thus the rupture front lags significantly the overpressure front ($R(t)\ll L(t)$). In the critically stressed limit, since $\lambda\gg1$, the equivalent shear load due to fluid injection can be approximated as a point force, 
\begin{equation}\label{eq:point-force}
    f_\text{cons}\Delta p(r,t) \approx f_\text{cons}\Delta P(t) \delta^{dirac}(r)/2\pi r,\text{ with  }\Delta P(t)=\Delta p_*\int_0^\infty E_1\left(\frac{r^2}{4\alpha t}\right)2\pi r\text{d}r=4\pi \alpha t \Delta p_*,
\end{equation}
whereas in the marginally pressurized limit, its asymptotic form comes simply from expanding the exponential integral function for small values of its argument as $\lambda\ll1$: $f_\text{cons}\Delta p(r,t) \approx -f_\text{cons}\Delta p_*\left[\text{ln}\left(r^2/4\alpha t\right)+\gamma\right]$. Substituting the previous asymptotic forms for the fluid-injection `forces' into the rupture propagation condition \eqref{eq:propagation-condition-Coulomb-friction}, leads to the following asymptotes for the amplification factor \cite{Saez_Lecampion_2022}: 
\begin{equation}\label{eq:lambda-asymptotes}
    \lambda\approx 
        \begin{cases}
        1/\sqrt{2\mathcal{T}} & \text{for critically stressed faults, }\mathcal{T}\ll 1,\\
        \frac{1}{2}\exp\left(\left[2-\gamma-\mathcal{T}\right]/2\right) & \text{for marginally pressurized faults, } \mathcal{T}\sim 10.
    \end{cases}
\end{equation}
Comparing the previous asymptotes with the exact solution \eqref{eq:analytical-solution-Coulomb-friction} suggests that the asymptotic approximation \eqref{eq:lambda-asymptotes} is accurate up to 5\% in the critically stressed and marginally pressurized regimes, for $\mathcal{T}\lessapprox 0.16$ and $\mathcal{T}\gtrapprox 2$, respectively.

\subsection{The constant fracture energy model}\label{sec:constant-fracture-energy}

Consider the simple case in which the fracture energy $G_c$ is constant (and so the fracture toughness $K_c$). Although this is an idealized scenario, it accounts already for one of the main ingredients of the slip-weakening friction model, that is, a finite fracture energy. At the same time, the constant fracture energy model allows us to examine quickly the different regimes of propagation that emerge from the competition between the three distinct terms that compose the front-localized energy balance \eqref{eq:final-energy-balance}. Furthermore, a constant fracture energy model will result to be an excellent approximation of some important rupture regimes in the slip-weakening model.

\subsubsection{Scaling and structure of the solution}

Similarly to the case of Coulomb's friction, let us define an amplification factor in the form
\begin{equation}\label{eq:lambda-time-dependent}
    \lambda(t)=\frac{R(t)}{L(t)}.
\end{equation}
Unlike the solution of the previous section that is self-similar, here the introduction of a finite fracture energy breaks the self-similarity of the problem and makes the solution for $\lambda$ be now time-dependent. 

Non-dimensionalization of the front-localized energy balance shows that the solution for the amplification factor $\lambda$ can be written as
\begin{equation}\label{eq:lambda-structure-Gc-constant}
    \lambda\left(\mathcal{T}_r,\mathcal{K}(t)\right),
\end{equation}
where $\mathcal{T}_r$ and $\mathcal{K}$ are two dimensionless parameters with the physical meanings that we explain below. Interestingly, the dependence of $\lambda$ on time is only in the second parameter $\mathcal{K}$. 

The first parameter $\mathcal{T}_r$ is a dimensionless number in the form
\begin{equation}\label{eq:residual-stress-injection-parameter}
    \mathcal{T}_r=\frac{f_r \sigma_0^\prime-\tau_0}{f_r\Delta p_*},
\end{equation}
which turns out to be identical to the stress-injection parameter of the constant friction model (equation \eqref{eq:T-definition}) except that the constant friction coefficient $f_{\text{cons}}$ is now the residual friction coefficient $f_r$. For this reason, we name it as the `residual' stress-injection parameter.

$\mathcal{T}_r$ quantifies the combined effect of the two equivalent shear loads that drive the propagation of the `fractured' slipping patch, namely, the uniform stress $f_r \sigma_0^\prime-\tau_0$ and the distributed load associated with fluid injection $f_r\Delta p(r,t)$ whose intensity is $f_r\Delta p_*$. The uniform stress is equal to the difference between the residual fault strength under ambient conditions $f_r \sigma_0^\prime$ and the initial shear stress $\tau_0$. Note that depending on the sign of $f_r \sigma_0^\prime-\tau_0$, $\mathcal{T}_r$ can be either positive or negative. Moreover, as noted by first Garagash and Germanovich \cite{Garagash_Germanovich_2012} in their two-dimensional model, the sign of $f_r \sigma_0^\prime-\tau_0$ is expected to strongly affect the overall stability of the fault response. According to Garagash and Germanovich, if the condition $f_r \sigma_0^\prime<\tau_0$ is satisfied ($\mathcal{T}_r$ negative), ruptures may ultimately run away dynamically and never stop within the limits of such a homogeneous and infinite fault model. Conversely, when $f_r \sigma_0^\prime>\tau_0$ ($\mathcal{T}_r$ positive), ruptures would propagate ultimately in a quasi-static, stable manner. Assuming for now that the ultimate stability condition of Garagash and Germanovich \cite{Garagash_Germanovich_2012} holds in the circular rupture configuration, we consider only ultimately quasi-static cases in this section and, therefore, values for $\mathcal{T}_r$ that are strictly positive. We will soon show that this assumption is indeed satisfied.

Let us now find the limiting values of $\mathcal{T}_r$. As lower bound, $\mathcal{T}_r$ can be as small as possible ($\mathcal{T}_r\to0$) when $f_r \sigma_0^\prime\to\tau_0$. Since in this limit the fault is approaching the ultimate unstable condition of Garagash and Germanovich \cite{Garagash_Germanovich_2012}, we refer to it as the `nearly unstable' limit. As upper bound, similarly to the case of constant friction, the maximum value of $\mathcal{T}_r$ is set by the minimum possible magnitude of $\Delta p_*$, which in turn relates to the minimum amount of overpressure that is required to activate fault slip. This is given by the approximate relation $f_p\Delta p_c\approx f_p\sigma_0^\prime-\tau_0$, where $f_p$ is the peak friction coefficient and $\Delta p_c$ is the characteristic overpressure of the fluid source (equation \eqref{eq:pc-definition}). By substituting this previous relation into \eqref{eq:residual-stress-injection-parameter}, we find the sought upper bound to be $\mathcal{T}_r\lessapprox 4\pi (\sigma_0^\prime-\tau_0/f_r)/(\sigma_0^\prime-\tau_0/f_p)$. Since the ratio $f_r/f_p$ is always between 0 and 1 and the maximum value of the upper bound is obtained when $f_r/f_p\to 1$, we obtain such a maximum upper bound as $\mathcal{T}_r\lessapprox10$ (again, the factor $4\pi$). Given that in this limit the upper bound is still related to the minimum amount of overpressure that is required to activate a frictional rupture, we still denominate it as marginally pressurized limit. Note that the upper bound limit has a quite similar meaning than the one of $\mathcal{T}$ in the constant friction model. Conversely, the lower bound limit of $\mathcal{T}_r$ has no longer the interpretation of a critically stressed fault as for $\mathcal{T}$.

The second parameter of the constant fracture energy model, $\mathcal{K}$, corresponds to a time-dependent dimensionless toughness. It can be either defined using the stress scales of the nearly unstable or marginally pressurized limits depending on the proper regime characterizing the fault response. These two choices are:
\begin{equation}\label{eq:dimensionless-toughnesses}
    \mathcal{K}_{nu}(t)=\frac{K_{c}}{\left(f_r\sigma_0^\prime-\tau_0\right)\sqrt{R(t)}},\;\text{and}\quad
    \mathcal{K}_{mp}(t)=\frac{K_{c}}{f_r\Delta p_*\sqrt{R(t)}},
\end{equation}
for the nearly unstable ($\mathcal{T}_r\ll1$) and marginally pressurized ($\mathcal{T}_r\sim10$) regimes, respectively. Note that in \eqref{eq:dimensionless-toughnesses}, the explicit dependence of $R$ on time has been emphasized as it provides the time dependence of $\mathcal{K}$. 

The dimensionless toughness $\mathcal{K}$ quantifies the relevance of the fracture energy in the near-front energy balance at a given time $t$. Since for any physically admissible solution in which the injection is continuous, the rupture radius must increase monotonically with time, the solution will
always evolve from a large-toughness regime ($\mathcal{K}\gg1$) to a small-toughness regime ($\mathcal{K}\ll1$). Moreover, the effect of the fracture energy in the energy balance can be ultimately neglected as the dimensionless toughness effectively vanishes ($\mathcal{K}\to0$) in the limit $R\to\infty$ (or $t\to\infty$). We denominate this ultimate solution as the zero-toughness or zero-fracture-energy solution. Since the effect of the fracture energy becomes irrelevant in this limit, such asymptotic solution is self-similar ($\lambda$ in \eqref{eq:lambda-structure-Gc-constant} becomes time-independent).

Finally, the transition between the large- and small-toughness regimes is characterized by the following rupture length scales (obtained by setting $\mathcal{K}_{nu}=1$ and $\mathcal{K}_{mp}=1$, respectively),
\begin{equation}\label{eq:constant-fracture-energy-rupture-length-scales}
    R_{nu}^*=\left(\frac{K_c}{f_r \sigma_0^\prime-\tau_0}\right)^2\;\text{and}\quad
    R_{mp}^*=\left(\frac{K_c}{f_r \Delta p_*}\right)^2.
\end{equation}
Note that the two dimensionless toughnesses in \eqref{eq:dimensionless-toughnesses} are of course not independent as they are two choices of one same parameter. They are indeed related through the residual stress-injection parameter $\mathcal{T}_r$ as
\begin{equation}
    \mathcal{K}_{mp}=\mathcal{T}_r\mathcal{K}_{nu}.
\end{equation}

\subsubsection{General and ultimate zero-fracture-energy solutions}

\begin{figure}
    \centering
    \includegraphics[width=15cm]{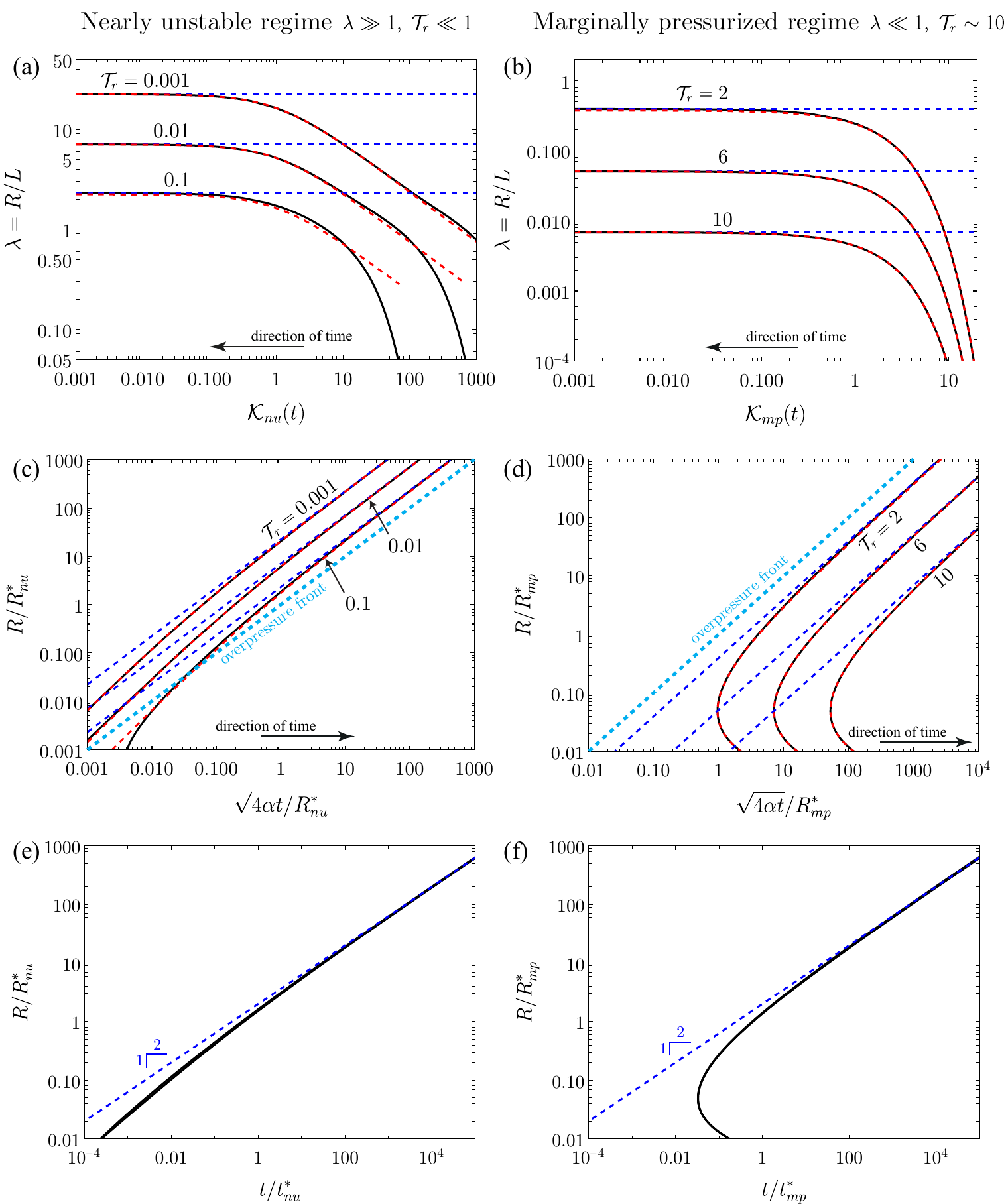}
    \caption{The constant fracture energy model. (Left) Nearly unstable regime ($\lambda\gg1$, $\mathcal{T}_r\ll1$) and (right) marginally pressurized regime ($\lambda\ll1$, $\mathcal{T}_r\sim10$). (a, b) Amplification factor $\lambda$ as a function of dimensionless toughness $\mathcal{K}(t)$ for different values of $\mathcal{T}_r$. (c, d) Normalized rupture radius $R$ as a function of the normalized squared root of time or position of the overpressure front $L(t)=\sqrt{4\alpha t}$. (e, f) Normalized rupture radius $R$ as a function of dimensionless time $t$. All curves tend to collapse when using the latter scaling. Legends: black solid lines are the general solution of the constant fracture energy model; red dashed lines correspond to the asymptotes for $\lambda$ \eqref{eq:lambda-asymptotes-Gc-constant} in (a, b), and the asymptotes for the normalized rupture radius \eqref{eq:asymptote-rupture-radius-nearly-unstable} and \eqref{eq:asymptote-rupture-radius-marginally-pressurized} in (c) and (d), respectively; blue dashed lines represent the constant residual friction, ultimate zero-fracture-energy solution.}\label{fig:constant-fracture-energy}
\end{figure}

Considering the scaling of the previous section plus the definition of the following non-dimensional integral:
\begin{equation}\label{eq:non-dimensional-integral}
    \Psi(\lambda) = \int_{0}^{1}\frac{E_1\left(\lambda\eta\right)}{\sqrt{1-\eta^{2}}}\eta\textrm{d}\eta,
\end{equation}
the front-localized energy balance \eqref{eq:final-energy-balance} can be written in dimensionless form as
\begin{equation}\label{eq:dimensionless-energy-balance}
    \frac{\Psi(\lambda)}{\mathcal{T}_r} - 1 =
    \frac{\sqrt{\pi}}{2}\mathcal{K}_{nu},\;\text{and}\quad
    \Psi(\lambda) - \mathcal{T}_r =
    \frac{\sqrt{\pi}}{2}\mathcal{K}_{mp},
\end{equation}
in the nearly unstable ($\mathcal{T}_r\ll1$) and marginally pressurized ($\mathcal{T}_r\sim10$) regimes, respectively. Note that the non-dimensional integral $\Psi$ is identical to the one in equation \eqref{eq:propagation-condition-Coulomb-friction} and can be thus evaluated analytically to obtain the left-hand side of equation \eqref{eq:analytical-solution-Coulomb-friction}. Moreover, the limiting behaviors of such integral are: $\Psi(\lambda) \approx 1/2\lambda^2 + O(\lambda^{-4})$ when $\lambda\gg1$, and $\Psi(\lambda) \approx 2-\gamma-\text{ln}\left(4\lambda^2\right)+ O(\lambda^{2})$ when $\lambda\ll1$. Using the previous asymptotic expansions and assuming similarly to the constant friction model that $\lambda\gg1$ when $\mathcal{T}_r\ll1$, and $\lambda\ll1$ when $\mathcal{T}_r\sim10$, we derive from equations \eqref{eq:dimensionless-energy-balance}, the following closed-form asymptotic expressions for the amplification factor:
\begin{equation}\label{eq:lambda-asymptotes-Gc-constant}
    \lambda\approx 
        \begin{cases}
        1/\sqrt{\left(2+\sqrt{\pi}\mathcal{K}_{nu}\right)\mathcal{T}_r} & \text{for nearly unstable faults, }\mathcal{T}_r\ll 1,\\
        \frac{1}{2}\exp\left[\left(2-\gamma-\mathcal{T}_r-\mathcal{K}_{mp}\sqrt{\pi}/2\right)/2\right] & \text{for marginally pressurized faults, } \mathcal{T}_r\sim 10,
    \end{cases}
\end{equation}
where the dependence of both $\mathcal{K}$ and $\lambda$ on time has been omitted for simplicity.

The full solution of the model given by equations \eqref{eq:dimensionless-energy-balance} together with the asymptotics \eqref{eq:lambda-asymptotes-Gc-constant} are shown in figures \ref{fig:constant-fracture-energy}a and \ref{fig:constant-fracture-energy}b. 
The direction of time in these plots goes from right to left as the dimensionless toughness $\mathcal{K}(t)$ decreases with time. Moreover, since ultimately ($t\to\infty$, $R\to\infty$) the dimensionless toughness is negligibly small ($\mathcal{K}\to0$), equations \eqref{eq:dimensionless-energy-balance}a and \eqref{eq:dimensionless-energy-balance}b become both identical to the rupture propagation condition of the constant friction model, equation \eqref{eq:propagation-condition-Coulomb-friction}, as long as the constant friction coefficient $f_\text{cons}$ is now understood as the residual one $f_r$. The solution of the constant friction model \eqref{eq:analytical-solution-Coulomb-friction} with $f_{\text{cons}}=f_r$ is also displayed in figures \ref{fig:constant-fracture-energy}a and \ref{fig:constant-fracture-energy}b. It is now clear how the constant fracture energy solution approaches asymptotically the constant residual friction solution as $\mathcal{K}\to0$. This can be also seen in the asymptotics \eqref{eq:lambda-asymptotes-Gc-constant} that become identical to \eqref{eq:lambda-asymptotes} when $\mathcal{K}=0$. 

The previous result has an important implication: the constant friction model analyzed in \cite{Saez_Lecampion_2022} can be now interpreted in two distinct manners: as an scenario in which the friction coefficient does not significantly weaken ($f_{\text{cons}}\approx f_p$), or as the ultimate asymptotic solution of a model with constant fracture energy provided that $f_{\text{cons}}=f_r$. In the former, the fracture energy $G_c=0$ by definition. In the latter, the effect of the non-zero fracture energy in the rupture-front energy balance is to leading order negligible compared to the other two terms that drive the propagation of the rupture. In addition, because the integral $\Psi(\lambda)$ is strictly positive and in the ultimate asymptotic regime $\mathcal{K}\to0$, the near-front energy balance \eqref{eq:dimensionless-energy-balance} admits ultimate quasi-static solutions only if $\mathcal{T}_r>0$. Negative values of $\mathcal{T}_r$ which are equivalent to the condition $f_r\sigma_0^\prime<\tau_0$ may be thus related to ultimately unstable solutions, not accounted for the quasi-static energy balance. This result supports our assumption that the ultimate stability condition of Garagash and Germanovich \cite{Garagash_Germanovich_2012} holds in the circular rupture configuration. 

We now recast the solution of the constant fracture energy model in a perhaps more intuitive way, as the evolution of the rupture radius with time: 
\begin{equation}\label{eq:lambda-as-R-vs-L}
    R(t)=\lambda(\mathcal{T}_r,\mathcal{K}(t))\cdot L(t).
\end{equation}
Recalling that $L(t)=\sqrt{4\alpha t}$ and noting that 
\begin{equation}\label{eq:K-as-function-R}
    \mathcal{K}_{nu}=(R/R_{nu}^*)^{-1/2},\;\text{and}\quad \mathcal{K}_{mp}=(R/R_{mp}^*)^{-1/2},
\end{equation}
we solve equations \eqref{eq:dimensionless-energy-balance}a and \eqref{eq:dimensionless-energy-balance}b for $R/R^*$ as a function of the normalized squared root of time $\sqrt{4\alpha t}/R^*$ and $\mathcal{T}_r$, where $R^*$ represents the characteristic rupture length scale of either the nearly unstable or marginally pressurized regime (equation \eqref{eq:constant-fracture-energy-rupture-length-scales}).

This version of the solution is displayed in figures \ref{fig:constant-fracture-energy}c and \ref{fig:constant-fracture-energy}d. In these plots, the normalized square root of time can be also interpreted as the normalized position of the overpressure front $L(t)$. Indeed, the thicker dashed line corresponds to the current position of the overpressure front. Slip fronts propagating above this line represent cases in which the rupture front outpaces the overpressure front. We observe that such a situation is a common feature of nearly unstable faults ($\mathcal{T}_r\ll1$), being the analog regime of critically stressed faults in the constant friction model. Moreover, taking into account \eqref{eq:lambda-as-R-vs-L} and \eqref{eq:K-as-function-R}, the asymptotics \eqref{eq:lambda-asymptotes-Gc-constant} can be recast as the following implicit equations for the normalized rupture radius $R/R^*$ as a function of time:
\begin{equation}\label{eq:asymptote-rupture-radius-nearly-unstable}
    \sqrt{4\alpha t}/R_{nu}^*=\frac{R}{R_{nu}^*}\left[\left(2+\frac{\sqrt{\pi}}{\sqrt{R/R_{nu}^*}}\right)\mathcal{T}_r\right]^{1/2}
\end{equation}
for nearly unstable faults, and 
\begin{equation}\label{eq:asymptote-rupture-radius-marginally-pressurized}
    \sqrt{4\alpha t}/R_{mp}^*=\frac{2 \left(R/R_{mp}^*\right)}
    {\exp\left[\left(2-\gamma-\mathcal{T}_r-\left(\sqrt{\pi}/2\right)\left(R/R_{mp}^*\right)^{-1/2}\right)/2\right]}
\end{equation}
for marginally pressurized faults.

Note that the transition from the large-toughness ($\mathcal{K}\gg1$) to small-toughness ($\mathcal{K}\ll1$) regime in figures \ref{fig:constant-fracture-energy}c and \ref{fig:constant-fracture-energy}d occurs along the vertical axis when $R/R_{nu}^*\sim1$ and $R/R_{mp}^*\sim1$, respectively. The characteristic time at which this transition occurs can be approximated by using the constant residual friction solution or, what is the same, the ultimate zero-fracture-energy solution,  $\lambda_r=\lambda\left(\mathcal{T}_r,\mathcal{K}=0\right)$, which yields
\begin{equation}\label{eq:time-scales-transition-Gc-constant}
    t_{nu}^*\approx\frac{1}{\alpha\lambda_r^2}\left(\frac{K_c}{f_r \sigma_0^\prime-\tau_0}\right)^4,\; \text{and}\quad t_{mp}^*\approx\frac{1}{\alpha\lambda_r^2}\left(\frac{K_c}{f_r\Delta p_*}\right)^4.
\end{equation}
$\lambda_r$ can be estimated from the asymptotes presented in equation \eqref{eq:lambda-asymptotes} for both nearly unstable ($\mathcal{T}_r\ll1$) and marginally pressurized ($\mathcal{T}_r\sim10$) faults, provided that $\mathcal{T}$ is replaced by $\mathcal{T}_r$. Normalizing time by the previous characteristic times naturally tends to collapse all solutions for every value of $\mathcal{T}_r$ as displayed in figures \ref{fig:constant-fracture-energy}e and \ref{fig:constant-fracture-energy}f, where the power law 1/2 reflects the diffusively self-similar property of the ultimate zero-fracture-energy solution.

Finally, it seems worth mentioning that the solution for marginally pressurized faults is nonphysical at times in which the rupture is small, $R/R_{mp}^*\lessapprox0.05$ (see figures \ref{fig:constant-fracture-energy}d and \ref{fig:constant-fracture-energy}f). This could be related either to the occurrence of a dynamic instability or to a rupture size that is too small comparing to realistic process zone sizes. Indeed, the solution constructed here for the case of a constant fracture energy has the inherent limitations of LEFM theory. First, it does not account for the initial stage in which the process zone is under development ($R<\ell_*$) and, second, it is an approximate solution that relies on the small-scale yielding assumption ($R\gg\ell_*$). Both limitations are overcome in the next section by solving numerically the governing equations of the coupled initial boundary value problem for slip-weakening friction.

\section{Scaling analysis, map of rupture regimes and ultimate stability condition}\label{sec:scaling-map-regimes}

\subsection{Scaling analysis}

The scaling of the slip-weakening problem comes directly from the two-dimensional linear-weakening model of Garagash and Germanovich \cite{Garagash_Germanovich_2012}, which is also valid for the exponential-weakening version of the friction law. We summarize the scaling as follows:
\begin{equation}\label{eq:scaling-slip-weakening}
    \bar{t}=\frac{t}{R_w^2/\alpha},\quad 
    \bar{r}=\frac{r}{R},\quad 
    \bar{\tau}=\frac{\tau}{f_p\sigma_0^\prime},\quad 
    \bar{\delta}=\frac{\delta}{\delta_w},\quad 
    \Delta \bar{p}=\frac{\Delta p}{\Delta p_*},
\end{equation}
where the bar symbol represents dimensionless quantities, $\delta_w$ is the slip weakening scale, and $R_w$ is an elasto-frictional rupture length scale, given respectively by (see also Uenishi and Rice \cite{Uenishi_Rice_2002})
\begin{equation}\label{eq:slip-scale-rupture-length-scale-slip-weakening}
    \delta_w=\frac{f_p}{f_p-f_r}\delta_c,\;\text{and}\quad R_w=\frac{\mu}{\left(f_p-f_r\right)\sigma_0^\prime}\delta_c.
\end{equation}
In the previous equation, $(f_p-f_r)\sigma_0^\prime/\delta_c$ is the so-called slip-weakening rate \cite{Uenishi_Rice_2002}.

Nondimensionalization of the governing equations of the model using the previous scaling shows that the normalized fault slip $\bar{\delta}$ depends in addition to dimensionless space $\bar{r}$ and time $\bar{t}$, on the following three dimensionless parameters:
\begin{equation}\label{eq:non-dimensional-parameters}
    \mathcal{S}=\frac{\tau_0}{f_p\sigma_0^\prime},\quad
    \mathcal{P}=\frac{\Delta p_*}{\sigma_0^\prime},\quad
    \mathcal{F}=\frac{f_r}{f_p}.
\end{equation}
The first parameter $\mathcal{S}$ is the pre-stress ratio or sometimes called, stress criticality. It is the quotient between the initial shear stress $\tau_0$ and the initial static fault strength $f_p\sigma_0^\prime$. The pre-stress ratio $\mathcal{S}$ quantifies how close to frictional failure the fault is under ambient (pre-injection) conditions. The range of values for $\mathcal{S}$ is naturally
\begin{equation}
    0\leq\mathcal{S}<1,
\end{equation}
being zero when the fault has no initial shear stress whatsoever, and one when the fault is critically stressed or about to fail under ambient conditions, $\tau_0\to f_p\sigma_0^\prime$.

The second parameter $\mathcal{P}$ is the overpressure ratio, which quantifies the intensity of the injection $\Delta p_*$ with regard to the initial effective normal stress $\sigma_0^\prime$. The range of possible values for $\mathcal{P}$ is determined as follows. Its upper bound comes from the maximum possible amount of overpressure that in our model corresponds to an scenario in which the fault interface is about to open: $\Delta p_c\approx\sigma_0^\prime$, where $\Delta p_c=4\pi \Delta p_*$ (equation \eqref{eq:pc-definition}). On the other hand, the lower bound of $\mathcal{P}$ comes from the minimum amount of overpressure that is required to activate fault slip: $f_p\Delta p_c\approx f_p\sigma_0^\prime-\tau_0$. By replacing the previous approximate relations into $\mathcal{P}=\Delta p_*/\sigma_0^\prime$, we obtain the sought range of values for $\mathcal{P}$ in an approximate sense as
\begin{equation}
    10^{-1}(1-\mathcal{S})\lessapprox\mathcal{P}\lessapprox10^{-1},
\end{equation}
where $\approx10^{-1}$ comes from the factor $1/4\pi$. Finally, the third parameter, the residual-to-peak friction ratio $\mathcal{F}$ is such that
\begin{equation}
    0\leq\mathcal{F}\leq1.
\end{equation}
$\mathcal{F}$ is zero when there is a total loss of frictional resistance upon the passage of the rupture front, a situation that is unlikely to occur for stable, slow slip, as oppose to fast slip in which thermally-activated dynamic weakening mechanisms could make the fault reach quite low values for $\mathcal{F}$ \cite{Rice_2006}. On the other hand, $\mathcal{F}$ is equal to one when the friction coefficient does not weaken at all, which corresponds indeed to the particular case of Coulomb's friction $f_\text{cons}=f_p$.

Finally, it will result useful to define the residual stress-injection parameter $\mathcal{T}_r$ of the constant fracture energy model, equation \eqref{eq:residual-stress-injection-parameter}, as a combination of the three dimensionless parameters of the slip-weakening model,
\begin{equation}\label{eq:Tr-function-S-P-F}
    \mathcal{T}_r=
    \frac{1-\mathcal{S}/\mathcal{F}}{\mathcal{P}}.
\end{equation}
In addition, one can also define a stress-injection parameter based on the peak value of friction $f_p$ instead of the residual one. Such a parameter reads as
\begin{equation}\label{eq:peak-stress-injection-parameter}
    \mathcal{T}_p=\frac{f_p \sigma_0^\prime-\tau_0}{f_p\Delta p_*}=
    \frac{1-\mathcal{S}}{\mathcal{P}}.
\end{equation}
We denominate $\mathcal{T}_p$ as the `peak' stress-injection parameter. The latter is indeed the maximum possible value of the residual stress-injection parameter $\mathcal{T}_r$ (when $\mathcal{F}=1$), so that
\begin{equation}
    \mathcal{T}_r\leq\mathcal{T}_p.
\end{equation}
As a final comment on the scaling, when comparing the fault response for each version of the friction law, the results that we present in the next sections are particularly valid under the assumption that both friction laws are characterized by the same slip weakening scale $\delta_c$ (see figure \ref{fig:friction-laws}). Alternatively, one could compare the effect of both friction laws under a different condition such as, for instance, an equal fracture energy $G_c$, or any other criterion. In the case of equal $G_c$, the characteristic slip weakening scales would be related as $\delta_{c,\text{lin}}=2\delta_{c,\text{exp}}$. Our results can be then easily re-scaled using the previous relation as the dimensionless solution remains unchanged, and the same could be done with any other criterion.

\subsection{Map of rupture regimes and ultimate stability condition}
  
\begin{figure}
    \centering
    \includegraphics[width=11cm]{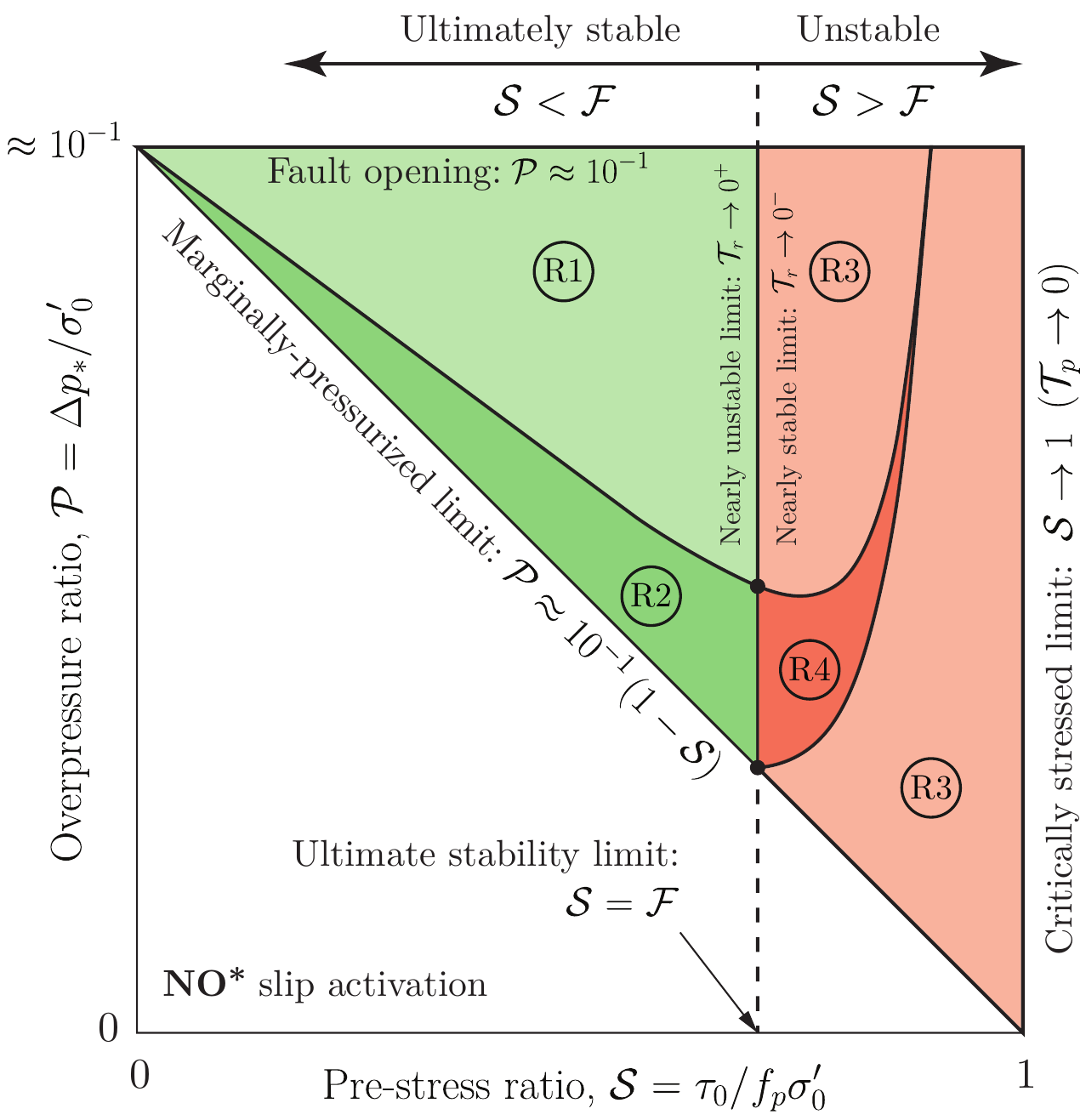}
    \caption{Map of rupture regimes for linear slip-weakening model (adapted from figure 11 in \cite{Garagash_Germanovich_2012}). (R1) Unconditonally stable fault slip. (R2) Quasi-static slip up to the nucleation of a dynamic rupture, followed by arrest and then purely quasi-static slip. (R3) Quasi-static slip until the nucleation of a run-away dynamic rupture. (R4) Quasi-static slip up to the nucleation of a dynamic rupture, followed by arrest and then re-nucleation of a run-away dynamic rupture. $^*$The condition of no slip is established in the approximate sense discussed in Appendix \ref{appendix:line-source}.}
    \label{fig:map-regimes}
\end{figure}

Given the similarity of the scaling between our three-dimensional axisymmetric rupture model and the two-dimensional plane-strain model of Garagash and Germanovich \cite{Garagash_Germanovich_2012}, we find, not surprisingly, that the map of regimes of fault behavior in our model is essentially the same as in the two-dimensional problem \cite{Garagash_Germanovich_2012}. Figure \ref{fig:map-regimes} summarizes the map of regimes in the parameter space composed by $\mathcal{S}$, $\mathcal{P}$ and $\mathcal{F}$. Moreover, as anticipated when examining the constant fracture energy model, the ultimate stability condition of Garagash and Germanovich \cite{Garagash_Germanovich_2012} holds in the circular rupture case. Therefore, mixed-mode circular ruptures will propagate ultimately ($t\to\infty$, $R\to\infty$) in a quasi-static, stable manner, if any of the following three equivalent conditions is satisfied:
\begin{equation}\label{eq:ultimate-stability-conditions}
    f_r\sigma_0^\prime>\tau_0 \iff
    \mathcal{S}<\mathcal{F} \iff
    \mathcal{T}_r>0.
\end{equation}
Notably, the residual stress-injection parameter $\mathcal{T}_r$ must be strictly positive. Else, ruptures will propagate ultimately in an unstable, dynamic manner. In the latter case, dynamic ruptures will run away and never stop within the limits of such a homogeneous and infinite fault model. Since in this work, we are mainly interested in the propagation of quasi-static slip, the regimes of major interest are the ones corresponding to ultimately stable ruptures and the quasi-static nucleation phase preceding dynamic ruptures.
We examine both scenarios in what follows.

\section{Ultimately stable ruptures}\label{sec:stable-ruptures}

Figure \ref{fig:ultimately-stable-ruptures} displays the propagation of the slip front in the case of ultimately stable ruptures: $f_r\sigma_0^\prime>\tau_0$. Without loss of generality, we fix the residual-to-peak friction ratio $\mathcal{F}=0.7$ and examine for both the linear- and exponential-weakening friction laws the parameter space for $\mathcal{P}$ and $\mathcal{S}<0.7$. The case of an overpressure ratio $\mathcal{P}=0.05$ is shown in figures \ref{fig:ultimately-stable-ruptures}a and \ref{fig:ultimately-stable-ruptures}b. For all values of $\mathcal{S}$ in these figures, we obtain ruptures that propagate in a purely quasi-static manner without any dynamic excursion, that is, the regime R1 in figure \ref{fig:map-regimes}. Figures \ref{fig:ultimately-stable-ruptures}c and \ref{fig:ultimately-stable-ruptures}d show, on the other hand, the case of a lower overpressure ratio $\mathcal{P}=0.035$. For this value of $\mathcal{P}$, we observe the occurrence of the regime R2 for the linear-weakening case and both regimes R1 and R2 for the exponential-weakening case. The regime R2 corresponds to a situation in which a dynamic rupture nucleates, arrest, and is then followed by purely quasi-static slip.

\subsection{Early-time Coulomb's friction stage and localization of the process zone}

\begin{figure}
    \centering
    \includegraphics[width=15cm]{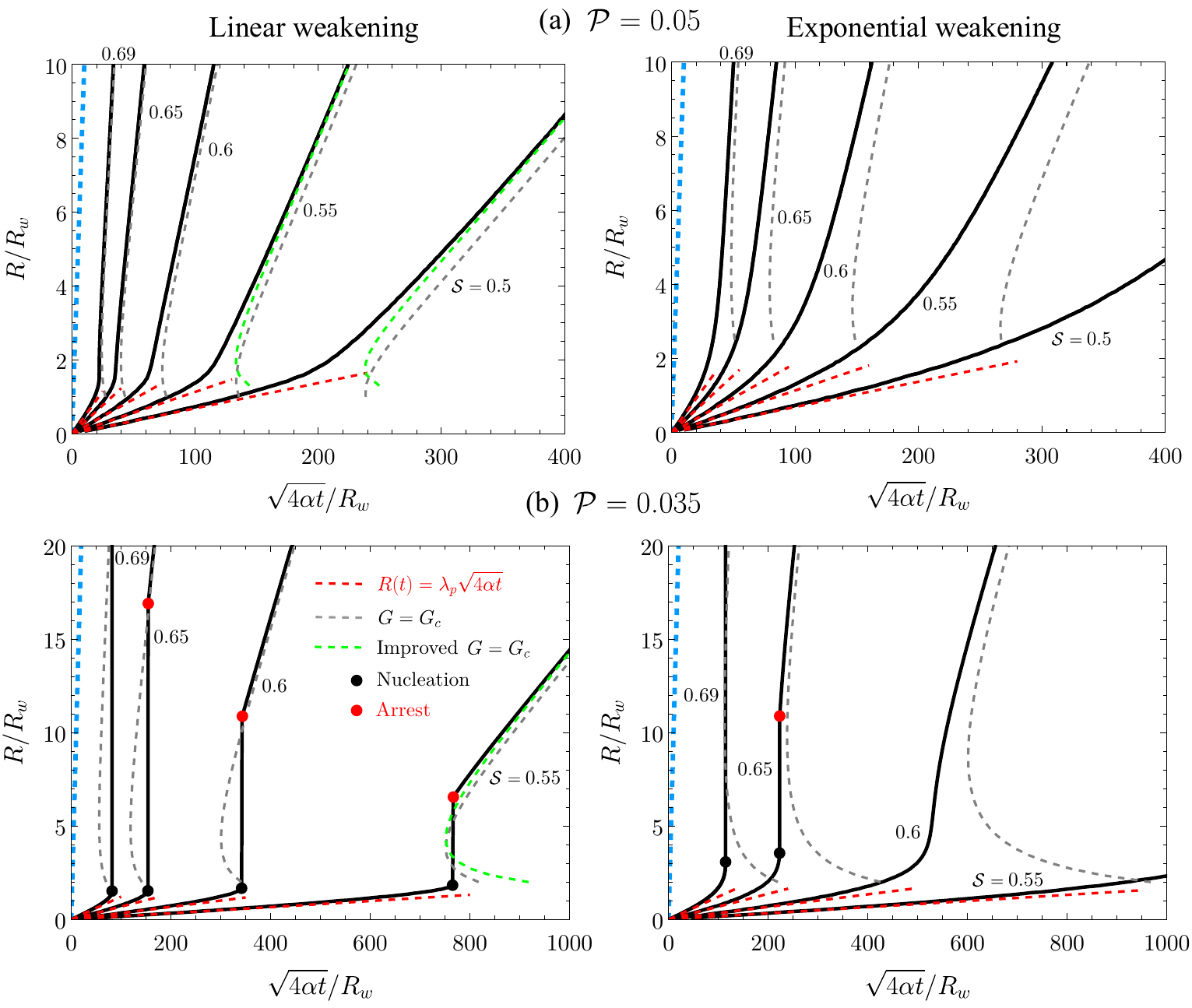}
    \caption{Ultimately stable faults, $\mathcal{S}<\mathcal{F}=0.7$, at distances $R/R_w<20$. Normalized rupture radius versus square root of dimensionless time for (left) linear and (right) exponential weakening versions of the friction law. (a) $\mathcal{P}=0.05$ and (b) $\mathcal{P}=0.035$. Red dashed lines correspond to the analytical constant friction solution considering the peak friction coefficient. Gray dashed lines correspond to the solution of the front-localized energy balance, $G=G_c$. Green dashed lines correspond to an improved version of the front-localized energy balance as explained in the main text, shown here for only a few cases. Light blue lines represent the position of the overpressure front $L(t)=\sqrt{4\alpha t}$. Note that for all cases, $R(t)\ll L(t)$.}
    \label{fig:ultimately-stable-ruptures}
\end{figure}

It is clear from figure \ref{fig:ultimately-stable-ruptures} that at early times and for both regimes (R1 and R2), the propagation of the slip front is well approximated by the Coulomb's friction model, $f_\text{cons}=f_p$. The rupture radius thus evolves in this stage as
\begin{equation}\label{eq:R-with-peak-friction}
    R(t)\approx\lambda_p L(t),
\end{equation}
where $\lambda_p$ is the amplification factor given by equation \eqref{eq:analytical-solution-Coulomb-friction}  considering the peak stress-injection parameter $\mathcal{T}_p$, and $L(t)=\sqrt{4\alpha t}$ is the position of the overpressure front as usual. This early-time Coulomb's friction similarity solution is meant to be valid while the friction coefficient does not decrease significantly throughout the slipping region, as shown for a few cases in the examples of figures \ref{fig:ultimately-unstable-ruptures-profiles}d and \ref{fig:ultimately-unstable-ruptures-profiles}e, when looking at the spatial distribution of the friction coefficient for the earliest times ($t_1$ and $t_2$). Note that in figure \ref{fig:ultimately-stable-ruptures}, we also include the evolution of the overpressure front with time (light blue dashed line). In the spatial range covered by this figure, the slip front of ultimately stable ruptures always lags the overpressure front ($R(t)\ll L(t)$) as the corresponding values of $\mathcal{T}_p$ are well into the marginally pressurized regime ($\mathcal{T}_p\sim10$).


Now, beyond this early stage, the propagation of the slip front starts departing from the Coulomb's friction similarity solution while the slipping region experiences further weakening of friction. At this point, a dynamic instability could nucleate, arrest, and be followed by aseismic slip (examples in figures \ref{fig:ultimately-stable-ruptures}c and \ref{fig:ultimately-stable-ruptures}d), within a relatively narrow region of the parameter space (R2 in figure \ref{fig:map-regimes}). More generally, ruptures will propagate in a purely quasi-static manner (examples in figures \ref{fig:ultimately-stable-ruptures}a and \ref{fig:ultimately-stable-ruptures}b, regime R1 in figure \ref{fig:map-regimes}). Either way, when this transition happens, the rupture radius becomes greater than the rupture length scale $R_w$, which is around the same order than the process zone size $\ell_*$ for the linear weakening law (see, for an example, figure \ref{fig:ultimately-unstable-ruptures-profiles}f). In the case of the exponential weakening case, figure \ref{fig:ultimately-stable-ruptures} shows that the transition is smoother and occurs later than for the linear weakening case, whereas the localization of the process zone occurs also at a later time and for rupture lengths that seem to be many times or even an order of magnitude greater than the elasto-frictional length scale $R_w$ (see, also, an example in figure \ref{fig:ultimately-unstable-ruptures-profiles}f). Furthermore, starting from this point, the process zone has fully developed and thus a proper fracture energy $G_c$ can be calculated. In this way, we can now examine the evolution of the rupture front through the near-front energy balance \eqref{eq:final-energy-balance}, to an accuracy set by the small-scale yielding approximation.


\subsection{Front-localized energy balance, large- and small-toughness regimes}

Using equation \eqref{eq:Gc-approximation} in combination  with \eqref{eq:p-solution}, \eqref{eq:linear-weakening-friction} and \eqref{eq:exponential-weakening-friction}, we obtain an expression for the fracture energy of the slip-weakening model as
\begin{equation}\label{eq:slip-weakening-fracture-energy}
G_c \approx 
\kappa\left(f_p-f_r\right)\delta_c\left[\sigma_0^\prime-\Delta p_*E_{\text{1}}\left(\lambda^2\right)\right]
\end{equation}
with $\lambda(t)=R(t)/L(t)$ as usual, and $\kappa$ is a coefficient equal to $1/2$ for the linear weakening case, and $1$ for the exponential law, reflecting that the fracture energy of the latter is twice the fracture energy of the former at equal $\delta_c$. 

Introducing the scaling of the slip weakening model \eqref{eq:scaling-slip-weakening} into equation \eqref{eq:final-energy-balance}, one can nondimensionalize the front-localized energy balance in the following form,
\begin{equation}\label{eq:slip-weakening-energy-balance-dimensionless}
    \mathcal{F}\mathcal{P}\Psi\left(\lambda\right) +
    \left(\mathcal{S}-\mathcal{F}\right)
    =
    \frac{\sqrt{\pi}}{2}\sqrt{2\kappa}\left(1-\mathcal{F}\right)\frac{\sqrt{1-\mathcal{P}E_1\left(\lambda^2\right)}}{\sqrt{R/R_w}},
\end{equation}
where $\Psi(\lambda)$ is the non-dimensional integral \eqref{eq:non-dimensional-integral} whose evaluation is know analytically in \eqref{eq:analytical-solution-Coulomb-friction}. As expected from the scaling of the problem, equation \eqref{eq:slip-weakening-energy-balance-dimensionless} shows that the normalized rupture radius $R/R_w$ depends in addition to dimensionless time $L(t)/R_w$ (which is implicit in $\lambda$) on the three dimensionless parameters of the model: $\mathcal{S}$, $\mathcal{P}$ and $\mathcal{F}$.  

Considering that for any physically admissible quasi-static solution, the rupture radius must increase monotonically with time, we solve equation \eqref{eq:slip-weakening-energy-balance-dimensionless} by imposing $R/R_w$ and then calculating $L(t)/R_w$ for a given combination of dimensionless parameters. The solution of the front-localized energy balance \eqref{eq:slip-weakening-energy-balance-dimensionless} is shown in figure \ref{fig:ultimately-stable-ruptures} together with the full numerical solutions. We observe that in the linear weakening case, the near-front energy balance yields a good approximation of the full numerical solution already for $R\gtrapprox 2 R_w$. On the other hand, in the exponential decay case, a good approximation is reached only after $R\gtrapprox 10 R_w$. The difference is due to the fact that the localization of the process zone is less sharp and takes longer for the exponential decay in comparison to the linear weakening case, as exemplified in figure \ref{fig:ultimately-unstable-ruptures-profiles}f. Moreover, the approximate nature of the energy-balance solution is of course due to the finite size of the process zone as opposed to the infinitesimal size required by LEFM. Indeed, in the two-dimensional problem, a correction due to the finiteness of the process zone for the linear weakening case was considered by Garagash and Germanovich \cite{Garagash_Germanovich_2012} based on the work on cohesive tensile crack propagation due to uniform far-field load by Dempsey \textit{et al}. \cite{Dempsey_Tan_2010}, providing a solution with improved accuracy. Figure \ref{fig:ultimately-stable-ruptures} shows the results of this correction for a few cases in our model. The solution gets slightly better despite our frictional shear crack is circular and as such, the pre-factors in the scaling relations of the two-dimensional problem should differ from ours.

The front-localized energy balance allows us notably to examine the evolution of the rupture radius beyond the spatial range covered by figure
\ref{fig:ultimately-stable-ruptures}, without the need of calculating the full numerical solutions. Note that the energy-balance solution is not only a good approximation over this spatial range but will also become an exact asymptotic solution in the LEFM limit $\ell_*/R\to0$. Solutions for $10\leq R/R_w\leq10^3$ are displayed in figure \ref{fig:intermediate-distances-slip-weakening}. In particular, figure \ref{fig:intermediate-distances-slip-weakening}a shows that the higher the pre-stress ratio is, the faster the rupture propagates. Similarly, figure \ref{fig:intermediate-distances-slip-weakening}b displays that the more intense injection is (higher overpressure ratio), the faster the rupture propagates too. Both effects are intuitively expected and consistent with the definition and effect of the stress-injection parameter in the constant friction model (section \ref{sec:Coulomb-friction}). Moreover, initially ($R/R_w\leq10$), we observe in figure \ref{fig:ultimately-stable-ruptures} that the rupture front lags the overpressure front ($\lambda<1$) as faults are governed at early times by Coulomb's friction with a peak stress-injection parameter $\mathcal{T}_p$ well into the marginally pressurized regime ($\mathcal{T}_p\sim10$). Nonetheless, as the rupture accelerates due to the further weakening of friction, figure \ref{fig:intermediate-distances-slip-weakening} shows that at later times, the slip front may end up outpacing the overpressure front ($\lambda>1$). We examine the conditions leading to such behavior in what follows.

\subsubsection{Nearly unstable faults, $\lambda(t) \gg 1$}

\begin{figure}
    \centering
    \includegraphics[width=15cm]{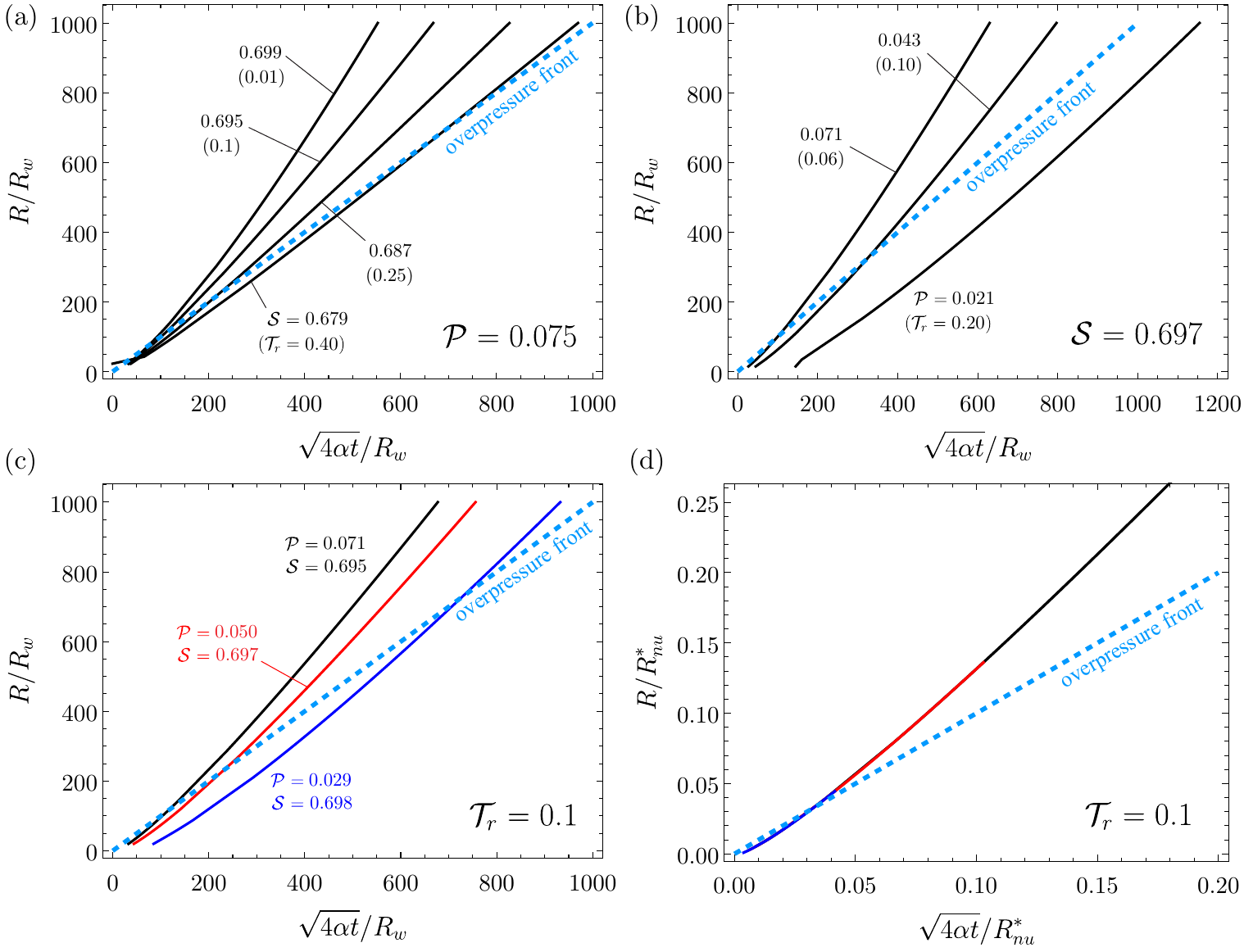}
    \caption{Ultimately stable ruptures, $\mathcal{S}<\mathcal{F}=0.7$, at distances $10\leq R/R_w\leq 10^3$ for the linear-weakening friction law. Normalized rupture radius $R/R_w$ versus square root of dimensionless time $\sqrt{4\alpha t}/R_w$ for: (a) $\mathcal{P}=0.075$ and different values of $\mathcal{S}$ (values of $\mathcal{T}_r$ indicated between brackets); (b) $\mathcal{S}=0.697$ and different values of $\mathcal{P}$; (c) different combinations of $\mathcal{P}$ and $\mathcal{S}$ for the same $\mathcal{T}_r=0.1$; and (d) same as (c) but re-scaling both axes using $R_{nu}^*$. Note that all curves collapse under the new scaling in the latter plot.}
    \label{fig:intermediate-distances-slip-weakening}
\end{figure}


When $\lambda(t)\gg1$, the term $\Delta p_*E_{\text{1}}\left(\lambda^2\right)$ in \eqref{eq:slip-weakening-fracture-energy} can be neglected as the overpressure within the process zone is vanishingly small. Hence, the fracture energy $G_c$ becomes approximately constant and simply equal to
\begin{equation}\label{eq:fracture-energy-nearly-unstable}
    G_c\approx \kappa\left(f_p-f_r\right)\delta_c \sigma_0^\prime.
\end{equation}
At these length scales and in this regime, the problem becomes now identical to the constant fracture energy model that we extensively analyzed in section \ref{sec:constant-fracture-energy}. All the results and insights obtained for nearly unstable faults in  that model are therefore inherited here. In fact, the rupture front will always outpace the overpressure front provided that the fault responds in the nearly unstable regime as quantified by the residual stress-injection parameter $\mathcal{T}_r\ll1$, which is indeed intentionally the case of all the examples shown in figure \ref{fig:intermediate-distances-slip-weakening}. 

Introducing \eqref{eq:fracture-energy-nearly-unstable} into \eqref{eq:dimensionless-toughnesses}a via \eqref{eq:K-equivalent}, and then \eqref{eq:dimensionless-toughnesses}a into \eqref{eq:dimensionless-energy-balance}a, leads to the dimensionless form of the front-localized energy balance in the nearly unstable regime,
\begin{equation}\label{eq:nearly-unstable-energy-balance-slip-weakening}
    \frac{1}{\mathcal{T}_r} \Psi\left(\lambda\right) - 1 =
    \frac{\sqrt{\pi}}{2}\sqrt{2\kappa}\frac{1}{\sqrt{R/R_{nu}^*}},
\end{equation}
with
\begin{equation}\label{eq:Rnu-slip-weakening}
    \frac{R_{nu}^*}{R_w}=\left(\frac{f_p\sigma_0^\prime-f_r\sigma_0^\prime}{f_r \sigma_0^\prime-\tau_0}\right)^2=\left(\frac{1-\mathcal{F}}{\mathcal{F}-\mathcal{S}}\right)^2.
\end{equation}
Equations \eqref{eq:nearly-unstable-energy-balance-slip-weakening} and \eqref{eq:Rnu-slip-weakening} can be also obtained by neglecting the term $\mathcal{P} E_{\text{1}}\left(\lambda^2\right)$ in \eqref{eq:slip-weakening-energy-balance-dimensionless} and then dividing the latter by $\mathcal{F}-\mathcal{S}$. What is interesting to highlight, is that when the overpressure across the process zone becomes approximately constant (and so the fracture energy $G_c$), the mathematical structure of the solution for the rupture front in the slip weakening model changes and it is no longer dependent on three but only one single dimensionless number, the residual stress-injection parameter $\mathcal{T}_r$. 

The new scaling is exemplified in figures \ref{fig:intermediate-distances-slip-weakening}c and \ref{fig:intermediate-distances-slip-weakening}d. The former figure shows the evolution of the rupture front as given by equation \eqref{eq:slip-weakening-energy-balance-dimensionless} for different combinations of $\mathcal{S}$ and $\mathcal{P}$ that are all characterized by the same value of the residual stress-injection parameter, $\mathcal{T}_r=0.1$. After re-scaling the solution using the characteristic rupture length scale $R_{nu}^*$ \eqref{eq:Rnu-slip-weakening}, figure \ref{fig:intermediate-distances-slip-weakening}d displays how all curves in figure \ref{fig:intermediate-distances-slip-weakening}c collapse under the new, constant-fracture-energy scaling. Moreover, by using the asymptotic behavior of $\Psi\left(\lambda\right)$ for large $\lambda$, equation \eqref{eq:asymptote-rupture-radius-nearly-unstable} provides an implicit equation for the normalized rupture radius $R/R_{nu}^*$ as a function of the normalized square root of time $\sqrt{4\alpha t}/R_{nu}^*$ and the residual stress-injection parameter $\mathcal{T}_r$, provided that $R_{nu}^*$ is replaced by \eqref{eq:Rnu-slip-weakening}.

\subsubsection{Marginally pressurized faults, $\lambda(t) \ll 1$}

A similar reasoning can be considered now for the case in which $\lambda(t)\ll1$. Here, the overpressure within the process zone can be taken as approximately constant and equal to the overpressure at the fluid source, $\Delta p_c$ \eqref{eq:pc-definition}, as the rupture radius is much smaller than the pressurized zone, $R(t)\ll L(t)$. Therefore, we can approximate the fracture energy \eqref{eq:slip-weakening-fracture-energy} as
\begin{equation}\label{eq:fracture-energy-marginally-pressurized}
    G_c\approx \kappa\left(f_p-f_r\right)\delta_c\left(\sigma_0^\prime-\Delta p_c\right),    
\end{equation}
which is constant as well. We recall that $\Delta p_c$ is a rough approximation of the fluid-source overpressure as discussed in Appendix \ref{appendix:line-source}. Again, all the results and insights from the constant fracture energy model are inherited now in this regime. Particularly, the so-called marginally pressurized regime as quantified by the residual stress-injection parameter ($\mathcal{T}_r\sim10$) is the one related to $\lambda(t)\ll1$. We recall that this marginally pressurized regime is not defined exactly as the one emerging during the early-time Coulomb's friction stage. The latter is defined by the condition $f_p\Delta p_c\approx f_p\sigma_0^\prime-\tau_0$, where as the former relates to the residual friction coefficient instead, $f_r\Delta p_c\approx f_r\sigma_0^\prime-\tau_0$. We use the same name for these two regimes, yet there is this subtle difference between them. 

By introducing \eqref{eq:fracture-energy-marginally-pressurized} into \eqref{eq:dimensionless-toughnesses}b via \eqref{eq:K-Kc}, and then \eqref{eq:dimensionless-toughnesses}b into \eqref{eq:dimensionless-energy-balance}b, we obtain the dimensionless form of the front-localized energy balance in the marginally-pressurized regime,
\begin{equation}\label{eq:marginally-pressurized-energy-balance-slip-weakening}
    \Psi\left(\lambda\right) 
    - \mathcal{T}_r =
    \frac{\sqrt{\pi}}{2}\sqrt{2\kappa}\frac{1}{\sqrt{R/R_{mp}^*}},
\end{equation}
with
\begin{equation}\label{eq:Rmp-slip-weakening}
    \frac{R_{mp}^*}{R_w}=\frac{\left(f_p-f_r\right)^2\sigma_0^\prime\left(\sigma_0^\prime-\Delta p_c\right)}{\left(f_r \Delta p_*\right)^2}
    \approx\left(\frac{1-\mathcal{F}}{\mathcal{F}\mathcal{P}}\right)^2\left(1-10\mathcal{P}\right).
\end{equation}
In the latter equation, we have approximated the factor $4\pi$ by $10$ as usual in the marginally pressurized limit. Alternatively, equations \eqref{eq:marginally-pressurized-energy-balance-slip-weakening} and \eqref{eq:Rmp-slip-weakening} can be derived by approximating the term $\mathcal{P} E_{\text{1}}\left(\lambda^2\right)\approx 4\pi \mathcal{P}$ in \eqref{eq:slip-weakening-energy-balance-dimensionless} and then dividing the latter by $\mathcal{F}\mathcal{P}$. Moreover, by using the asymptotic behavior of $\Psi\left(\lambda\right)$ for small $\lambda$, equation \eqref{eq:asymptote-rupture-radius-marginally-pressurized} provides an implicit equation for the normalized rupture radius $R/R_{mp}^*$ as a function of the normalized square root of time $\sqrt{4\alpha t}/R_{mp}^*$ and the residual stress-injection parameter $\mathcal{T}_r$, provided that $R_{mp}^*$ is replaced by \eqref{eq:Rmp-slip-weakening}.

The meaning of the rupture length scales $R_{nu}^*$ and $R_{mp}^*$ are the same as in the constant fracture energy model. Essentially, when $R\ll R^*$, the fracture energy plays a dominant role in the near-front energy balance. This is, the large-toughness regime. On the other hand, when $R\gg R^*$, the fracture energy becomes increasingly less relevant in the rupture-front energy budget, corresponding to the small-toughness regime. Hence, likewise in the constant fracture energy model, unconditionally stable ruptures in the slip-weakening model will always transition from a large-toughness to a small-toughness regime. This transition is shown in figures \ref{fig:constant-fracture-energy}c and \ref{fig:constant-fracture-energy}d for both nearly unstable and marginally pressurized faults, respectively, with $R_{nu}^*$ and $R_{mp}^*$ as in equations \eqref{eq:Rnu-slip-weakening} and \eqref{eq:Rmp-slip-weakening}.

\subsection{Ultimate zero-fracture-energy similarity solution}

By taking the ultimate limit $R/R^*\to\infty$ in equations \eqref{eq:nearly-unstable-energy-balance-slip-weakening} and \eqref{eq:marginally-pressurized-energy-balance-slip-weakening}, it is evident that the fracture-energy term in the energy balance (the right-hand side) vanishes for both the nearly unstable ($\mathcal{T}_r\ll1$) and marginally pressurized regimes ($\mathcal{T}_r\sim10$). Hence, in this limit, equations \eqref{eq:nearly-unstable-energy-balance-slip-weakening} and \eqref{eq:marginally-pressurized-energy-balance-slip-weakening} become simply
\begin{equation}\label{eq:ultimate-propagation-condition-slip-weakening}
    \Psi(\lambda)=\mathcal{T}_r,
\end{equation}
which is exactly the rupture propagation condition of the constant friction model (equation \eqref{eq:propagation-condition-Coulomb-friction}) but with a constant friction coefficient $f_\text{cons}$ equal to the residual one $f_r$. The self-similar constant friction model is therefore the ultimate asymptotic solution of the slip-weakening model, provided that $f_\text{cons}=f_r$. 
The transition from the small-toughness regime to the constant residual friction solution, also denominated as ultimate zero-fracture-energy solution, is shown in figures \ref{fig:constant-fracture-energy}c and \ref{fig:constant-fracture-energy}d for both nearly unstable and marginally pressurized faults, respectively. Note that one could alternatively define a dimensionless toughness $\mathcal{K}$ for both regimes ($\lambda(t)\gg1$ and $\lambda(t)\ll1$) as in section \ref{sec:constant-fracture-energy} (equation \eqref{eq:dimensionless-toughnesses}) to show the same type of transition than in figures \ref{fig:constant-fracture-energy}a-b, since the solution for the amplification factor can be written as $\lambda\left(\mathcal{T}_r,\mathcal{K}(t)\right)$. Such a dimensionless toughness will always decrease with time and ultimately tend to zero, $\mathcal{K}\to0$.

Finally, using the constant residual friction solution $\lambda_r$, one can estimate as done in section \ref{sec:constant-fracture-energy} (equation \eqref{eq:time-scales-transition-Gc-constant}) the transition timescales between the large-toughness and small-toughness regimes, which results in 
\begin{equation}\label{eq:time-scales-transition-slip-weakening}
    t_{nu}^*\approx\frac{\left(R_{nu}^*\right)^2}{\alpha\lambda_r^2},\; \text{and}\quad t_{mp}^*\approx\frac{\left(R_{mp}^*\right)^2}{\alpha\lambda_r^2}.
\end{equation}

\section{The nucleation phase preceding a dynamic rupture}\label{sec:nucleation-phase}

\begin{figure}
    \centering
    \includegraphics[width=15cm]{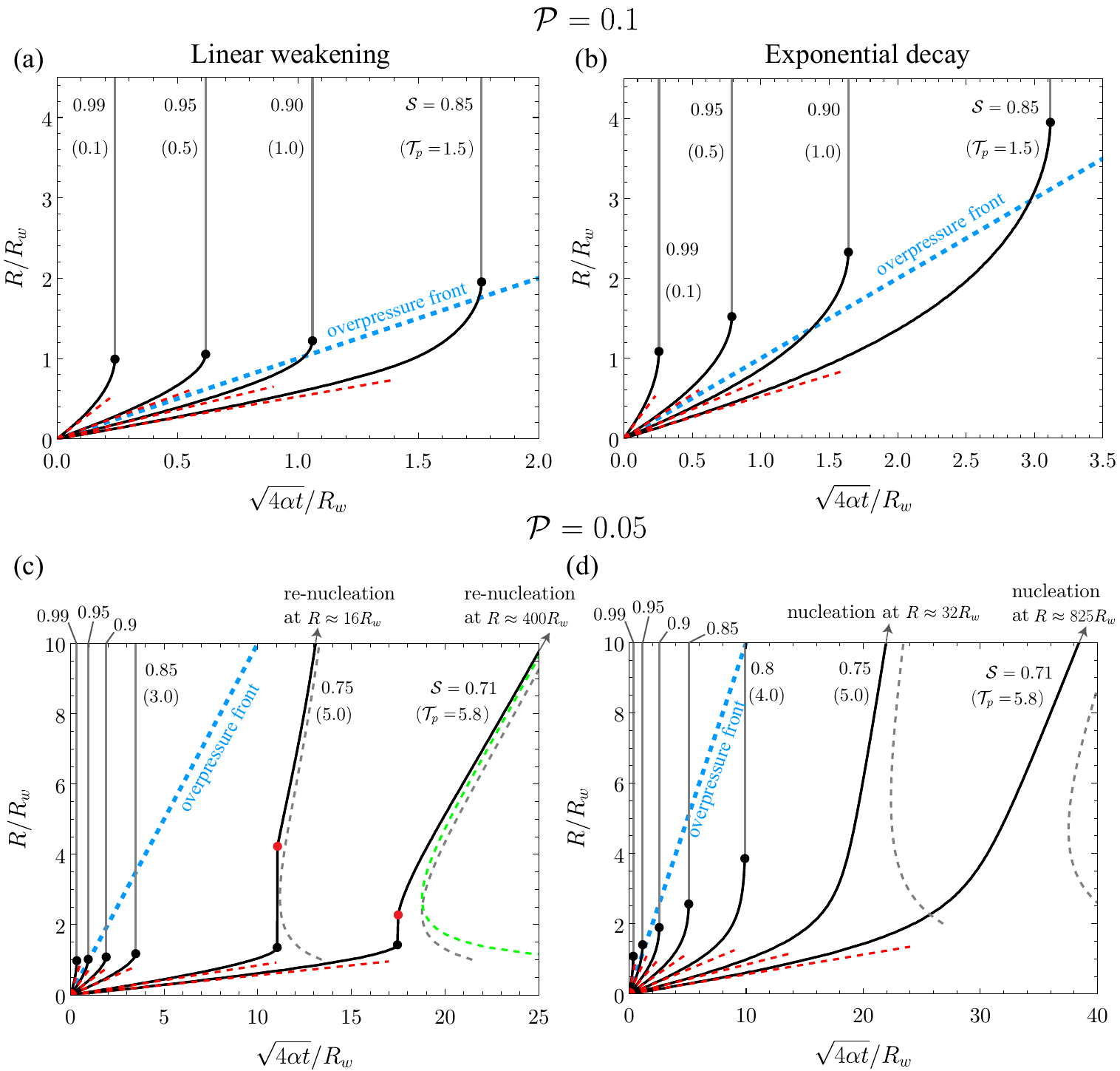}
    \caption{Ultimately unstable faults, $\mathcal{S}>\mathcal{F}=0.7$. Normalized rupture radius versus square root of dimensionless time for (left) linear weakening and (right) exponential decay versions of the slip weakening friction law. (a, b) $\mathcal{P}=0.1$ and (c, d) $\mathcal{P}=0.05$. Black and red circles indicate the nucleation and arrest of a dynamic rupture, respectively. Red dashed lines correspond to the analytical Coulomb's friction model considering the peak friction coefficient. In (c) and (d), gray dashed lines represent the near-front energy balance solution. In (c), the green dashed line corresponds to an improved energy-balance solution as explained in the main text.}
    \label{fig:ultimately-unstable-ruptures}
\end{figure}
\begin{figure}
    \centering
    \includegraphics[width=15.6cm]{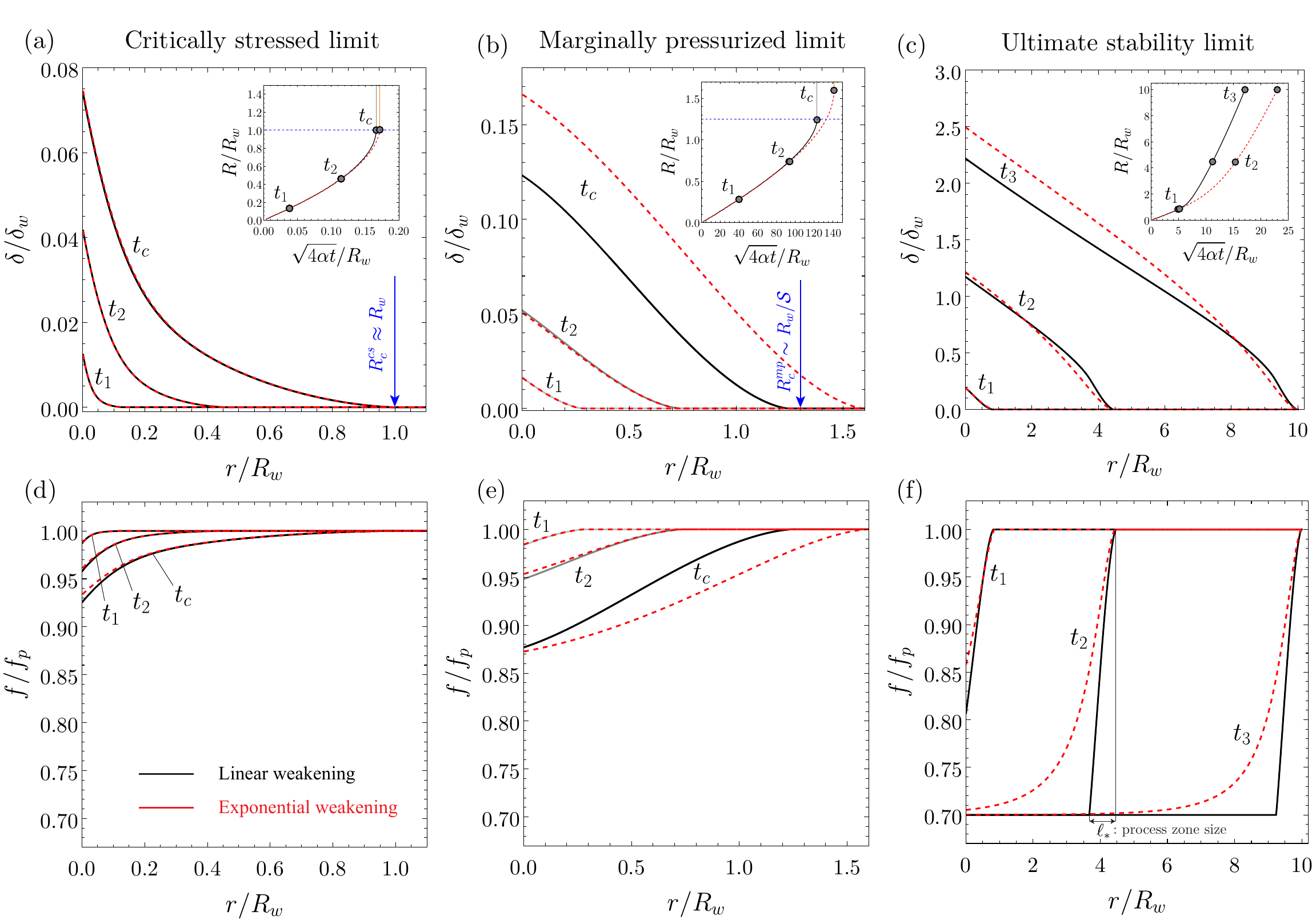}
    \caption{(a, b, c) Normalized spatial distribution of slip at the times indicated in the insets and commented in the main text. Insets: normalized rupture radius as a function of the normalized squared root of time. Blue arrows in (a) and (b) represent the theoretical prediction for the nucleation radius in the critically stressed and marginally pressurized limits. (d, e, f) Spatial profile of the normalized friction coefficient at the same times indicated previously. (Left) Critically stressed limit, $\mathcal{S}=0.995$, $\mathcal{P}=0.1$, $\mathcal{F}=0.7$, and associated $\mathcal{T}_p=0.05$. (Center) Marginally pressurized limit, $\mathcal{S}=0.8$, $\mathcal{P}=0.02$, $\mathcal{F}=0.7$, and associated $\mathcal{T}_p=10$. (c, f) Ultimate stability limit, $\mathcal{S}=0.71$, $\mathcal{P}=0.075$, and $\mathcal{F}=0.7$.}
    \label{fig:ultimately-unstable-ruptures-profiles}
\end{figure}

Figure \ref{fig:ultimately-unstable-ruptures} displays the case of ultimately unstable ruptures: $f_r\sigma_0^\prime<\tau_0$. Again, without loss of generality, we fix the residual-to-peak friction ratio as $\mathcal{F}=0.7$, and examine the parameter space for $\mathcal{P}$ and now $\mathcal{S}>0.7$, for both versions of the slip-weakening friction law. The case of an overpressure ratio $\mathcal{P}=0.1$ which corresponds to an injection that is about to open the fault, is shown in figures \ref{fig:ultimately-unstable-ruptures}a and \ref{fig:ultimately-unstable-ruptures}b. For all values of $\mathcal{S}$ in these figures, we observe the nucleation of a dynamic rupture that runs away and never stop within the limits of our model, this is, the regime R3 in figure \ref{fig:map-regimes}. On the other hand, the case of a lower overpressure ratio $\mathcal{P}=0.05$ is shown in figures \ref{fig:ultimately-unstable-ruptures}c and \ref{fig:ultimately-unstable-ruptures}d. For the linear weakening model, we observe the occurrence of both regimes R3 and R4 of figure \ref{fig:map-regimes}. The latter corresponds to cases in which a dynamic rupture nucleates, propagates and arrests, with a new rupture instability nucleating afterwards on the same fault, which is ultimately unstable (run-away). Moreover, in figure \ref{fig:ultimately-unstable-ruptures}d for the exponential weakening version of the friction law, we do not observe the regime R4, at least for the numerical solutions we include in this figure.

\subsection{Early-time Coulomb's friction stage and acceleration towards rupture instability}\label{sec:early-time-Coulomb-friction}

Similarly to the case of ultimately stable ruptures, here, unstable ruptures are also well-described by the Coulomb's friction model at early times (see figure \ref{fig:ultimately-unstable-ruptures}). Therefore, the rupture radius evolves approximately as equation \eqref{eq:R-with-peak-friction}, with $\lambda_p$ given by equation \eqref{eq:analytical-solution-Coulomb-friction} considering the peak stress-injection parameter $\mathcal{T}_p$. Moreover, figure \ref{fig:ultimately-unstable-ruptures} also shows that during the nucleation phase, the slip front may largely outpace the overpressure front ($\lambda_p\gg1$) when faults are critically stressed as quantified by the peak stress-injection parameter ($\mathcal{T}_p\ll1$), or significantly lag the overpressure front ($\lambda_p\ll1$) when faults are marginally pressurized ($\mathcal{T}_p\sim10$). Figures \ref{fig:ultimately-unstable-ruptures-profiles}d and \ref{fig:ultimately-unstable-ruptures-profiles}e display, on the other hand, the spatial distribution of the friction coefficient for critically stressed and marginally pressurized cases, respectively. We can clearly observe that at the Coulomb's friction stage, $f/f_p\approx 1$ throughout most of the slipping region. Note that in this stage, we could also approximate the spatio-temporal evolution of fault slip in the critically stressed and marginally pressurized regimes, using the analytical asymptotic expressions derived by Sáez \textit{et al.} \cite{Saez_Lecampion_2022} (equation 25 and 26 in \cite{Saez_Lecampion_2022}), provided that $f_\text{cons}=f_p$. The same can be done for the ultimate zero-fracture-energy solution of ultimately stable ruptures, with $f_\text{cons}=f_r$.

Now, beyond this early-time stage, the propagation of the slip front starts departing from the Coulomb's friction similarity solution due to the further weakening of friction. The latter can be seen in figures \ref{fig:ultimately-unstable-ruptures-profiles}d and \ref{fig:ultimately-unstable-ruptures-profiles}e for intermediate times ($t_2$) and times close to nucleation ($t_c$). The slip front accelerates indeed towards the nucleation of a dynamic rupture. Figure \ref{fig:ultimately-unstable-ruptures} shows that the rupture radius at the instability time increases with decreasing pre-stress ratio $\mathcal{S}$ and increasing overpressure ratio $\mathcal{P}$, for both versions of the friction law. This is consistent with the extensive analysis on earthquake nucleation provided by Garagash and Germanovich \cite{Garagash_Germanovich_2012} for the two-dimensional, linear weakening model. Moreover, figure \ref{fig:ultimately-unstable-ruptures} also displays that one of the main effects of the exponential weakening version of the friction law is to smooth the transition of the rupture towards the dynamic instability with regard to the linear law. Furthermore, the exponential law retards the instability time, and generally increases the critical radius for the rupture to become unstable. Such effect of the exponential law becomes stronger when the pre-stress ratio $\mathcal{S}$ decreases towards its minimum value in the ultimately unstable case, this is, the ultimate stability limit $\mathcal{S}=\mathcal{F}$.

Given the importance of the nucleation radius to characterize the maximum size that quasi-static ruptures can afford in the ultimately unstable case, we calculate in the next sections theoretical bounds for it, following the procedure of Uenishi and Rice \cite{Uenishi_Rice_2002} and Garagash and Germanovich \cite{Garagash_Germanovich_2012}. This corresponds to an extension of their results from the two-dimensional (mode II or III) fault model to the three-dimensional circular (modes II+III) configuration.

\subsection{Theoretical bounds for the nucleation radius}

\subsubsection{Critically stressed and marginally pressurized limits}

As shown in Appendix \ref{sec:generalized-eigenproblem} and \ref{sec:regular-eigenproblem}, at the time of instability $t_c$, the time derivative of the quasi-static elastic equilibrium throughout the slipping region takes the form of the following eigenvalue problem for both the critically stressed ($\mathcal{T}_p\ll1$) and marginally pressurized ($\mathcal{T}_p\sim10$) regimes:
\begin{equation}\label{eq:eigen-value-equation}
    \frac{1}{2\pi}\int_{0}^{1}F\left(\bar{r},\bar{\xi}\right)\frac{\partial \bar{v}(\bar{\xi})}{\partial\bar{\xi}}\mathrm{d}\bar{\xi} = 
    \beta\bar{v}(\bar{r}),
\end{equation}
where $\bar{v}=v/v_\text{rms}$ is the normalized slip rate distribution (with $v_\text{rms}$ given by equation \eqref{eq:v-rms-definition}), and $\beta$ is the eigenvalue
\begin{equation}\label{eq:eigenvalue}
    \beta=\frac{R}{R_w}\cdot
    \begin{cases}
        1 & \text{for critically stressed faults }\mathcal{T}_p\ll1\\
        \nicefrac{\tau_0}{f_p\sigma_0^\prime} & \text{for marginally pressurized faults }\mathcal{T}_p\sim10.
    \end{cases}
\end{equation}
In the previous equations, the dependence of $\bar{v}$ and $R$ on the instability time $t_c$ has been omitted for simplicity. Moreover, equations \eqref{eq:eigen-value-equation} and \eqref{eq:eigenvalue} are valid not only for the linear weakening friction law \eqref{eq:linear-weakening-friction}, but also for the exponential one \eqref{eq:exponential-weakening-friction}. This is due to both the critically stressed and marginally pressurized limits are characterized by small slip at the nucleation time, $\delta(r=0,t_c)\ll\delta_c$ (see, for example, figure \ref{fig:ultimately-unstable-ruptures-profiles}). It takes just a simple Taylor expansion to show that in this range of slip, the exponential weakening version of the friction law is, to first order in $\delta/\delta_c$, asymptotically equal to the linear weakening case. This also means that the residual branch of the linear weakening law does not need to be considered in such stability analysis.

The solution of \eqref{eq:eigen-value-equation} for the eigenvalues and eigenfunctions is calculated in Appendix \ref{sec:solution-eigenproblem}. This is done by discretizing the linear integral operator on the left-hand side of \eqref{eq:eigen-value-equation} via a collocation boundary element method using piece-wise ring `dislocations' of constant slip rate. The most important result is the smallest eigenvalue $\beta_1$, which was shown by Uenishi and Rice \cite{Uenishi_Rice_2002} for the two-dimensional problem to give the critical nucleation radius. We find (see Table \ref{table:eigen-values})
\begin{equation}\label{eq:eigenvalue-beta1}
    \beta_1\approx1.003,
\end{equation}
which is interestingly, for all practical purposes, approximately equal to one. Taking hereafter $\beta_1\approx1$, the nucleation radius is recovered from equation \eqref{eq:eigenvalue} as
\begin{equation}\label{eq:nucleation-radius-critically-stressed}
    R_c^{cs}\approx R_w
\end{equation}
for critically stressed faults ($\mathcal{T}_p\ll1$, vertical line $\mathcal{S}=1$ on the right side of figure \ref{fig:map-regimes}), and
\begin{equation}\label{eq:nucleation-radius-marginally-pressurized}
    R_c^{mp}\approx \frac{f_p\sigma_0^\prime}{\tau_0}R_w=\frac{R_w}{\mathcal{S}},
\end{equation}
for marginally pressurized faults ($\mathcal{T}_p\sim10$, inclined line $\mathcal{P}\approx 10^{-1}(1-\mathcal{S})$ in figure \ref{fig:map-regimes}).

The theoretical estimates \eqref{eq:nucleation-radius-critically-stressed} and \eqref{eq:nucleation-radius-marginally-pressurized} are compared to numerical solutions that are representative of each limiting regime in figures \ref{fig:ultimately-unstable-ruptures-profiles}a and \ref{fig:ultimately-unstable-ruptures-profiles}b, respectively. In these figures, the blue arrows indicate the theoretical radii at the instability time $t_c$. We highlight that the critically stressed nucleation radius \eqref{eq:nucleation-radius-critically-stressed} is a proper asymptote that is always reached in the limit $\tau_0\to f_p\sigma_0^\prime$, up to the numerical approximation made for the eigenvalue \eqref{eq:eigenvalue-beta1}. On the other hand, the marginally pressurized nucleation radius \eqref{eq:nucleation-radius-marginally-pressurized} can be defined only in an approximate sense due to the reasons explained in Appendix \ref{appendix:line-source}. Although this approximation seems to be quite accurate for the linear weakening law (see figure \ref{fig:ultimately-unstable-ruptures-profiles}b), the exponential decay version does not seem to follow this trend. This is likely due to the additional assumption of small slip that the exponential weakening law requires in order to be well approximated by a linear relation. In the example of figure \ref{fig:ultimately-unstable-ruptures-profiles}b, slip does not seem to be small enough. Because of the approximate nature of the marginally pressurized limit, it is challenging to find the model parameters that will result in sufficiently small slip at the nucleation time for the linear approximation of the exponential weakening law to be valid. Furthermore, equations \eqref{eq:nucleation-radius-critically-stressed} and \eqref{eq:nucleation-radius-marginally-pressurized} suggest that the minimum possible nucleation radius is the one associated with critically stressed faults \eqref{eq:nucleation-radius-critically-stressed}, whereas the greatest possible nucleation radius can be as large as infinity for marginally pressurized faults \eqref{eq:nucleation-radius-marginally-pressurized}, in the limit of zero pre-stress $\tau_0\to0$. Yet such a limit corresponds indeed to an ultimately stable rupture, specifically, a case in which the fault is about to open (top left corner of figure \ref{fig:map-regimes}), so that the dynamic rupture will eventually arrest and then propagate ultimately in a quasi-static manner.

Finally, as shown in Appendix \ref{appendix:eigenvalue-problem}, the nucleation radius in the critically stressed limit \eqref{eq:nucleation-radius-critically-stressed} is independent of the specific form of the spatio-temporal evolution of pore pressure, that is, equation \eqref{eq:nucleation-radius-critically-stressed} is also valid for other type of fluid injections than the constant volumetric rate considered in this study. On the other hand, the nucleation radius in the marginally pressurized limit \eqref{eq:nucleation-radius-marginally-pressurized} is, under certain conditions (see details in Appendix \ref{sec:regular-eigenproblem}), also independent of the injection scenario. 
Moreover, the critically stressed nucleation radius \eqref{eq:nucleation-radius-critically-stressed} is itself an extension of the nucleation length of Uenishi and Rice \cite{Uenishi_Rice_2002} (found also previously by Campillo and Ionescu \cite{Campillo_Ionescu_1997} under different assumptions) from their two-dimensional fault model to the three-dimensional axisymmetric configuration. Since for the shear mixed-mode (II+III) rupture, the circular front shape is strictly valid only when $\nu=0$, our results could be used in combination with perturbation techniques such as the work of Gao \cite{Gao_1988} to characterize the corresponding non-circular slipping region at the nucleation time of a shear rupture for $\nu\neq0$. Indeed, since the work of Gao \cite{Gao_1988} is based on linear elastic fracture mechanics (valid in the small-scale yielding limit) and the nucleation radius \eqref{eq:nucleation-radius-critically-stressed} (and \eqref{eq:nucleation-radius-marginally-pressurized}) is smaller than the process zone size, one should rather consider a variational approach as the one proposed recently by Lebihain \textit{et al}. \cite{Lebihain_Roch_2022} for cohesive cracks based on the perturbation of crack face weight functions. The approach of Gao \cite{Gao_1988} would be still useful to characterize non-circularity in the nearly stable limit of the next section. This would provide an alternative to the work of Uenishi \cite{Uenishi_2018} who considered an energy approach and fixed the rupture shape to an ellipse. An elliptical rupture shape may be a very good approximation for a shear rupture \cite{Saez_Lecampion_2022}, yet not necessarily the actual equilibrium shape. Finally, for a tensile (mode I) rupture, the nucleation radius \eqref{eq:nucleation-radius-critically-stressed} is valid for any value of $\nu$, as long as the load driving the rupture growth is peaked around the crack center and axisymmetric in magnitude. Further details about this generalization of our results can be found in Appendix \ref{appendix:eigenvalue-problem}.

\subsubsection{Nearly stable limit}

Figures \ref{fig:ultimately-unstable-ruptures}c and \ref{fig:ultimately-unstable-ruptures}d show that the nucleation radius of ultimately unstable ruptures becomes very large, $R(t_c)\gg R_w$, when approaching the ultimate stability condition $f_r \sigma_0^{\prime}\to \tau_0$ (vertical line $\mathcal{S}=\mathcal{F}$ in figure \ref{fig:map-regimes}). Specifically, figure \ref{fig:ultimately-unstable-ruptures}c displays a case of large re-nucleation radius (regime R4) in the linear weakening model for $\mathcal{S}=0.71$ ($\mathcal{F}=0.7$), whereas figure \ref{fig:ultimately-unstable-ruptures}d shows an example of large nucleation radius for the exponential weakening model and the same parameters than before. In the two-dimensional model, Garagash and Germanovich \cite{Garagash_Germanovich_2012} not only found this same behavior but provided also an asymptote for the nucleation length in this limit, that we also derive here for the circular rupture model. 

First, let us note that the condition $R(t_c)\gg R_w$ implies also that $R(t_c)\gg \ell_*$, since the process zone size $\ell_*$ for the linear weakening model is roughly around the same order than the elasto-frictional lengthscale $R_w$, and about an order of magnitude larger in the case of the exponential weakening law. Hence, we can invoke the front-localized energy balance, equation \eqref{eq:final-energy-balance}. Indeed, figures \ref{fig:ultimately-unstable-ruptures}c and \ref{fig:ultimately-unstable-ruptures}d display such energy-balance solution for some ruptures that are on their way to become unstable. On the other hand, near the ultimate stability limit, $R(t_c)$ is also much bigger than the radius of the overpressure front $L(t_c)$, such that $\lambda(t_c)\gg1$. Therefore, we can approximate the equivalent shear load associated with the fluid source as a point force via equation \eqref{eq:point-force}. The corresponding axisymmetric stress intensity factor for such a point force comes from resolving the integral of the left-hand side of equation \eqref{eq:final-energy-balance} considering \eqref{eq:point-force} and $f_\text{cons}=f_r$, which gives
\begin{equation}
    K_{p}= \frac{f_r \Delta P(t)}{(\pi R)^{3/2}}
\end{equation}
with $\Delta P(t)=4\pi\alpha t \Delta p_*$. Substituting the previous equation into \eqref{eq:final-energy-balance} leads to following form of the front-localized energy balance,
\begin{equation}\label{eq:point-force-energy-balance}
    \underbrace{\frac{f_r \Delta P(t)}{(\pi R)^{3/2}}}_{K_{p}} +
    \underbrace{\frac{2}{\sqrt{\pi}}\left[\tau_{0}-f_{r}\sigma_{0}^{\prime}\right]\sqrt{R}}_{K_{\tau}}
    =
    K_c,
\end{equation}
where the fracture toughness $K_c=\sqrt{2\mu G_c}$ is, after combining equations \eqref{eq:slip-scale-rupture-length-scale-slip-weakening} and \eqref{eq:fracture-energy-nearly-unstable}, equal to
\begin{equation}
    K_c=(f_p-f_r)\sigma_0^\prime\sqrt{2\kappa R_w}.
\end{equation}
We recall that the coefficient $\kappa$ is equal to 1/2 for the linear weakening friction law, and 1 for the exponential weakening case. Moreover, the fracture toughness $K_c$ is constant due to the negligible overpressure within the process zone when $\lambda\gg1$.

By differentiating equation \eqref{eq:point-force-energy-balance} with respect to time and then dividing by $\dot{R}$ on both sides, one can show upon taking the limit at the nucleation time: $R\to R_c$ and $\dot{R}\to\infty$, that $K_{p}=K_{\tau}/3$. Substituting this previous relation into \eqref{eq:point-force-energy-balance} allows us to eliminate $\Delta P(t_c)$ and so the instability time $t_c$ in the equation, leading to the sought critical nucleation radius:
\begin{equation}\label{eq:nucleation-radius-stability-limit}
    \frac{R_c^\infty}{R_w}\simeq
    \frac{9\pi\kappa}{32}\left(\frac{f_p\sigma_0^\prime-f_r\sigma_0^\prime}{\tau_0-f_p\sigma_0^\prime}\right)^2=
    \frac{9\pi\kappa}{32}\left(\frac{1-\mathcal{F}}{\mathcal{S}-\mathcal{F}}\right)^2,\text{ when } f_r\sigma_0^{\prime} \to \tau_0.
\end{equation}
Note that the previous equation is a proper asymptote due to the small-scale yielding approximation. In addition, the relation $K_{p}=K_{\tau}/3$ plus the previous expression for $R_c^\infty$ can provide together an expression for the nucleation time $t_c$. Indeed, expressions for the instability time $t_c$ in the critically stressed and marginally pressurized limits might be also obtainable analytically, via asymptotic analysis as conducted by Garagash and Germanovich \cite{Garagash_Germanovich_2012}, yet we do not attempt to pursue this route in this paper.

\section{Discussion}\label{sec:discussion}

\subsection{Frustrated dynamic ruptures and unconditionally stable slip: the two propagation modes of injection-induced aseismic slip}

In our model, injection-induced aseismic slip can be the result of either a frustrated dynamic rupture that did not reach the required size to become unstable, or the propagation of slip that is unconditionally stable. Whether injection-induced aseismic ruptures occur in one regime or the other, depends primarily on the ultimate stability condition of Garagash and Germanovich \cite{Garagash_Germanovich_2012}, that we demonstrated here to be applicable to the circular rupture configuration as well (equation \eqref{eq:ultimate-stability-conditions}).

\subsubsection{Unconditionally stable ruptures}

\begin{figure}
    \centering
    \includegraphics[width=14cm]{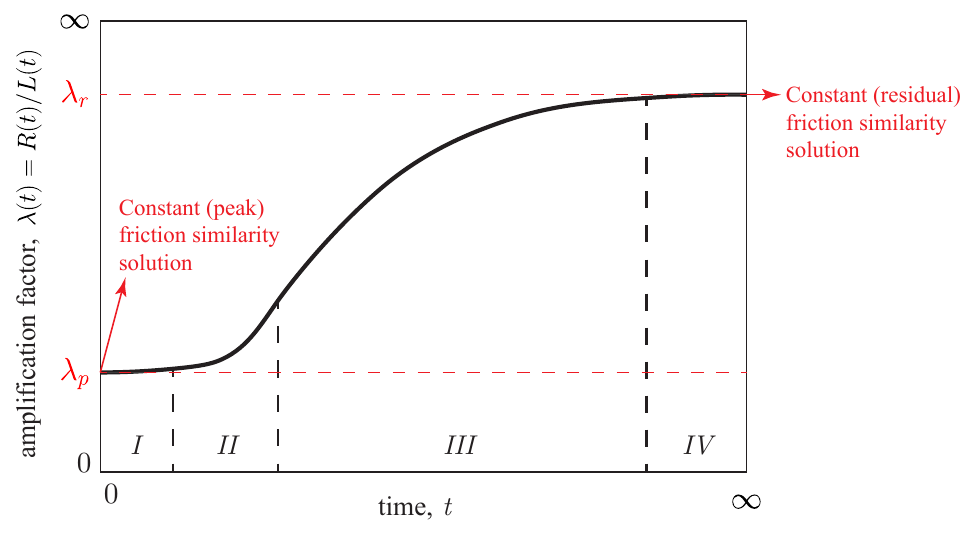}
    \caption{Schematic solution for unconditionally stable ruptures undergoing four distinct stages in time. Stage (I), Coulomb's friction similarity solution, $f_\text{cons}=f_p$. Stage (II), acceleration due to frictional weakening and localization of the process zone. Stage (III), rupture is governed by an energy balance of the Griffith's type, $G=G_c$, transitioning from a large-toughness to a small-toughness regime. Stage (IV), ultimate constant residual friction similarity solution, $f_\text{cons}=f_r$, also denominated ultimate zero-fracture-energy solution. Note that in the case of aseismic slip as a frustrated dynamic instability, stage I is always present, while stages II and III might be experienced to different extents depending on how large the nucleation radius is compared to the process zone size.}
    \label{fig:schematic-solution}
\end{figure}

When the initial shear stress $\tau_0$ is lower than the in-situ residual fault strength, $\tau_0<f_r\sigma_0^\prime$, faults tend to produce mostly unconditionally stable ruptures (regime R1 in figure \ref{fig:map-regimes}), except for a relatively narrow range of parameters where the nucleation of a dynamic rupture occurs, followed by arrest and purely quasi-static slip (regime R2 in figure \ref{fig:map-regimes}). We found that unconditionally stable ruptures evolve always between two similarity solutions (see figure \ref{fig:schematic-solution}). At early times (stage I), they behave as being governed by Coulomb's friction, that is, a constant friction coefficient equal to the peak value $f_p$. During this initial stage, fault slip is self-similar in a diffusive manner and is governed by one single dimensionless number: the peak stress-injection parameter $\mathcal{T}_p$. After, the response of the fault gets more complex in stages II and III yet ultimately, slip recovers the same type of similarity at very large times (stage IV). In this ultimate regime, the rupture behaves as if it were governed by a constant friction coefficient equal to the residual one $f_r$, and depends also on one single dimensionless number: the residual stress-injection parameter $\mathcal{T}_r$. An interesting characteristic in both limiting regimes is that the rupture propagates as having zero fracture energy $G_c$. While at early times $G_c=0$ in an absolute sense as the process zone has not developed yet, at large times the contribution of the finite fracture energy to the rupture-front energy balance is to leading order negligible compared to the other terms that drive the propagation of the rupture. Furthermore, the two similarity solutions are equivalent to the analytical solution for a constant friction coefficient derived in \cite{Saez_Lecampion_2022}, as long as the so-called stress-injection parameter $\mathcal{T}$ in \cite{Saez_Lecampion_2022} is replaced by $\mathcal{T}_p$ at early times and $\mathcal{T}_r$ at large times, which are then associated with constant amplification factors $\lambda_p$ and $\lambda_r$, respectively, as shown in figure \ref{fig:schematic-solution}. This is a key finding of our work as it puts the results of the former constant-friction model of Sáez \textit{et al.} \cite{Saez_Lecampion_2022} in a more complete picture of the problem of injection-induced aseismic slip.

In between the two similarity solutions, fault slip undergoes two subsequent stages. First, after departing from the Coulomb's friction solution, the rupture accelerates due to frictional weakening (stage II). The details of the friction law matter here as the rupture radius is of the same order than the process zone size. The exponential weakening law tends to slow down the propagation of slip and smooth the acceleration phase with regard to the linear weakening case when considering the same $\delta_c$ in both laws. Fault slip depends in addition to dimensionless space and time, on three non-dimensional parameters: the pre-stress ratio $\mathcal{S}$, the overpressure ratio $\mathcal{P}$, and the residual-to-peak friction ratio $\mathcal{F}$. The higher the initial shear stress on the fault is (higher $\mathcal{S}$) or the more intense the injection is (higher $\mathcal{P}$), the faster the rupture propagates. Note that this dependence of the rupture speed on $\mathcal{S}$ and $\mathcal{P}$ is embedded in both the peak stress-injection parameter $\mathcal{T}_p$ and residual stress-injection parameter $\mathcal{T}_r$. Therefore, it is a general feature present in all stages of injection-induced aseismic slip.

In a subsequent stage, once the process zone has adequately localized, the evolution of the slip front is well approximated by the rupture-front energy balance (stage III). The details of how the friction coefficient weakens from its peak value towards its residual value no longer matter in relation to the position of the slip front or the rupture speed. The only two important quantities here associated with the friction law are: the amount of fracture energy that is dissipated near the rupture front, and the residual friction coefficient $f_r$. Moreover, the fracture energy is approximately constant (albeit of different magnitude) in the two end-member cases of nearly unstable ($\lambda(t)\gg1$) and marginally pressurized faults ($\lambda(t)\ll1$), with the amplification factor $\lambda$ depending on only two dimensionless numbers: a time-dependent dimensionless toughness $\mathcal{K}(t)$ and the residual stress-injection parameter $\mathcal{T}_r$. A constant fracture energy model as the one introduced in section \ref{sec:constant-fracture-energy} is thus sufficient to capture the dynamics of the slip front for the two end-members. The dimensionless toughness $\mathcal{K}(t)$ quantifies the relevance of the dissipation of fracture energy in the rupture-front energy balance, which decreases monotonically with time. The rupture speed thus increases with time as the diminishing effect of the fracture energy offers less `opposition' for the rupture to advance. Eventually, $\mathcal{K}(t)\to0$ when $t\to\infty$ and the rupture reaches asymptotically the large-time similarity solution (stage IV), where the only information about the friction law that matters is $f_r$. 

Finally, the residual stress-injection parameter $\mathcal{T}_r$ plays a crucial role in stages III and IV. When faults are near the ultimate stability limit ($\mathcal{T}_r\to0$) thus responding in the so-called nearly unstable regime ($\mathcal{T}_r\ll1$), the slip front always outpace the overpressure front $\lambda(t)\gg1$, even though at early times (stage I) the rupture front would likely lag the overpressure front $\lambda(t)\ll1$. Conversely, when faults operate in the so-called marginally pressurized regime ($\mathcal{T}_r\sim10$), the slip front will always move much slower than the overpressure front, $\lambda(t)\ll1$, over the entire lifetime of the rupture: the slip front will never outpace the overpressure front.

\subsubsection{Aseismic slip as a frustrated dynamic instability}

When the initial shear stress $\tau_0$ is greater than the in-situ residual fault strength, $\tau_0>f_r\sigma_0^\prime$, faults will always host a dynamic event, sometimes even more than one (regime R4 in figure \ref{fig:map-regimes}) if injection is sustained for sufficient time. The maximum size that aseismic ruptures can reach before becoming unstable is as small as the elasto-frictional length scale $R_w$ for faults that are critically stressed, and as large as infinity for faults that are either marginally pressurized and about to open, or near the ultimate stability limit (so-called nearly stable faults). The spatial range for a fault to exhibit aseismic slip as a frustrated dynamic instability is therefore extremely broad. Moreover, similarly to the case of unconditionally stable ruptures, the quasi-static nucleation phase is governed at early times by the Coulomb’s friction similarity solution with $f_\text{cons}=f_p$ (stage I in figure \ref{fig:schematic-solution}). Aseismic ruptures can therefore move much faster than the diffusion of pore pressure right after the start of fluid injection when faults are critically stressed according to the peak stress-injection parameter ($\mathcal{T}_p\ll1$, $\lambda_p\gg1$), or propagate much slower than that when faults operate in the so-called marginally pressurized regime ($\mathcal{T}_p\sim10$, $\lambda_p\ll1$).  

After, depending on how large the nucleation radius is with regard to the process zone size, aseismic ruptures may be able to either partially or fully explore stages II and III on their way to reach their critical unstable size. In stage II, the rupture behavior is the same as described in the previous section for unconditionally stable ruptures. Moreover, if the critical nucleation radius is sufficiently large compared to the process zone size, the rupture could transit stage III where the propagation of the slip front is well approximated by an energy balance of the Griffith's type, similarly to stage III of unconditionally stable ruptures. 

Finally, it is worth noting that critically stressed faults and marginally pressurized faults nucleate dynamic ruptures with very little decrease of the friction coefficient, far from reaching the residual friction value over the entire slipping region at the instability time. Conversely, nearly stable ruptures undergo dynamic nucleation in a `crack-like' manner, that is, with a small process zone where the fracture energy is dissipated and the remaining much larger part of the `fracture' (slipping area) is at the residual friction level. In between these two limiting behaviors of dynamic rupture nucleation, a continuum of instabilities are spanned in our model; a result already found by Garagash and Germanovich \cite{Garagash_Germanovich_2012} and also present in heterogeneous, mechanically-loaded slip-weakening frictional interfaces \cite{Castellano_Lorez_2023}.

\subsection{Laboratory experiments}

Laboratory experiments of injection-induced fault slip on rock samples where a finite rupture grows along a pre-existing interface \cite{Passelegue_Almakari_2020,Cebry_Ke_2022} have recently confirmed some insights predicted by theory. For instance, the meter-scale experiments of Cebry \textit{et al.} \cite{Cebry_Ke_2022} showed that the closer to frictional failure the fault is before the injection starts, the faster aseismic slip propagates. A somewhat similar observation was made previously by Passelègue \textit{et al.} \cite{Passelegue_Almakari_2020} through a set of centimeter-scale experiments and using a rupture-tip energy balance argument. This general feature of injection-induced aseismic slip can be particularly seen in the closed-form expression for the rupture speed of critically stressed faults responding in the Coulomb's friction stage: $V_r=\left[\left(f_p\Delta p_*/\left(f_p\sigma_0^\prime-\tau_0\right)\right)\left(\alpha/2t\right)\right]^{1/2}$ \cite{Saez_Lecampion_2022}. This formula displays, in addition to the previous stress state dependence, some other general and quite intuitive features of injection-induced aseismic slip: the more intense the injection is, or the higher the fault hydraulic diffusivity is; the faster the rupture propagates. The latter dependencies remain to be seen in the laboratory, as well as many other aspects of injection-induced fault slip. We discuss a few of them in the context of published experiments in what follows. Notably, our results provide a means for characterizing the conditions under which distinct regimes and stages of injection-induced aseismic slip are expected to emerge under well-controlled conditions in the laboratory, where the validation of the relevant physics incorporated in our model can be potentially realized.

For example, the type of experiments conducted by Passelègue \textit{et al.}'s \cite{Passelegue_Almakari_2020} 
 could provide important insights into the aseismic slip phase preceding dynamic ruptures. Indeed, the nucleation radius of critically stressed faults, equation \eqref{eq:nucleation-radius-critically-stressed}, which is itself the circular analog of Uenishi and Rice's nucleation length \cite{Uenishi_Rice_2002}, 
has been estimated under somewhat similar laboratory conditions (confining pressures $\sim 100$ MPa) in $R_{c}^{cs}\sim1$ m \cite{Uenishi_Rice_2002}. Variations of this nucleation radius could be reasonably expected due to uncertainties mostly in the critical slip-weakening scale $\delta_c$. Moreover, a nucleation length of about $1$ m has been recently measured by Cebry \textit{et al.} \cite{Cebry_Ke_2022} during meter-scale experiments of fluid injection with fault normal stresses of about $4$ MPa. Since the experiments of Passelègue \textit{et al.} were carried out in a similar rock, if ones corrects the nucleation length of Cebry \textit{et al.} \cite{Cebry_Ke_2022} by the effective normal stresses that are representative of Passelègue \textit{et al.}'s experiments, we obtain roughly $R_{c}^{cs}\sim10$ cm. Since the critically stressed nucleation radius \eqref{eq:nucleation-radius-critically-stressed} is the minimum possible nucleation size of injection-induced dynamic ruptures in our model, and given that Passelègue \textit{et al.}'s cylindrical rock samples have a diameter of 4 cm, we expect that their aseismic ruptures might have likely been operating in the Coulomb's friction phase (stage I) and perhaps some excursion in the acceleration phase towards a dynamic instability (stage II), yet never reached the onset of a macroscopic dynamic rupture in the sample. Moreover, as the initial shear stress was set to be 90 percent of the `in-situ' static fault strength, it is likely that the initial shear stress was greater than the residual strength of the fault thus further supporting the ultimately unstable condition assumed for these experiments.

Another interesting set of fluid injection experiments are the ones reported by Cebry \textit{et al.} \cite{Cebry_Ke_2022}. Their 3-meters long, quasi-one-dimensional fault, allowed them to observed not only the nucleation of injection-induced dynamic ruptures, but also some details of the quasi-static phase preceding such instabilities
. Due to the elasticity and fluid flow boundary conditions in their experimental setup, it is not possible to make direct quantitative comparisons with our three-dimensional model, nor with two-dimensional plane-strain models \cite{Garagash_Germanovich_2012}
. An unbounded rupture may have certainly propagated before fault slip reached the shortest side of the sample, yet most of the measurements were conducted starting from this moment. 
In spite of these differences, it is interesting to note at least two experimental observations that are qualitatively consistent with our model: i) the aseismic slip front continuously decelerates during fluid injection except for the moment right before instabilities occur, and ii) the transition from self-arrested to run-away dynamic ruptures seems to have occurred in relation to the ultimate stability condition \eqref{eq:ultimate-stability-conditions}. This type of experiments are well suited to be analyzed via dynamic numerical modeling to provide the possibility for direct quantitative comparisons.

Finally, using rock analog materials with reduced shear modulus such as PMMA \cite{Cebry_McLaskey_2021,Gori_Rubino_2021} could provide important insights into injection-induced aseismic slip by reducing the elasto-frictional length scale $R_w$ in approximately one order of magnitude. This ``widens'' the observable spatial range of the problem thus providing the chance to explore larger-scale processes and regimes that would be otherwise difficult to observe in the laboratory using rock samples. For instance, stage III, where injection-induced aseismic slip is governed by an energy balance of the Griffith's type, or perhaps even stage IV, where ultimately stable ruptures behave as having nearly zero fracture energy, could be potentially investigated in this type of experimental settings. The front-localized energy balance of dynamic ruptures has been extensively studied on both dry \cite{Svetlizky_Fineberg_2014,Paglialunga_Passelegue_2022} and fluid-lubricated frictional interfaces \cite{Bayart_Svetlizky_2016b,Paglialunga_Passelgue_2023}. Yet the same kind of energy balance for injection-induced slow slip, which is determined by the competition of three distinct factors (equation \eqref{eq:final-energy-balance}), remains to be investigated experimentally. Indeed, stages III and IV might be the relevant regimes for subsurface processes 
such as the reactivation in shear of fractures by fluid injection for geo-energy applications, and natural earthquake-related phenomena where coupled fluid flow and aseismic slip processes are thought to play an important role (e.g., seismic swarms, aftershock sequences and slow earthquakes).

\subsection{In-situ experiments}

Fluid injection experiments in shallow natural faults have recently provided important insights into the mechanics of injection-induced fault slip \cite{Guglielmi_Cappa_2015,Cappa_Guglielmi_2022}. Yet laboratory experiments under well-controlled conditions are likely better positioned than in-situ experiments to validate the physics of injection-induced fault slip due to the further uncertainties naturally present in the field (e.g., heterogeneities of stress, strength, fracture/fault geometrical complexities, among others), in-situ experiments do have various benefits such as, for instance, covering a larger decameter scale and providing more realistic field conditions, particularly those ones allowing the growth of fully three-dimensional unbounded ruptures initiated from a localized fluid source, as the one considered in this study. Our results thus provide the opportunity to make quantitative comparisons in these cases.

Among them, the experiments of Guglielmi \textit{et al.} \cite{Guglielmi_Cappa_2015} have been particularly impactful as they were able to measure for the first time not only micro-seismicity and injection-well fluid pressure and volume rate history (as done previously in large-scale field experiments \cite{Scotti_Cornet_1994,Cornet_Helm_1997}), but also the history of induced fracture slip and opening at the injection point/interval. This unique dataset has been analyzed via dynamic numerical modelling by various researchers \cite{Guglielmi_Cappa_2015,Bhattacharya_Viesca_2019,Cappa_Scuderi_2019,Larochelle_Lapusta_2021}. The multiplicity of proposed models that have fitted the data suggests indeed that more spatially distributed measurements, as done more recently \cite{Cappa_Guglielmi_2022}, could further help to constrain the underlying physical processes operating behind these experiments. For the purpose of illustrating the application of our model, hereafter we focus on the modeling work of Bhattacharya and Viesca \cite{Bhattacharya_Viesca_2019} for the unique reason that they assumed a slip-weakening fault model which makes the application of our results more straightforward.

Let us first estimate the three dimensionless parameters of our model: $\mathcal{S}$, $\mathcal{P}$, and $\mathcal{F}$ \eqref{eq:non-dimensional-parameters}. Considering the best-fit model parameters of Bhattacharya and Viesca \cite{Bhattacharya_Viesca_2019}: $\mu=11.84$ GPa, $\delta_c=0.37$ mm, $f_p=0.6$, $f_r=0.42$, $\sigma_0^\prime$=5.08 MPa, and $\tau_0=2.41$ MPa; we can directly compute $\mathcal{S}\approx0.79$ and $\mathcal{F}=0.7$. Note that the estimated initial shear stress of $2.41$ MPa is greater than the in-situ residual fault strength $f_r\sigma_0^\prime\approx2.1$ MPa (or alternatively $\mathcal{S}>\mathcal{F}$ \eqref{eq:ultimate-stability-conditions}), so that the injection-induced rupture is inferred to be ultimately unstable. Because no macroscopic dynamic rupture occurred during the experiment, the propagation mode of aseismic slip must be the one of a frustrated dynamic instability. Let us now estimate the remaining dimensionless parameter $\mathcal{P}$. To do so, we need to estimate first the injection intensity $\Delta p_*$, equation \eqref{eq:p-solution}. We approximate the injection history via a constant-volume-rate injection characterized by the same total volume of fluid injected during the experiment. This gives roughly $Q\sim40$ l/min. The fault intrinsic permeability $k$ is widely recognized to have increased during this test \cite{Guglielmi_Cappa_2015,Cappa_Scuderi_2019,Bhattacharya_Viesca_2019}. It was estimated by Bhattacharya and Viesca \cite{Bhattacharya_Viesca_2019} between $k_\text{min}=0.8\times10^{-12}$ m$^2$ and $k_\text{max}=1.3\times10^{-12}$ m$^2$. Consider, for instance, the average $\tilde{k}=1.05\times10^{-12}$ m$^2$. By assuming a fluid dynamic viscosity of $\eta\sim10^{-3}$ Pa$\cdot$s and fault-zone width $w=0.2$ m \cite{Bhattacharya_Viesca_2019}, the characteristic wellbore overpressure, equation \eqref{eq:pc-definition}, is $\Delta p_c\approx3.17$ MPa, which is very close to the actual, nearly constant wellbore overpressure measured at the latest part of the injection, $\sim 3$ MPa (see figure 2a in \cite{Bhattacharya_Viesca_2019}). Following this approximation for the fluid injection, we obtain an injection intensity $\Delta p_*\approx0.25$ MPa, which in turn yields $\mathcal{P}\approx0.05$. By considering either the minimum or maximum fault permeability, we would obtain $\mathcal{P}\approx0.065$ and $\mathcal{P}\approx0.04$, respectively; the higher the permeability, the lower the injection intensity.

To understand under what regime aseismic slip may have developed during the initial Coulomb's friction stage (either critically stressed or marginally pressurized regime), let us calculate the peak stress-injection parameter $\mathcal{T}_p$. Given the known values for $\mathcal{S}$ and $\mathcal{P}$, we obtain via \eqref{eq:peak-stress-injection-parameter} $\mathcal{T}_p\approx4.22$, which is well into the marginally pressurized regime, with an associated amplification factor $\lambda_p\approx0.12$ (equation \eqref{eq:lambda-asymptotes}). By considering $k_\text{min}$ and $k_\text{max}$ instead, we would obtain just a modest change in $\mathcal{T}_p$ approximately equal to $3.20$ and $5.22$, respectively. Furthermore, we can estimate the critical nucleation radius in the marginally pressurized regime via equation \eqref{eq:nucleation-radius-marginally-pressurized}. For that, we must calculate first the elasto-frictional lengthscale $R_w$, equation \eqref{eq:slip-scale-rupture-length-scale-slip-weakening}. Given the best-fit model parameters, $R_w\approx4.79$ m, and the nucleation radius is then $R_c^{mp}\approx6.06$ m. Considering the size of the aseismic rupture estimated by Bhattacharya and Viesca \cite{Bhattacharya_Viesca_2019} at the final time analyzed in their work, $t_f=1378$ s (see inset of figure 3a in \cite{Bhattacharya_Viesca_2019}), one could expect that the aseismic rupture was quite close to become unstable. Note that in \cite{Bhattacharya_Viesca_2019}, their original circular rupture solutions were modified to appear elliptical considering the perturbative approach for circular shear cracks of Gao \cite{Gao_1988}, who gives an aspect ratio $a/b=1/(1-\nu)$ (to first order in $\nu$), where $a$ and $b$ are the semi-major and semi-minor axes of the elliptical rupture front. The calculations of Gao \cite{Gao_1988} involve a planar circular crack whose shape is perturbed under uniform shear load and constant energy release rate along the front. Perhaps, a better correction could be done by considering the asymptotic behavior in the marginally pressurized regime obtained in \cite{Saez_Lecampion_2022}, $a/b=(3-\nu))/(3-2\nu)$, at least in the Coulomb's friction stage where the previous equation is valid. This would lead to ruptures that are less elongated than the ones considered by Bhattacharya and Viesca \cite{Bhattacharya_Viesca_2019}. For example, considering a Poisson's ratio $\nu=0.25$, the marginally-pressurized aspect ratio is $a/b=1.1$, whereas Gao's aspect ratio is $a/b\approx1.33$. The latter was indeed found to be the asymptotic behavior in the critically stressed regime ($\mathcal{T}_p\ll1$) of the Coulomb's friction stage \cite{Saez_Lecampion_2022}. 

Let us assume as a first estimate, that the rupture propagates with Coulomb's friction until the final time $t_f=1378$ s. In this scenario, the rupture radius would be at this time simply $R_f=\lambda_p\sqrt{4\alpha t_f}$, and the corresponding accumulated slip at the injection point, $\delta_f=(8/\pi)(f_p\Delta p_*/\mu)R_f$ (equation 27 in \cite{Saez_Lecampion_2022}). To perform the previous calculations, we need to estimate the fault hydraulic diffusivity $\alpha=k/\eta S$. Considering the storage coefficient $S$ estimated in $2.2\times10^{-8}$ Pa$^{-1}$ \cite{Bhattacharya_Viesca_2019}, we obtain a hydraulic diffusivity $\alpha\approx0.048$ m$^2$/s. The final rupture radius and accrued slip are then $R_f\approx2.01$ m and $\delta_f\approx0.07$ mm, respectively. Variations in fault permeability considering $k_\text{min}$ and $k_\text{max}$ would result in $R_f\approx2.91$ m and $1.35$ m, and $\delta_f\approx0.12$ mm and $0.04$ mm, respectively. In any case, the previous quantities do not account for the total slip measured at the injection point which is an order of magnitude higher ($\sim 1$ mm) and estimated rupture radius $\sim 5$ m. Moreover, there is a clear acceleration of slip in the final part of the injection (see figure 3a in \cite{Bhattacharya_Viesca_2019}) which in our model could only come from frictional weakening (stage II). We therefore calculate the evolution of the rupture front and slip at the injection point numerically. 

Our numerical solution shows that the nucleation of a dynamic rupture occurs at $R_c\approx6.08$ m, which is in excellent agreement with the theoretical nucleation radius for marginally pressurized faults calculated previously. The calculated accrued slip at the center of the rupture at the instability time is $0.35$ mm, which is about a third part of the actual measurement. On the other hand, the nucleation time is $t_c=5885$ s, which is several times longer than $t_f$. Indeed, at the time $t_f$ the rupture radius is just $2.17$ m in our numerical solution, not much larger than the Coulomb's friction approximation ($2.01$ m). The latter indicates that our model is indeed operating in stage I at $t_f$. The differences between our calculations and the ones of Bhattacharya and Viesca \cite{Bhattacharya_Viesca_2019} are due to time-history variations in permeability not accounted in our model, and the approximation of the fluid injection via an equivalent constant volume rate. Nevertheless, the theoretical nucleation radius is expected to hold as this quantity is relatively independent of the injection scenario. Finally, we note that a rupture propagating in the marginally pressurized regime is expected to be confined within the pressurized area even at the instability time. In this regard, the main difference with Bhattacharya and Viesca \cite{Bhattacharya_Viesca_2019} who suggested that the slip front outpaced the migration of fluids is merely a matter of definitions. In our case, the overpressure front $L(t)=\sqrt{4\alpha t}$ represents the radial distance from the injection point at which the overpressure is approximately 2 percent of the fluid-source overpressure, the latter being approximately $3$ MPa at the final time. In \cite{Bhattacharya_Viesca_2019}, various overpressure isobars are drawn. The one with lowest overpressure is at $0.5$ MPa, which is around $17$ percent of the fluid-source overpressure.

\subsection{A note on rate-and-state fault models: similarities and differences}

Laboratory-derived friction laws \cite{Dieterich_1979,Ruina_1983} are widely used in the earthquake modeling community to reproduce the entire spectrum of slip velocities on natural faults \cite{Scholz_2019}. These empirical friction laws capture the dependence of friction on slip rate and the history of sliding (via a state variable) as observed during velocity-step laboratory experiments on bare rock surfaces and simulated fault gouge \cite{Marone_1998}. In its simplest form, the rate-and-state friction coefficient is expressed as
\begin{equation}\label{eq:r-s-friction}
    f(v,\theta)=f_0 + a\text{ln}\left( \frac{v}{v_0} \right) + b \text{ln}\left( \frac{v_0\theta}{d_c}\right),
\end{equation}
where $f_0$ is the friction coefficient at a reference slip rate $v_0$ and state variable $\theta_0=d_c/v_0$ with units of time, $d_c$ is a characteristic slip `distance' for the evolution of $\theta$, which is usually thought to be an order of magnitude smaller than $\delta_c$ of the slip weakening model \cite{Cocco_Bizzarri_2002,Uenishi_Rice_2002}, and $a$ and $b$ are the rate-and-state friction parameters, both positive and of order $10^{-2}$. An additional dynamical equation describing the evolution of the state variable $\theta$ is required. For the purpose of this discussion, we consider hereafter a widely used state-evolution equation known as aging law: $\dot{\theta}=1-v\theta/d_c$.

The similarities between frictional ruptures obeying slip-weakening and rate-and-state friction have been long recognized (see \cite{Cocco_Bizzarri_2002,Uenishi_Rice_2002,Ampuero_Rubin_2008,Garagash_Germanovich_2012} for example). Furthermore, in the context of injection-induced fault slip, some similarities were already recognized by Garagash and Germanovich \cite{Garagash_Germanovich_2012} particularly in relation to the nucleation of dynamic slip. Specifically, they noted that the nucleation length of critically stressed faults for linear slip-weakening friction or, what is the same, the one of Uenishi and Rice \cite{Uenishi_Rice_2002}, is identical to the nucleation length of rate-and-state faults for $a/b\ll1$ \cite{Rubin_Ampuero_2005}. On the other hand, the large nucleation length near the ultimate stability limit which is equal (except by a pre-factor of order one) to the one of Andrews \cite{Andrews_1976b}, is identical to the nucleation length of rate-and-state faults when approaching the velocity-neutral limit $a/b\to1$ \cite{Rubin_Ampuero_2005}. The equivalence between the previous nucleation lengths is obtained by recasting the slip-weakening friction law in terms of the rate-and-state parameters, this is, replacing the peak to residual strength drop $(f_p-f_r)\sigma_0^\prime$ by $b\sigma_0^\prime$, and the stress drop $\tau_0-f_r\sigma_0^\prime$ by $(b-a)\sigma_0^\prime$ \cite{Rubin_Ampuero_2005,Ampuero_Rubin_2008}. 

In addition to the previous similarities, we discuss some additional ones now. For instance, the ultimate stability condition \eqref{eq:ultimate-stability-conditions} 
has been observed to determine in rate-and-state models whether velocity-weakening faults ($a/b<1$) produce either self-arrested or run-away ruptures \cite{Norbeck_Horne_2018}. Also, the same stability condition has been observed to determine whether velocity-strengthening faults ($a/b>1$) host either purely quasi-static slip or a dynamic instability \cite{Dublanchet_2019}. Both results are essentially the same as predicted by slip-weakening fault models. 
Indeed, the ultimate stability condition is somehow expected to occur in rate-and-state models (both velocity weakening and velocity strengthening) in the vicinity of the velocity-neutral limit ($a/b\to1$), as the condition \eqref{eq:ultimate-stability-conditions} that the residual fault strength drops at nearly the same level of the shear stress present further ahead of the rupture is reminiscent of the case $a\approx b$ when analyzing the steady-state ($\dot{\theta}=0$) response of equation \eqref{eq:r-s-friction} to an incremental velocity step approximating crudely the passage of a rupture front.

Another similarity between slip-weakening and rate-and-state fault models relates to our constant-residual-friction similarity solution, or ultimate zero-fracture-energy regime. As already noted by Sáez \textit{et al}. \cite{Saez_Lecampion_2022} for the particular case of injection at constant volumetric rate in a one-dimensional fault (see section 5.2 in \cite{Saez_Lecampion_2022}), the ultimate regime of rate-and-state faults \cite{Dublanchet_2019} coincides with the solution for a constant friction coefficient \cite{Saez_Lecampion_2022}. It is clear now albeit in a three-dimensional model, that this friction coefficient corresponds to the residual value in a slip-weakening model and that this regime is characterized by negligible fracture energy. The latter is additionally consistent with the work of Garagash \cite{Garagash_2021} who found that slip transients driven by a point-force-like injection approaches a zero-toughness condition as an ultimate regime in a one-dimensional fault. Note that the limit of a point-force-like injection is reached in our three-dimensional model in the nearly unstable limit, but not in the marginally pressurized one. Moreover, the features that give rise to this ultimate asymptotic behavior seem rather general and we think are likely expected to hold under other types of fluid injection. As already shown in \cite{Saez_Lecampion_2022} (section 5.1), the spatio-temporal patterns of injection-induced aseismic slip are strongly influenced by the type of fluid injection (or injection rate history). Quantifying this effect is important and we will address it soon in a future study.

We would also like to highlight some differences between slip-weakening and rate-and-state fault models. Indeed, one of the main physical ingredients that rate-and-state friction would incorporate in our model is frictional healing. The recovery of the friction coefficient with the logarithm of time is a well-established phenomenon \cite{Dieterich_1979,Ruina_1983,Marone_1998} that is essential in physics-based models that attempt to reproduce earthquake cycles \cite{Tse_Rice_1986,Lapusta_Rice_2000}. Frictional healing would provide, for instance, the possibility of nucleating multiple dynamic events on the same fault segment. Yet we highlight that this is not a unique characteristic of rate-and-state friction in the sense that two events can also nucleate on the same slip-weakening fault segment (regime R4 in figure \ref{fig:map-regimes}). Another case in which the rate-and-state framework could be particularly useful is to model the reactivation of faults that are thought to be steadily creeping. In fact, from a Coulomb's friction perspective, rate-and-state faults are always at failure: the shear stress is at any time and over the entire fault extent equal to the fault strength. The initial stress state is a result of the history of sliding. At an initial time $t=0$, it will be defined by the initial distribution of slip rate $v_i$ and initial state variable $\theta_i$.

To illustrate this latter point and get some further insights into a rate-and-state fault model, consider our same hydro-mechanical model but with a rate-and-state friction coefficient. Due to the `always-failing' condition, $R(t)\to\infty$ in equation \eqref{eq:elastic-equilibrium} and the inequality \eqref{eq:shear-strength} becomes an equality. One possible way of considering the initial stress state is to assume that the initial slip velocity $v_i$ is uniform and it operates at the creep rate that one could further consider as the reference slip rate $v_0$. The initial stress state is then encapsulated in the initial state variable $\theta_i$. Assuming the latter as uniform as well, one can readily show by dimensional analysis that the slip rate (the primary unknown in this model together with the state variable) depends in addition to dimensionless space $r b\sigma_0^\prime /d_c\mu$ and time $t v_0/d_c$, on the following five non-dimensional parameters: $a/b$, $\Delta p_*/\sigma_0^\prime$, $\alpha/\alpha_c$, $f_0/b$ and $\theta_i/\theta_0$, where $\alpha_c=\mu^2d_cv_0/b^2\sigma_0^{\prime 2}$ is a characteristic diffusivity. We note that this rate-and-state version of our model has an increased complexity with two more dimensionless parameters than the slip-weakening model. $\Delta p_*/\sigma_0^\prime$ is indeed our same overpressure ratio $\mathcal{P}$. $a/b$ quantifies and degree of weakening ($a/b<1$) or strengthening ($a/b>1$) and, as discussed previously, it would relate to the the ultimate stability behavior of the fault. $f_0/b$ quantifies the constant part of the friction coefficient with regard to $b$ which in turn relates to the strength drop (i.e., the decay from the peak to the residual friction). Finally, $\theta_i/\theta_0$ is where the initial shear stress or pre-stress ratio $\mathcal{S}$ of the slip-weakening model would equivalently emerge. In fact, by multiplying equation \eqref{eq:r-s-friction} by $\sigma_0^\prime$ and then expressing it at the initial conditions $v_i$ and $\theta_i$, the resulting dimensionless form of such equation reads as $\text{ln}\left(\theta_i/\theta_0\right)=\left(f_0/b\right)\left(\tau_0/f_0\sigma_0^\prime-1\right)$. The dimensionless parameter $\theta_i/\theta_0$ can be then alternatively chosen as $\tau_0/f_0\sigma_0^\prime$, which is similar to the pre-stress ratio of the slip-weakening model. 

\section{Concluding remarks}

We have investigated the propagation of fluid-driven slow slip and earthquake nucleation
on a slip-weakening circular fault subjected to fluid injection at a constant volume rate. 
Despite some simplifying assumptions in our model, our investigation has revealed a very broad range of aseismic slip behaviors, from frustrated dynamic instabilities to unconditionally stable slip, from ruptures that move much faster than the diffusion of pore pressure to ruptures that move much slower than that. 
The circular fault geometry is likely the simplest one enabling quantitative comparisons with field observations thanks to its three-dimensional nature. 
It is thus also useful for preliminary engineering design of hydraulic stimulations in geo-energy applications. 
In addition to the effect of a non-zero Poisson's ratio that will elongate the shape of the rupture along the direction of principal shear, changes in lithologies as commonly encountered in the upper Earth's crust will alter the dynamics of an otherwise unbounded rupture as the one we have examined here. In particular, the effect of layering might promote containment of the reactivated fault surface within certain lithologies, similar to what is observed for hydraulic fractures \cite{Bunger_Lecampion_2017}. This effect may be important in some cases and would require further quantification.

\subsection*{CRediT authorship contribution statement}

\textbf{Alexis Sáez:} Conceptualization, Methodology, Software, Validation, Formal analysis, Investigation, Writing - Original Draft, Writing – review \& editing, Visualization, Funding acquisition. 

\noindent\textbf{Brice Lecampion:}  Conceptualization, Methodology,  Software, Writing – review \& editing, Funding acquisition. 

\subsection*{Declaration of competing interest}

The authors declare that they have no known competing financial interests or personal relationships that could have appeared to influence the work reported in this paper.

\subsection*{Funding}

The results were obtained within the EMOD project (Engineering model for hydraulic stimulation). The EMOD project benefits from a grant (research contract no. SI/502081-01) and an exploration subsidy (contract no. MF-021-GEO-ERK) of the Swiss federal office of energy for the EGS geothermal project in Haute-Sorne, canton of Jura, which is gratefully acknowledged. Alexis Sáez was partially funded by the Federal Commission for Scholarships for Foreign Students via the Swiss Government Excellence Scholarship.

\subsection*{Acknowledgements}

Alexis Sáez would like to thank François Passelègue for discussions about his own experiments.

\begin{appendices}

\counterwithin*{equation}{section}
\renewcommand\theequation{\thesection\arabic{equation}}

\counterwithin*{table}{section}
\renewcommand\thetable{\thesection\arabic{table}}

\counterwithin*{figure}{section}
\renewcommand\thefigure{\thesection\arabic{figure}}

\section{Eigenvalue problem at the instability time for unlimited linear weakening of friction}\label{appendix:eigenvalue-problem}
\subsection{Generalized eigenvalue problem}\label{sec:generalized-eigenproblem}

Following Uenishi and Rice \cite{Uenishi_Rice_2002} and Garagash and Germanovich \cite{Garagash_Germanovich_2012}, we extend their eigenvalue-based stability analysis, valid for unlimited linear weakening of friction under either in-plane shear (II) or anti-plane shear (III) mode of sliding, to the case of a circular rupture propagating under mixed-mode (II+III) conditions. 

Equilibrium dictates that within the slipping region $r\leq R(t)$, the fault strength $\tau_s$ \eqref{eq:shear-strength} must be locally equal to the fault shear stress $\tau$ \eqref{eq:elastic-equilibrium}. Equating the two previous equations and then differentiating with respect to time, leads to
\begin{equation}\label{eq:elastic-equilibrium-rate}
    v\left(r,t\right)\frac{\text{d}f}{\text{d} \delta}\sigma^\prime(r,t)+f(\delta)\frac{\partial \sigma^\prime(r,t)}{\partial t}= \frac{\partial \tau_0(r,t)}{\partial t} - \frac{\mu}{2\pi}\int_{0}^{R(t)}F\left(r,\xi\right)\frac{\partial v(\xi,t)}{\partial\xi}\mathrm{d}\xi,
\end{equation}
where $v=\partial\delta/\partial t$ is the fault slip rate. When differentiating the integral term previously, we have considered Leibniz’s integral rule and applied the condition $\partial\delta\left(R(t),t\right)/\partial r=0$ which guarantees non-singular shear stresses along the rupture front.

In our problem, the effective normal stress $\sigma^\prime(r,t)=\sigma_0^\prime-\Delta p(r,t)$ decreases from the initial uniform value $\sigma_0^\prime$ due to overpressure $\Delta p(r,t)$ associated with fluid injection, whereas the shear stress that would be present on the fault if no slip occurs is a uniform and constant value $\tau_0$. Nevertheless, for the sake of generality, we keep utilizing the generic terms $\sigma^\prime(r,t)$ and $\tau_0(r,t)$. Indeed, $\sigma^\prime(r,t)$ generally contains not only the initial effective normal stress and changes in pore pressure, but also possible changes in total normal stress from the far field, as $\sigma^\prime=\sigma-p$. Similarly, $\tau_0(r,t)$ could be a summation of both the initial shear stress and far field shear loading. Far field loads may be due to, for instance, tectonic forces and seasonal variations of stress, among many others. Note that $\sigma^\prime(r,t)$ and $\tau_0(r,t)$ are both axisymmetric in magnitude and at least one of them must be locally peaked around the origin in order to initiate slip at $r=0$ at a certain time $t=t_0$. Equation \eqref{eq:elastic-equilibrium-rate} is then valid at any time $t>t_0$ and, as mentioned in the main text, it assumes a Poisson's ratio $\nu=0$.

Let us scale equation \eqref{eq:elastic-equilibrium-rate} by introducing the following non-dimensional quantities: $\bar{r}=r/R$, $\bar{\xi}=\xi/R$, and $\Bar{v}=v/v_\text{rms}$, where 
\begin{equation}\label{eq:v-rms-definition}
    v_\text{rms}(t)=\sqrt{\frac{1}{R(t)}\int_0^{R(t)} v^2(r,t)\text{d}r}
\end{equation}
is the root mean square of the slip rate distribution, such that $\int_0^1\bar{v}^2(\bar{r})\text{d}\bar{r}=1$. Note that for the linear-weakening friction law \eqref{eq:linear-weakening-friction}, $\text{d}f/\text{d}\delta=-(f_p-f_r)/\delta_c$. The latter has the strong assumption that the residual friction coefficient $f_r$ has not been reached yet at any point within the rupture. Otherwise, wherever $f=f_r$, $\text{d}f/\text{d}\delta=0$. Moreover, for the exponential-weakening friction law \eqref{eq:exponential-weakening-friction}, the same expression for $\text{d}f/\text{d}\delta$ is approximately valid in the range of small slip $\delta\ll\delta_c$, to first order in $\delta/\delta_c$.

Considering the previous quantities plus the relation \eqref{eq:slip-scale-rupture-length-scale-slip-weakening} $\mu\delta_c=(f_p-f_r)\sigma_0^\prime R_w$, we nondimensionalize equation \eqref{eq:elastic-equilibrium-rate} to obtain
\begin{equation}\label{eq:elastic-equilibrium-rate-dimensionless}
    -\frac{R}{R_w}\bar{v}(\bar{r})\frac{\sigma^\prime(\bar{r}R)}{\sigma_0^\prime} + 
    \frac{R}{\mu v_\text{rms}}f(\delta)\frac{\partial \sigma^\prime(\bar{r}R)}{\partial t} = \frac{R}{\mu v_\text{rms}}\frac{\partial \tau_0(\bar{r}R)}{\partial t} - \frac{1}{2\pi}\int_{0}^{1}F\left(\bar{r},\bar{\xi}\right)\frac{\partial \bar{v}(\bar{\xi})}{\partial\bar{\xi}}\mathrm{d}\bar{\xi}.
\end{equation}
In the previous equation, we dropped the explicit dependence on time $t$ of the various variables for simplicity in the notation. 

Following Uenishi and Rice \cite{Uenishi_Rice_2002}, at the instability time $t_c$, the slip rate diverges all over the fault plane, such that the root mean square of the slip rate distribution $v_\text{rms}(t_c)\to\infty$. The only non-vanishing terms of \eqref{eq:elastic-equilibrium-rate-dimensionless} leads to the following generalized eigenvalue problem:
\begin{equation}\label{eq:generalized-eigenvalue-problem}
    \frac{R}{R_w}\bar{v}(\bar{r})\frac{\sigma^\prime(\bar{r}R)}{\sigma_0^\prime} = 
    \frac{1}{2\pi}\int_{0}^{1}F\left(\bar{r},\bar{\xi}\right)\frac{\partial \bar{v}(\bar{\xi})}{\partial\bar{\xi}}\mathrm{d}\bar{\xi},
\end{equation}
which corresponds to the mixed-mode, circular rupture version of equation (14) in \cite{Garagash_Germanovich_2012}. 

Given a normalized distribution of effective normal stress at the instability time, $\sigma^\prime(r,t_c)/\sigma_0^\prime$, equation \eqref{eq:generalized-eigenvalue-problem} can be solved to obtain the corresponding generalized eigenvalues and eigenfunctions. Moreover, what is more important is to calculate the smallest eigenvalue that would be related to the instabilities we observe in the full numerical solutions. In our problem, $\sigma^\prime(r,t_c)$ is set by the distribution of overpressure at the time of instability, which does not allow us to obtain a purely analytical insight as the instability time is generally unknown and, more importantly, information about one of the problem parameters, the pre-stress ratio $\mathcal{S}$, is lost when deriving the eigen problem \eqref{eq:generalized-eigenvalue-problem}. The only scenario in which equation \eqref{eq:generalized-eigenvalue-problem} is independent of $t_c$ is when $\sigma^\prime(r,t_c)$ is uniform, which in turn leads to a regular eigenvalue problem. Furthermore, in the particular case of $\sigma^\prime(r,t)=\sigma_0^\prime$, we obtain the circular rupture version of the eigenvalue problem of Uenishi and Rice (equation (12) in \cite{Uenishi_Rice_2002}), which will give the corresponding universal nucleation radius of their problem.

\subsection{Eigenvalue problem in the critically stressed and marginally pressurized limits}\label{sec:regular-eigenproblem}

Let us come back to our particular problem where $\sigma^\prime(r,t)=\sigma_0^\prime-\Delta p(r,t)$ and assume a rather general but self-similar injection scenario such that the overpressure can be written in the similarity form: $\Delta p(r,t)=\Delta p_w(t)\Pi(\xi)$, where $\Delta p_w(t)$ is the overpressure at the fluid source, and $\Pi(\xi)$ is the spatial distribution of overpressure with the properties: $\Pi(0)=1$ and $\Pi(\infty)\to0$. Note that such self-similar injection scenario is possible only if one assumes a line source of fluids such that no length scale associated with the fluid source is introduced into the problem. For a discussion about the line-source approximation, see Appendix \ref{appendix:line-source}. Introducing the previous relations for the effective normal stress into \eqref{eq:generalized-eigenvalue-problem}, the generalized eigenvalue problem becomes
\begin{equation}\label{eq:generalized-eigenvalue-problem-fluid-driven-problem}
    \frac{R}{R_w}\bar{v}(\bar{r})\left(1-\frac{\Delta p_w}{\sigma_0^\prime}\Pi\left(\lambda\bar{r}\right)\right) = 
    \frac{1}{2\pi}\int_{0}^{1}F\left(\bar{r},\bar{\xi}\right)\frac{\partial \bar{v}(\bar{\xi})}{\partial\bar{\xi}}\mathrm{d}\bar{\xi},
\end{equation}
where $\lambda(t_c)=R(t_c)/\sqrt{4\alpha t_c}$ is the so-called amplification factor at the instability time. We recall that the dependence of the various variables in the previous equation on $t_c$ is omitted for simplicity.

Consider now the critically stressed limit: $\tau_0\to f_p\sigma_0^\prime$, where the rupture front largely outpaces the overpressure front at the time of instability, such that $\lambda(t_c)\gg1$. In view of the properties of $\Pi(\xi)$, the term $(\Delta p_w/\sigma_0^\prime)\Pi(\lambda\bar{r})\ll1$ so that, if neglected, equation \eqref{eq:generalized-eigenvalue-problem-fluid-driven-problem} further simplifies to
\begin{equation}\label{eq:eigenvalue-problem-critically-stressed}
    \frac{R}{R_w}\bar{v}(\bar{r}) = 
    \frac{1}{2\pi}\int_{0}^{1}F\left(\bar{r},\bar{\xi}\right)\frac{\partial \bar{v}(\bar{\xi})}{\partial\bar{\xi}}\mathrm{d}\bar{\xi}.
\end{equation}
The previous equation is a regular eigenvalue problem. It corresponds indeed to the circular rupture version of the eigenvalue problem of Uenishi and Rice \cite{Uenishi_Rice_2002}. Derivation of equation \eqref{eq:eigenvalue-problem-critically-stressed} can be alternatively done by following the reasoning of Garagash and Germanovich \cite{Garagash_Germanovich_2012} that in the critically stressed limit, the effective normal stress over the slipping region is largely unchanged so that $\sigma^\prime(r,t_c)\approx\sigma_0^\prime$, except for a very small region of approximate size $\sqrt{4\alpha t_c}$ near the rupture center that at spatial scales in the order of the rupture size can be neglected. Replacing $\sigma^\prime(r,t_c)\approx\sigma_0^\prime$ into \eqref{eq:generalized-eigenvalue-problem} leads equivalently to \eqref{eq:eigenvalue-problem-critically-stressed}.

Let us now examine the marginally pressurized limit: $f_p\Delta p_w\approx f_p\sigma_0^\prime-\tau_0$, where the rupture front significantly lags the overpressure front at the instability time, so that $\lambda(t_c)\ll1$. We refer to Appendix C for a discussion about the marginally pressurized limit and its relation to the line-source approximation. Particularly, we note that the property $\Pi(0)=1$ cannot be rigorously defined but, still, it can be established in an order of magnitude sense. Furthermore, for injection at constant volumetric rate, the prefactor is quite close to one for all practical purposes (see figure \ref{fig:line-source-approximation}b). It is therefore convenient for practical applications to define the marginally pressurized limit in an approximate sense. Hence, we approximate the fluid overpressure within the rupture as $\Delta p_w\Pi(\lambda\bar{r}) \approx \sigma_0^\prime-\tau_0/f_p$. After substituting the previous relation into \eqref{eq:generalized-eigenvalue-problem-fluid-driven-problem}, we obtain the following regular eigenvalue problem for marginally pressurized cases,
\begin{equation}\label{eq:eigenvalue-problem-marginally-pressurized}
    \frac{R}{R_w}\frac{\tau_0}{f_p\sigma_0^\prime}\bar{v}(\bar{r}) = 
    \frac{1}{2\pi}\int_{0}^{1}F\left(\bar{r},\bar{\xi}\right)\frac{\partial \bar{v}(\bar{\xi})}{\partial\bar{\xi}}\mathrm{d}\bar{\xi}.
\end{equation}

It is important to mention that the critically stressed and marginally pressurized limits are both characterized by small slip at the instability time: $\delta(r=0,t_c)\ll\delta_c$. This is observed in our numerical solutions and was also established by Garagash and Germanovich \cite{Garagash_Germanovich_2012} in the two-dimensional problem. The latter is very important since it implies that the approximation of the exponential-weakening friction law by a linear relation is valid, as well as the assumption of unlimited linear-weakening of friction (never reaching the residual strength of the fault). Finally, as a last comment, we have established the eigenvalue problems in both limits for a general (self-similar) injection scenario, not restricted to the constant-volumetric rate case that we solve in the main text. However, in the marginally pressurized limit, we have implicitly assumed that the overpressure at the fluid source at the instability time is approximately equal to the overpressure at the time of activation of slip. Such approximation is reasonable in the case of constant-volumetric rate as the increase of overpressure at the fluid source is slowly logarithmic (see figure \ref{fig:line-source-approximation}b) and assumed to be approximately constant, equal to $\Delta p_c$, for practical applications. This approximation has to be carefully considered when dealing with other injection scenarios (see, for instance, \cite{Garagash_Germanovich_2012,Ciardo_Lecampion_2019,Ciardo_Rinaldi_2021}).

\subsection{Numerical solution of the regular eigenvalue problem}\label{sec:solution-eigenproblem}

In the critically stressed and marginally pressurized limits, the eigen equations \eqref{eq:eigenvalue-problem-critically-stressed} and \eqref{eq:eigenvalue-problem-marginally-pressurized} can be recast as:
\begin{equation}\label{eq:eigenvalue-problem-both-regimes}
    \frac{1}{2\pi}\int_{0}^{1}F\left(\bar{r},\bar{\xi}\right)\frac{\partial \bar{v}(\bar{\xi})}{\partial\bar{\xi}}\mathrm{d}\bar{\xi} =
    \beta\bar{v}(\bar{r}),
\end{equation}
with the eigenvalue 
\begin{equation}
    \beta = \frac{R}{R_w}\cdot
    \begin{cases}
        1 & \text{for critically stressed faults }\mathcal{T}_p\ll1\\
        \nicefrac{\tau_0}{f_p\sigma_0^\prime} & \text{for marginally pressurized faults }\mathcal{T}_p\sim10.
    \end{cases}
\end{equation}
We calculate the eigenvalues $\beta_k$ and eigenfunctions $\bar{v}_k$ of \eqref{eq:eigenvalue-problem-both-regimes}, with $k=1,2,...,\infty$, by discretizing the linear integral operator on the left-hand side via a collocation boundary element method employing ring `dislocations' with piece-wise constant slip rate. The details of such implementation can be found in the supplementary material of \cite{Saez_Lecampion_2023}. For the numerical calculations, it is convenient to express the discretized form of the eigen equation \eqref{eq:eigenvalue-problem-both-regimes} in matrix-vector form as
\begin{equation}\label{eq:eigenvalue-problem-discretized}
    \mathbf{E}\bar{\boldsymbol{v}}_k=\beta_k \bar{\boldsymbol{v}}_k,
\end{equation}
where $\bar{\boldsymbol{v}}_k\in\mathbb{R}^N$ are the discretized eigenfunctions with $k=1,2,...,N$, where $N$ is the number of ring-dislocation elements, and $\mathbf{E}\in\mathbb{R}^{N\times N}$ is a non-dimensional matrix that is equivalent to the collocation boundary element matrix of a circular shear crack of unit radius and unit shear modulus (see \cite{Saez_Lecampion_2023}). By collocation boundary element matrix, we mean that the product between $\mathbf{E}$ and a given vector representing a discretized slip distribution $\boldsymbol{\delta}$, would give as a result the corresponding discretized shear stress distribution $\boldsymbol{\tau}$ that is in quasi-static equilibrium with $\boldsymbol{\delta}$, in an infinite and otherwise unstressed solid. 

\begin{figure}
    \centering
    \includegraphics[width=10cm]{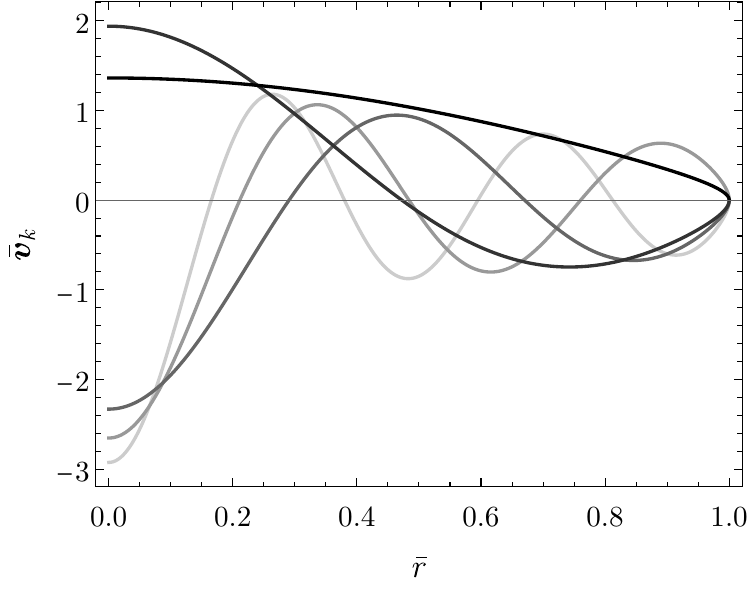}
    \caption{Normalized eigenfunctions $\bar{\boldsymbol{v}}_k$ for $N=10000$.}
    \label{fig:eigen-functions}
\end{figure}

We solve the discretized eigen equation \eqref{eq:eigenvalue-problem-discretized} with the standard Wolfram Mathematica functions \textit{Eigenvalues} and \textit{Eigenvectors} which can be instructed to search only for the smallest eigenvalues and their corresponding eigenvectors. We do not intend here to conduct an extensive analysis of the eigenvalues and eigenfunctions as our unique goal in this work is to determine the smallest eigenvalue that we expect to give the nucleation radii for the critically stressed and marginally pressurized regimes. Nevertheless, we do report the first five (smaller) eigenvalues and their eigenvectors in table \ref{table:eigen-values} and figure \ref{fig:eigen-functions}, respectively. The eigenfunctions are normalized such that $\int_0^1\bar{\boldsymbol{v}}_k^2(\bar{r})\text{d}\bar{r}=1$, meaning that the normalized eigenvectors from \eqref{eq:eigenvalue-problem-discretized} must be divided by $\sqrt{1/N}$. It is interesting to note that the smallest eigenvalue $\beta_1$ is for all practical purposes equal to $1$. Also, we note that the eigenfunctions are not orthogonal as the matrix $\mathbf{E}$ is non-symmetric.

\begin{table}
\begin{centering}
\begin{tabular}{cccc}
\toprule 
\multirow{2}{*}{$k$} & \multicolumn{3}{c}{Number of boundary elements $N$}\tabularnewline
\cmidrule{2-4} \cmidrule{3-4} \cmidrule{4-4} 
 & 100 & 1000 & 10000\tabularnewline
\midrule
\midrule 
1 & 0.998912 & 1.002648 & 1.003018\tabularnewline
\midrule 
2 & 2.551356 & 2.561554 & 2.562539\tabularnewline
\midrule 
3 & 4.111055 & 4.128215 & 4.129803\tabularnewline
\midrule 
4 & 5.671338 & 5.696629 & 5.698843\tabularnewline
\midrule 
5 & 7.230949 & 7.265725 & 7.268573\tabularnewline
\bottomrule
\end{tabular}
\par\end{centering}
\caption{Eigenvalues $\beta_k$ as a function of the number of boundary elements $N$.\label{table:eigen-values}}
\end{table}

\subsection{Universal nucleation radius of Uenishi and Rice for tensile and shear circular rupture instabilities}\label{sec:Uenishi-Rice-extension}

The eigen equation \eqref{eq:eigenvalue-problem-critically-stressed} for the critically stressed limit corresponds to the penny-shaped version of the eigen equation of Uenishi and Rice \cite{Uenishi_Rice_2002}. The nucleation radius \eqref{eq:nucleation-radius-critically-stressed} in the main text, is therefore the nucleation radius of a dynamic instability in the conditions analyzed by Uenishi and Rice \cite{Uenishi_Rice_2002}. In our circular configuration, the tectonic shear loading that drives the quasi-static phase of the rupture must be considered to be unidirectional, locally peaked around $r=0$, and axisymmetric in magnitude. Moreover, this result is not only valid for a mixed-mode (II+III) shear rupture (with $\nu=0$), but also for a cohesive tensile (mode I) crack. In this later case, the nucleation radius is valid for any value of $\nu$ as long as the shear modulus $\mu$ is replaced by $E^\prime/2$, where $E^\prime=E/(1-\nu^2)$ is the plane-strain Young's modulus. Note that similarly to the shear rupture case, here the far-field tensile load driving the quasi-static growth of the mode-I rupture must be also locally peaked at $r=0$ (in order to initiate fracture yielding at the origin) and axisymmetric in magnitude.

\section{Numerical solver for slip-dependent friction}
We calculate the numerical solutions of the slip-weakening model via a fully-implicit boundary-element-based solver with an elasto-plastic-like interfacial constitutive law \cite{Saez_Lecampion_2022,Saez_Lecampion_2023}. The details of the three-dimensional version of the solver were presented by S\'aez \textit{et al} \cite{Saez_Lecampion_2022} for the particular case of Coulomb's friction, whereas the necessary modifications to solve in a more efficient manner for the special case of axisymmetric, circular shear ruptures were presented more recently by S\'aez and Lecampion \cite{Saez_Lecampion_2023}. Here, we extend our axisymmetric solver to a case in which the friction coefficient is an arbitrary function of slip. Such extension is indeed relatively straightforward and requires just changes in the integration of the constitutive interfacial law at the collocation point level, plus deriving the proper consistent tangent operator. We present these two changes in sections \ref{subsec:integration-interface} and \ref{subsec:CTO} respectively, plus some calculation and implementation details in section \ref{subsec:calculation-details}.

\subsection{Integration of the constitutive interfacial law}\label{subsec:integration-interface}

The integration of the constitutive interfacial law in three dimensions was described in section 2.2.3 of S\'aez \textit{et al} \cite{Saez_Lecampion_2022} for the case of Coulomb's friction. With reference to this latter section and following the notation in \cite{Saez_Lecampion_2022}, the extension to slip-dependent friction requires no further modification than just expressing the former constant friction coefficient $f$ as a function of the magnitude of the shear vector of plastic displacement discontinuity at each collocation point, $f(\lVert \boldsymbol{d}_s^p \rVert)$, with $\boldsymbol{d}_s^p=(d_1^p,d_2^p)^\top$, and $d_1^p$ and $d_2^p$ the two shear components of plastic displacement discontinuity at a certain collocation point. The latter are expressed in the local reference system of the triangular boundary elements. Furthermore, in the axisymmetric configuration of interest in this work, the extension is even simpler since the shear part of the displacement discontinuity vector is a scalar (the direction of slip is fixed and known). Hence, the friction coefficient is simply expressed as $f(d_s^p)$, where the subscript `s' denotes the shear component and `p' the plastic part of the displacement discontinuity.

With the previous considerations in mind, one can show that when frictional sliding occurs ($\Delta \gamma>0$), the system of equations (10)-(14) of \cite{Saez_Lecampion_2022} leads to the following implicit equation for the plastic multiplier $\Delta \gamma$,
\begin{equation}\label{eq:plastic-multiplier}
    \Delta \gamma = \frac{|t_s^{\text{trial}}|-f\left(d_s^{p,n}-\Delta \gamma\cdot\text{sgn}\left(t_s^{\text{trial}}\right)\right)t_n^{\text{trial}}}{k_s},
\end{equation}
where $d_s^{p,n}$ is the `plastic' (frictional) slip at the previous time step $n$, $t_s^{\text{trial}}$ and $t_n^{\text{trial}}$ are the shear and normal components of the elastic-trial traction vector $\boldsymbol{t}^{\text{trial}}=-\mathbf{D}\cdot \boldsymbol{d}^{n+1}$ of the elastic predictor-plastic corrector algorithm adopted in \cite{Saez_Lecampion_2022}, with $\boldsymbol{d}^{n+1}$ the total (elastic + plastic) displacement discontinuity vector at the current time step $n+1$ (coming from the global Newton-Raphson scheme that solves the quasi-static elastic equilibrium), and $k_s$ the shear component of the diagonal elastic stiffness matrix $\mathbf{D}$. We recall our adopted geomechanics convention of positive stresses in compression. At a given iteration of the global Newton-Raphson scheme, equation \eqref{eq:plastic-multiplier} is solved at every collocation point via a Newton-Raphson procedure, using as initial guess $\Delta \gamma=0$.

\subsection{The consistent tangent operator}\label{subsec:CTO}

The global Newton–Raphson iterations of the fully-implicit time integration scheme \cite{Saez_Lecampion_2022,Saez_Lecampion_2023} require the calculation of the so-called consistent tangent operator $\mathbf{C}_{TO}$, which depends on the specific constitutive interfacial law under consideration. For the case of Coulomb's friction, the consistent tangent operator has been derived analytically for both fully 3D (see Appendix A in \cite{Saez_Lecampion_2022}) and axisymmetric (see Supplemental Material 2 in \cite{Saez_Lecampion_2023}) cases. Here, we derive the proper axisymmetric operator for the case in which the friction coefficient is an arbitrary function of slip.

Using again the same notation than in \cite{Saez_Lecampion_2022}, we define the tangent operator as $\mathbf{C}_{TO}=-\partial\Delta\boldsymbol{t}^{\prime}/\partial\Delta\boldsymbol{d}$, where $\Delta\boldsymbol{t}^{\prime}$ is the increment of the effective traction vector and $\Delta\boldsymbol{d}$ is the increment of the displacement discontinuity vector. Note that the consistent tangent operator $\mathbf{C}_{TO}$ is a block diagonal matrix of size $2N\times2N$, where $N$ is the number of collocation points, and is composed by squared blocks $\mathbf{C}_{TO}^{i}$ of size $2\times2$, where $i=1,...,N$ is the collocation point index. Combining equations (10), (11) and (13) from \cite{Saez_Lecampion_2022}, one can obtain $\Delta\boldsymbol{t}^{\prime}$ as a function of $\Delta\boldsymbol{d}$: 
\begin{equation}
    \Delta\boldsymbol{t}^{\prime}=-\mathbf{D}\cdot\left[\Delta\boldsymbol{d}+\Delta\gamma\left(\Delta \boldsymbol{d}\right)\{\text{sgn}\left(t_s^\text{trial}\right),0\}^\top\right].
\end{equation}
Differentiation of the latter expression with respect to $\Delta\boldsymbol{d}$ leads to the following expression for the squared blocks that composed the tangent operator:
\begin{equation}\label{eq:CTO}
    \mathbf{C}_{TO}^{i}=\mathbf{D}-\mathbf{C}_{TO}^p\;\text{, with}\quad
    \mathbf{C}_{TO}^p=- \begin{pmatrix} k_s \text{sgn}\left(t_s^\text{trial}\right)  \\ 0 \end{pmatrix} \otimes \begin{pmatrix} \partial\Delta\gamma/\partial\Delta d_s \\ \partial\Delta\gamma/\partial\Delta d_n \end{pmatrix},
\end{equation}
where $\mathbf{C}_{TO}^p$ is the plastic part of the tangent operator and $\otimes$ is the tensor product. Note that if $\Delta\gamma=0$, that is, if the collocation point state is elastic or, in other words, no frictional slip occurs, then $\mathbf{C}_{TO}^p$ is a null matrix, and $\mathbf{C}_{TO}^{i}=\mathbf{D}$.

At this point, we just need the partial derivatives of the plastic multiplier $\Delta\gamma$ with respect to $\Delta \boldsymbol{d}$ to obtain the consistent tangent operator. To do so, we consider the incremental form of the consistency condition (see Appendix A in \cite{Saez_Lecampion_2022}), which in the case of slip-dependent friction $f(d_s^p)$ reads
\begin{equation}
    \frac{\partial \mathcal{F}}{\partial \boldsymbol{t}^\prime}\cdot\Delta \boldsymbol{t}^\prime+\frac{\partial \mathcal{F}}{\partial d_s^p}\cdot \Delta d_s^p=0.
\end{equation}
Using the previous equation in combination with equations (10), (11) and (13) in \cite{Saez_Lecampion_2022}, one obtains $\Delta \gamma$ as a function of $\Delta \boldsymbol{d}$, to finally calculate the sought partial derivatives
\begin{equation}
    \frac{\partial \Delta\gamma}{\partial \Delta d_s}=\frac{k_s \text{sgn}\left(t_s^\text{trial}\right)}{A},\;\text{and}\quad
    \frac{\partial \Delta\gamma}{\partial \Delta d_n}=-\frac{f\left(d_s^p\right) k_n}{A},
\end{equation}
where $A=f^\prime\left(d_s^p\right)t_n^\text{trial} \text{sgn}\left(t_s^\text{trial}\right)-k_s$, with $f^\prime$ the first derivative of the arbitrary function describing the dependence of friction on slip. Note that if the friction coefficient $f$ is constant, we effectively recover the axisymmetric consistent tangent operator for Coulomb's friction presented in Supplemental Material 2 of \cite{Saez_Lecampion_2023}.

\subsection{Some calculation and implementation details}\label{subsec:calculation-details}

We use the adaptive time-stepping scheme based on the rupture speed described in Supplemental Material 2 of \cite{Saez_Lecampion_2023}. The parameter $\beta$ that controls the number of elements that the front advances during one time step is fixed for most simulations as $2.5$, which results in a
front advancement of 2 to 3 elements per time step. To resolve properly the cohesive zone, we consider no less than $100$ elements covering the elasto-frictional length scale $R_w$. Verification tests for the numerical solver in the case of a constant friction coefficient were performed in \cite{Saez_Lecampion_2023}. Here, the solver is further verified for the slip-dependent friction case throughout the systematic match between the numerical solutions and the analytical asymptotic and approximate solutions derived in the main text for the different stages and regimes of the problem. With regard to numerical convergence, we consider that our Newton–Raphson scheme employed to solve every backward Euler time step converges when the relative increment of the L$^2$ norm of the displacement discontinuity vector (our primary unknown) between two consecutive iterations falls below $10^{-4}$. On the other hand, the tangent mechanical system at each Newton-Raphson iteration is solved using a biconjugate gradient stabilized iterative solver (BiCGSTAB) with a tolerance set to $10^{-4}$.

\section{A note on the line-source approximation of the fluid injection and the marginally pressurized limit}\label{appendix:line-source}

\begin{figure}
    \centering
    \includegraphics[width=15cm]{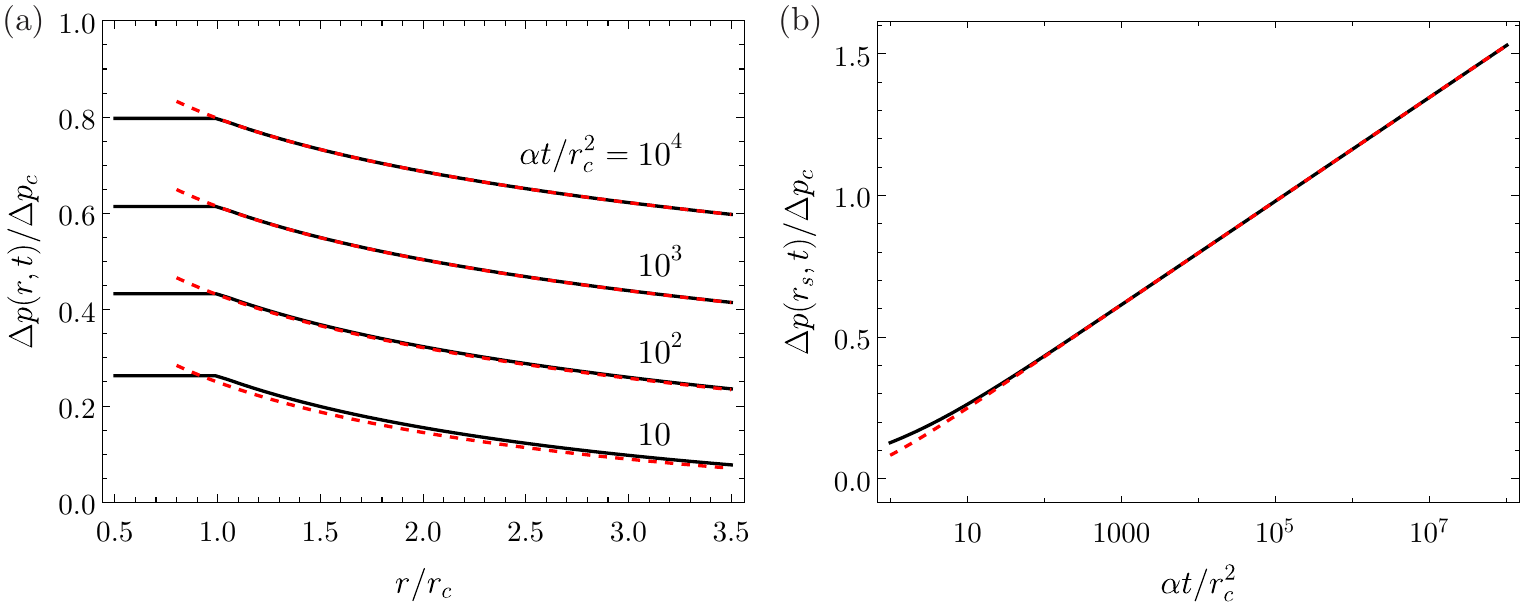}
    \caption{Finite source (solid black) versus line-source approximation (dashed red). (a) Spatial profile of normalized overpressure. (b) Normalized temporal evolution of the overpressure at the fluid source.}
    \label{fig:line-source-approximation}
\end{figure}

In our model, we idealize the fluid injection as a line source. Such approximation is of course valid for times $t\gg r_s^2/\alpha$, where $r_s$ is the characteristic size of the actual fluid source. This is graphically shown in figure \ref{fig:line-source-approximation}a, where the line-source approximation is compared to the solution for a finite circular source of radius $r_s$. The latter is calculated from the known solution in the Laplace domain (section 13.5, eq. 16, \cite{Carslaw_Jaeger_1959}) that we then invert numerically using the Stehfest's method \cite{Stehfest_1970}. Figure \ref{fig:line-source-approximation}a shows clearly how at large times the line-source and finite-source solutions become asymptotically equal at distances $r\geq r_s$. In particular, the overpressure at the fluid source can be approximated at large times by simply evaluating $\Delta p(r,t)$ (equation \eqref{eq:p-solution}) at $r=r_s$. By doing so, the argument of the exponential integral function is very small, $r_s^2/4\alpha t\ll1$, and the overpressure at the fluid source can be asymptotically approximated as
\begin{equation}\label{eq:log-behavior-line-source}
    \Delta p(r=r_s,t) \approx \frac{\Delta p_c}{4\pi} \left(-\gamma-\text{ln}\left(\frac{r_s^2}{4\alpha t}\right) \right),
\end{equation}
where $\gamma=0.577216...$ is the Euler-Mascheroni's constant.

Equation \eqref{eq:log-behavior-line-source} indicates that the overpressure at the fluid source increases logarithmically with time. This is further displayed in figure \ref{fig:line-source-approximation}b, where the temporal evolution of $\Delta p(r_s,t)$ is plotted for both the line-source and finite-source solutions. From this figure, we observe that the line-source approximation is already quite accurate for times $\alpha t/r_s^2\gtrapprox10$. Furthermore, figure \ref{fig:line-source-approximation}b shows that the characteristic overpressure $\Delta p_c$ (equation \eqref{eq:pc-definition}) is in the order of magnitude of the overpressure at the fluid source for a wide range of practically relevant times. Consider, for instance, the case of geo-energy applications where fluid injections are conducted through a wellbore of radius $r_s\sim 10$ cm. By assuming plausible values of hydraulic diffusivity in the range $10^{-5}$ to $1$ m$^{2}$/s, the characteristic time $r_s^2/\alpha$ takes values between $1000$ down to $0.01$ seconds which are much smaller than typical fluid injection duration in geo-energy applications. The large time limit is therefore commonly satisfied.

Note that we had already introduced $\Delta p_c$ in a previous work \cite{Saez_Lecampion_2023} with the purpose of defining the marginally pressurized limit in a form that is more convenient for practical applications than in \cite{Saez_Lecampion_2022}. We recall that the marginally pressurized limit is defined by the condition that the overpressure at the fluid source $\Delta p_w$ is just sufficient to activate fault slip, $f_p\Delta p_w\approx f_p\sigma_0^\prime-\tau_0$. As we have seen, we can approximate $\Delta p_w$ quite well through a line source, yet its magnitude is not constant but rather increases with time. This increase is nevertheless logarithmically slow, so one could think in approximating $\Delta p_w$ as constant and equal to the characteristic overpressure $\Delta p_c$. Indeed, the pre-factor in the order-of-magnitude relation $\Delta p_w \sim \Delta p_c$ is quite close to unity over a wide range of times (see figure \ref{fig:line-source-approximation}b). We therefore enforce $\Delta p_w \approx \Delta p_c$ with the aim of defining the marginally pressurized limit in the more practically convenient way: $f_p\Delta p_c\approx f_p\sigma_0^\prime-\tau_0$. In this way, we essentially avoid introducing the length scale of the fluid source $r_s$ into the problem that, we think, would unnecessary complexify the model and its practical applications. This subtle `assumption' is all over the main text. Moreover, because the so-called intensity of the injection is $\Delta p_*=\Delta p_c/4\pi$, the factor $4\pi$ is usually approximated by $10$.

\end{appendices}

\printbibliography

\end{document}